\documentclass[12pt,preprint]{aastex}

  \slugcomment{\today\ (submitted to ApJ November 28, 2007)}
  \shorttitle{PAH Variations Inside and Among Galaxies}
  \shortauthors{Galliano et al.}

\usepackage{color}
\definecolor{grey}{rgb}{0.5,0.5,0.5}

\newcommand{\sms}[1]{{\mbox{{\scriptsize #1}}}}

\newcommand{\ddiff}{{\;\rm d}}
\newcommand{\E}[1]{\times10^{#1}}

\newcommand{\refeq}[1]{Eq.~(\ref{#1})}

\newcommand{\reftab}[1]{Table~\ref{#1}}
\newcommand{\reffig}[1]{Fig.~\ref{#1}}
\newcommand{\reffigs}[1]{Figs.~\ref{#1}}

\newcommand{\refS}[1]{\S\ref{#1}}
\newcommand{\refapp}[1]{App.~\ref{#1}}

\newcounter{textlistctr}
\newcommand{\thetextlist}{, }
\newcommand{\textlist}[1]
            {\setcounter{textlistctr}{1}
             \renewcommand{\thetextlist}
             {{\it (\roman{textlistctr})}\stepcounter{textlistctr}}#1
              }

\newcounter{obsrefctr}
\newcommand{\obsref}[1]
           {\refstepcounter{obsrefctr}[\arabic{obsrefctr}]\label{#1}}
\newcommand{\refobs}[1]{[\ref{#1}]}

\newcommand{\ncode}[1]{{\tt #1}}

\newcommand{\peg}{\ncode{P\'EGASE}}

\newcommand{\mic}{\;\mu {\rm m}}

\newcommand{\upfov}{''{\rm pixel}^{-1}}

\newcommand{\ariii}{Ar$\,${\sc iii}}
\newcommand{\arii}{Ar$\,${\sc ii}}
\newcommand{\ci}{C$\,${\sc i}}
\newcommand{\cii}{C$\,${\sc ii}}

\newcommand{\hii}{H$\,${\sc ii}}

\newcommand{\neiii}{Ne$\,${\sc iii}}
\newcommand{\neii}{Ne$\,${\sc ii}}

\newcommand{\oi}{O$\,${\sc i}}

\newcommand{\siv}{S$\,${\sc iv}}
\newcommand{\siII}{Si$\,${\sc ii}}
\newcommand{\hmol}{H$_2$}

\newcommand{\ariiiline}{[\ariii]$_{8.99\mu m}$}
\newcommand{\ariiline}{[\arii]$_{6.98\mu m}$}
\newcommand{\ciiline}{[\cii]$_{158\mu m}$}

\newcommand{\neiiiline}{[\neiii]$_{15.56\mu m}$}
\newcommand{\neiiline}{[\neii]$_{12.81\mu m}$}

\newcommand{\oiline}{[\oi]$_{63\mu m}$}

\newcommand{\sivline}{[\siv]$_{10.51\mu m}$}
\newcommand{\siIIline}{[\siII]$_{34.82\mu m}$}
\newcommand{\COio}{$^{12}$CO(J$=$1$\rightarrow$0)$_{2.6mm}$}
\newcommand{\hmoloo}{\hmol$\,$0-0}

\newcommand{\hmolooiii}{\hmoloo$\,\rm S(3)_{9.7\mu m}$}

\newcommand{\arp}[1]{Arp$\;$#1}
\newcommand{\haro}[1]{Haro$\;$#1}
\newcommand{\IC}[1]{IC$\;$#1}
\newcommand{\IR}[1]{IR$\;$#1}

\newcommand{\M}[1]{M$\;$#1}
\newcommand{\mrk}[1]{Mrk$\;$#1}
\newcommand{\ngc}[1]{NGC$\;$#1}

\newcommand{\um}[1]{UM$\;$#1}

\newcommand{\orb}{Orion bar}
\newcommand{\smcb}{SMC$\;$B1$\#$1}
\newcommand{\smcn}{SMC$\;$N$\;$66}
\newcommand{\xxxdor}{30$\;$Doradus}
\newcommand{\cenA}{Centaurus$\;$A}
\newcommand{\hen}{He$\;$2-10}

\newcommand{\izw}{I$\;$Zw$\;$18}
\newcommand{\iizw}{II$\;$Zw$\;$40}
\newcommand{\sbs}{SBS$\;$0335-052}

\newcommand{\iras}{{\it IRAS}}
\newcommand{\iso}{{\it ISO}}
\newcommand{\isoST}{{\it Infrared Space Observatory}}
\newcommand{\spitz}{{\it Spitzer}}
\newcommand{\spitzST}{{\it Spitzer Space Telescope}}

\newcommand{\sofia}{{\it Sofia}}
\newcommand{\jwst}{{\it JWST}}

\newcommand{\akari}{{\it AKARI}}

\newcommand{\irs}{{\it Spitzer}/IRS}
\newcommand{\isocam}{{\it ISO}/CAM}
\newcommand{\isosws}{{\it ISO}/SWS}

\newcommand{\irasi}{\iras$_\sms{12$\mu m$}$}
\newcommand{\irasii}{\iras$_\sms{25$\mu m$}$}

\newcommand{\IRACiv}{IRAC$_\sms{8$\mu m$}$}

\newcommand{\MIPSi}{MIPS$_\sms{24$\mu m$}$}

  \newcommand{\frmet}{{\it Lorentzian method}}
  \newcommand{\nlmet}{{\it Spline method}}

  \newcommand{\ipah}[1]{{\rm I}_{#1}}
  \newcommand{\iPah}{{\rm I}_\sms{PAH}}
  \newcommand{\icont}{{\rm I}_\sms{cont}}
  
  \graphicspath{{.}}
  \DeclareGraphicsExtensions{.eps}

\begin{document}

   \title{VARIATIONS OF THE MID-IR AROMATIC FEATURES
          INSIDE AND AMONG GALAXIES}

  \author{Fr\'ed\'eric Galliano}
    \affil{Observational Cosmology Lab., Code 665,
           NASA Goddard Space Flight Center, 
           Greenbelt MD 20910, USA \\
           Department of Astronomy, University of Maryland, 
           College Park, MD 20742, USA}
    \email{galliano@astro.umd.edu}
  \author{Suzanne C. Madden}
    \affil{Service d'Astrophysique, L'Orme des Merisiers, CEA/Saclay,
           91191 Gif-sur-Yvette, France}
  \author{Alexander G. G. M. Tielens}
    \affil{NASA Ames Research Center, Mail Stop 245-3, Moffett Field, 
           CA 94035, USA}
  \author{Els Peeters}
    \affil{NASA Ames Research Center, Mail Stop 245-6, Moffett Field, 
           CA 94035, USA \\
    SETI Intstitute, 515 N. Whisman Rd, Mountain View, CA 94043, USA \\
    and Physics \& Astronomy Dept., University of Western Ontario,
        PAB 213, London ON N6A 3K7, Canada}
  \and
  \author{Anthony P. Jones}
    \affil{Institut d'Astrophysique Spatiale, Universit\'e de Paris XI,
           91405 Orsay, France}

  \begin{abstract}
    We present the results of a systematic study of mid-IR spectra of
    Galactic regions, Magellanic \hii\ regions, and galaxies of various
    types (dwarf, spiral, starburst), observed by the satellites \iso\ and
    \spitz.
    We study the relative variations of the 6.2, 7.7, 8.6 and $11.3\mic$
    features inside spatially resolved objects (such as \M{82}, \M{51}, \xxxdor,
    \M{17} and the \orb), as well as among 90 integrated spectra of 50 objects.
    Our main results are that the 6.2, 7.7 and $8.6\mic$ bands are
    essentially tied together, while the ratios between these bands and the 
    $11.3\mic$ band varies by one order of magnitude.
    This implies that the properties of the PAHs are remarkably universal 
    throughout our sample, and that the relative variations of the band ratios
    are mainly controled by the fraction of ionized PAHs.
    In particular, we show that we can rule out both the modification 
    of the PAH size distribution, and the mid-infrared extinction, as an 
    explanation of these variations.
    Using a few well-studied Galactic regions (including the spectral 
    image of the \orb), we give an empirical relation between the 
    $\ipah{6.2}/\ipah{11.3}$ ratio and the ionization/recombination ratio 
    $G_0/n_e\times\sqrt{T_\sms{gas}}$, therefore providing a useful
    quantitative diagnostic tool of the physical conditions in the regions
    where the PAH emission originates.
    Finally, we discuss the physical interpretation of the 
    $\ipah{6.2}/\ipah{11.3}$ ratio, on galactic size scales.
  \end{abstract}

  \keywords{dust --
            HII regions -- 
            ISM: structure --
            galaxies: dwarf, starburst --
            infrared: general
            }


\section{INTRODUCTION}
\label{sec:intro}

The reprocessing of stellar light by dust in the infrared (IR) is widely used
to probe embedded star formation. 
However, in the absence of other constraints, the physical properties which 
can usually be derived from an
almost-featureless grain continuum emission is limited to global quantities,
such as the dust mass and its average temperature.
The ubiquity of numerous mid-IR aromatic features, in a wide variety of 
astrophysical objects and environments, potentially provides more articulate 
diagnostics of the physical conditions.
Indeed, these features dominate the mid-IR spectra of evolved stars 
\citep[{e.g.}][]{blommaert05,kraemer06}, the cool ISM 
\citep[{e.g.}][]{abergel05,flagey06,povich07}, 
as well as whole galaxies \citep[{e.g.}][]{verma05,smith07}, that have been 
extensively observed by the \isoST\ (\iso), and are currently investigated 
with a higher sensitivity by the \spitzST.
In our Galaxy, one third of the stellar light is reprocessed by dust,
while this fraction can go up to $99\%$ and higher, in starburst 
galaxies.
At solar metallicity, roughly $15\%$ of the cooling is radiated through the 
most powerful mid-IR bands, centered at 3.3, 6.2, 7.7, 11.3 and 12.7$\mic$.

Historically, these emission features were attributed to very small grains 
($\simeq10\,$\AA),
transiently heated by single photon absorption, in order to account for
the independence of the color temperature with the distance from the 
illuminating star, in several reflection nebulae \citep{sellgren84}.
In parallel, the central wavelengths of these bands were recognized to coincide
with the vibrational modes of aromatic material \citep{duley81}.
These features are now commonly attributed to the molecular modes of 
Polycyclic Aromatic Hydrocarbons \citep[hereafter PAHs;][]{leger84,allamandola85,allamandola89}, which are planar molecules made of 
$\simeq10$ to 1000 carbon atoms, excited primarily by ultraviolet (UV) 
photons.
With silicate and carbon grains, PAHs are a main component of dust models
\citep{desert90,dwek97,draine01,zubko04}.
Their absorption efficiency has been modeled using astrophysical 
observations, laboratory measurements and quantum theory
\citep[in particular][]{desert90,joblin92,verstraete01,li01,mattioda05,mattioda05b,draine07,malloci07}.
In addition to being major radiative coolants of the interstellar medium (ISM),
PAHs are responsible for most of the photoelectric heating of 
the gas in photodissociation regions (hereafter PDRs) and the neutral 
interstellar medium, due to their high cumulative surface area 
\citep[{e.g.}][]{bakes94,hollenbach97}.
For the same reason, they probably play an important role in grain surface
chemistry \citep[{e.g.}][]{tielens87}.
In our Galaxy, they contain $\simeq15-20\%$ of the depleted carbon
\citep[with solar abundance constraints]{zubko04}.
As a consequence, they are believed to be part of the interstellar carbon
condensation chain \citep{cherchneff00,dartois05}.

From an extragalactic point of view, the luminosity of the $6.2\mic$ feature
can be used as a tracer of star formation \citep{peeters04}.
However, this tracer is biased by global parameters such as the ISM metallicity.
Indeed, PAHs are underabundant in low-metallicity galaxies 
\citep[{e.g.}][]{galliano03,galliano05,galliano08a,draine07b}.
There is a general correlation between the PAH-to-continuum intensity ratio 
and the ISM metallicity \citep{madden06,wu06,ohalloran06,smith07}, and 
consequently between the \IRACiv/\MIPSi\ broadband ratio and the metallicity 
\citep{engelbracht05}.
The origin of this trend has been attributed to radiative and mechanical
destruction mechanisms by \citet{madden06} and \citet{ohalloran06} respectively.
Conversely, from the detailed modeling of the spectral energy distribution 
(SED) of nearby galaxies, \citet{galliano08a} showed that the PAH-to-gas mass
ratio at different metallicities coincides with the relative amount of 
carbon dust condensed in the envelopes of low-mass stars, during the
Asymptotic Giant Branch phase (AGB).
This study suggests that PAHs are injected into the ISM by their 
progenitors, the AGB stars, several hundreds of Myr after the beginning of 
the star formation, when the gas has already been enriched by more massive 
stars.
This delay corresponds to the time needed for AGB stars to evolve off the 
main sequence.
Therefore, the delayed injection of AGB-condensed carbon dust into the
ISM offers a natural explanation for the paucity of PAHs in low-metallicity 
environments.

From a cosmological point of view, the large luminosity in these IR emission 
features coupled with the high sensitivity of \spitz\ has allowed 
the detection of PAHs in distant luminous infrared galaxies
out to redshift $z\simeq2$ \cite[{e.g.}][]{elbaz05,yan05,houck05}.
Hence, understanding what controls the properties of the aromatic features 
on large scales is required, in order to properly interpret broadband surveys.

The detailed characteristics of the mid-IR features, such as their shape, their
central wavelength or the intensity ratio between the different bands are 
known to vary \citep[see][]{peeters04b}.
Such variations are essentially due to modifications of the molecular structure
of the PAHs, in different astrophysical environments.
In particular, since each feature is attributed to a given vibrational 
mode, the ratio between these features will vary with quantities such as the
charge, the hydrogenation or the size and shape of the molecule.
The $3.3\mic$ PAH band arises from the radiative relaxation of CH 
stretching modes, while the $11.3\mic$ feature originates in the CH 
out-of-plane bending modes;
CC stretching modes are responsible for the features between 6 and $9\mic$;
CH in plane bending excitation produces part of the $8.6\mic$ band.
Now, laboratory studies and quantum calculations shows that the CC modes are
instrinsically weak in neutral PAHs, and become stronger when the PAHs are
ionized \citep{langhoff96,allamandola99,bauschlicher02,kim02}.
Therefore, the 6 to $9\mic$ bands will be much more intense for a PAH$^+$
than for a PAH$^0$, while it will be the opposite for the 3.3 and 
$11.3\mic$.
Consequently, the ratios between the CC and the CH feature intensities depend
on the charge of the PAHs, which is directly related to the physical 
conditions (e.g.\ intensity of the ionizing radiation field, electron 
density, etc.) in the environment where the emission is originating.

Evidence of variations between features in different astrophysical environments
have been reported by many authors.
For example, \citet{joblin96} showed that the $\ipah{8.6}/\ipah{11.3}$ ratio 
decreases with increasing distance from the exciting star, in the 
reflection nebulae \ngc{1333} 
--~where $\ipah{\lambda}$ is the integrated intensity of the feature centered 
at $\lambda\mic$.
\citet{hony01} found a good correlation between the 3.3 and $11.3\mic$ CH bands,
in a sample of Galactic \hii\ regions, YSOs, and evolved stars,
while they reported variations of $\ipah{6.2}/\ipah{11.3}$ by a factor of 5.
The observations of Galactic and Magellanic \hii\ regions, presented by 
\citet{vermeij02}, indicate that the ratios $\ipah{6.2}/\ipah{11.3}$, 
$\ipah{7.7}/\ipah{11.3}$ and $\ipah{8.6}/\ipah{11.3}$ are correlated.
Furthermore, they suggest a segregation between the values of these ratios 
in the Milky Way and those in the Magellanic Clouds.
\citet{bregman05} studied the variation of $\ipah{7.7}/\ipah{11.3}$ in three
reflection nebulae.
Assuming that this variation is controlled by the charge of the PAHs, they could
relate this band ratio to the ratio $G_0/n_e$ between the integrated intensity
of the UV field, $G_0$, and the electron density, $n_e$.
Similarly, \citet{compiegne07}, studying the detailed variations
of the mid-IR spectrum in the Horsehead nebula, attributed the high relative
strength of the $\ipah{11.3}$ feature to a high fraction of neutral 
PAHs, due
to the high ambient electron density.
On the contrary, \citet{smith07} studied the variation of 
$\ipah{7.7}/\ipah{11.3}$ coming from the nuclear regions of the SINGS legacy 
program galaxies.
They find that this ratio is relatively constant among pure starbursts, but
varies by a factor of 5 among galaxies having a weak AGN.
They interpret this effect as a selective destruction of the smallest PAHs
by the hard radiation arising from the accretion disk, ruling out the 
explanation in terms of ionization of the molecules, in these particular
environments.
This interpretation is also supported by the high $11.3\mic$ and the
weakness of the $3.3\mic$ band in the \akari\ spectrum of the giant
elliptical galaxy \ngc{1316} \citep{kaneda07}.

The previous considerations stress the diversity of the possible interpretation
of the mid-IR feature variations in galaxies.
We need to identify the main physical processes controling the PAH bands, if
we are to use them as diagnostic tools.
This is the aim of this paper.
It presents a quantitative analysis of mid-IR spectra (\iso\ and \spitz) of 
Galactic regions, low-metallicity dwarf galaxies, quiescent spirals, starbursts
and AGNs.
To achieve our goal, we focus on identifying the main trends between the 
various PAH features, at different spatial scales and in different environments.
We then link these variations to the physical conditions inside the studied 
region.
Preliminary results of this study were published in 
\citet{galliano04,galliano07c}.

The paper is organized as follows.
In \refS{sec:obs}, we present our sample and the data reduction.
We discuss the spectral decomposition of the mid-IR spectrum, in 
\refS{sec:decomp}.
The results of this decomposition are presented in \refS{sec:correl}; 
we study the various trends between the band ratios within galaxies, and among
different types of environment.
Then, in \refS{sec:explanation}, we provide a physical interpretation of these 
trends, when the structure of the ISM is resolved, and when it is not.
Finally, we summarize our conclusions in \refS{sec:concl}.


\section{OBSERVATIONS AND DATA REDUCTION}
\label{sec:obs}

  \subsection{The Sample}
  \label{sec:source}

In order to systematically study the properties of the mid-IR aromatic 
features, we include in our sample a wide variety of Galactic regions
and galaxies, covering a large range of metallicities and star formation 
activities.
We merge the \iso\ samples of starbursts and AGNs presented by 
\citet{laurent00}, spirals by \citet{roussel01}, 
Magellanic regions and dwarfs by \citet{madden06}, 
and complement them by low-metallicity sources observed with \spitz.
We add to this sample, the \isocam\ spectra of several Galactic regions, like 
\M{17} \citep{cesarsky96b}, \ngc{2023} \citep{abergel02}, \ngc{7027} 
\citep{persi99} and the \orb\ \citep{cesarsky00}, and the \isosws\ spectra 
($2.5-45\mic$)
of several compact \hii\ regions published by \citet{peeters02}.
The \smcb\ spectrum is the one presented by \citet{reach00}.

The global properties of the selected sources are presented in 
\reftab{tab:source}.
If relevant, the distances were homogenised to 
$H_\sms{0}=71\; h^{-2}\,\rm km\,s^{-1}\,Mpc^{-1}$.
\begin{deluxetable}{lrrrrrrrr}
  \tabletypesize{\footnotesize}
  \tablecolumns{9}
  \tablewidth{0pc}
  \tablecaption{General properties of the sample.}
  \tablehead{
    \colhead{Name} & \colhead{R.A.}    & \colhead{Dec.}                 & 
      \colhead{Aperture} & \colhead{Mid-IR}  & \colhead{Distance}       &
      \multicolumn{2}{c}{$12+\log({\rm O/H})$} & \colhead{Category}   \\
    \colhead{}    & \colhead{(J2000)} & \colhead{(J2000)}              & 
      \colhead{} &  \colhead{spectrograph}  & \colhead{(Mpc)}       & 
                            & \colhead{[ref.]} & }
  \startdata
    \haro{11}     & $00^h36^m52\fs5$  & $-33\degr33\arcmin19\arcsec$   &
      $10\arcsec$   &   \irs              & 92                       &
      7.9           & \obsref{ZHaro11}            & Dwarf \\
    \smcb         & $00^h45^m33\fs0$  & $-73\degr18\arcmin46\arcsec$   &
      $84\arcsec-36\arcsec$  & \isocam                 & 0.06          & 
      8.0           & \obsref{ZMC}                & Magellanic \\
    \ngc{253}     & $00^h47^m32\fs9$  & $-25\degr17\arcmin18\arcsec$   &
      $40\arcsec$ & \isocam                  & 3.3                      & 
      9.0           & \obsref{ZN253}              & SB/AGN \\
    \ngc{253}~p    & $00^h47^m32\fs9$  & $-25\degr17\arcmin18\arcsec$   &
      $10\arcsec$ & \isocam                  & 3.3                      & 
      9.0           & \refobs{ZN253}              & SB/AGN \\
    \ngc{253}~e    & $00^h47^m32\fs9$  & $-25\degr17\arcmin18\arcsec$   &
      $40\arcsec-10\arcsec$ & \isocam                  & 3.3                & 
      9.0           & \refobs{ZN253}              & SB/AGN \\
    \smcn          & $00^h59^m02\fs0$  & $-72\degr10\arcmin36\arcsec$   &
      $120\arcsec$ & \isocam                     & 0.06                     & 
      8.0           & \refobs{ZMC}               & Magellanic \\
    \ngc{520}      & $01^h24^m34\fs9$  & $+03\degr47\arcmin31\arcsec$   &
      $30\arcsec$ & \isocam                     & 27                       & 
      \nodata       &                             & Dwarf \\
    \ngc{613}      & $01^h34^m17\fs5$  & $-29\degr24\arcmin58\arcsec$   &
      $60\arcsec$ & \isocam                     & 19                       &
      9.2           & \obsref{ZN613}              & Spiral \\
    \ngc{613}~p      & $01^h34^m17\fs5$  & $-29\degr24\arcmin58\arcsec$   &
      $20\arcsec$ & \isocam                     & 19                       &
      9.2           & \refobs{ZN613}              & Spiral \\
    \ngc{891}      & $02^h22^m33\fs4$  & $+42\degr20\arcmin57\arcsec$   &
      $200\arcsec$ & \isocam                    & 9.6                      & 
      8.9           & \obsref{ZN891}              & Spiral \\
    \ngc{1068}     & $02^h42^m40\fs6$  & $-00\degr00\arcmin47\arcsec$   &
      $60\arcsec$ & \isocam                     & 15                       &
      9.0           & \obsref{ZN1068}             & SB/AGN \\
    \ngc{1068}~p     & $02^h42^m40\fs6$  & $-00\degr00\arcmin47\arcsec$   &
      $20\arcsec$ & \isocam                     & 15                       &
      9.0           & \refobs{ZN1068}             & SB/AGN \\
    \ngc{1068}~e     & $02^h42^m40\fs6$  & $-00\degr00\arcmin47\arcsec$   &
      $60\arcsec-20\arcsec$ & \isocam                 & 15                       &
      9.0           & \refobs{ZN1068}             & SB/AGN \\
    \ngc{1097}     & $02^h46^m19\fs1$  & $-30\degr16\arcmin28\arcsec$   &
      $100\arcsec$ & \isocam                    & 12                       &
      9.0           & \obsref{ZN1097}             & Spiral \\
    \ngc{1097}~p     & $02^h46^m19\fs1$  & $-30\degr16\arcmin28\arcsec$   &
      $40\arcsec$ & \isocam                    & 12                       &
      9.0           & \refobs{ZN1097}             & Spiral \\
    \ngc{1140}     & $02^h54^m33\fs5$  & $-10\degr01\arcmin44\arcsec$   &
      $20\arcsec$ & \isocam                    & 25                       & 
      8.0           & \obsref{heckman}            & Dwarf \\
    \ngc{1365}     & $03^h33^m35\fs6$  & $-36\degr08\arcmin23\arcsec$   &
      $100\arcsec$ & \isocam                   & 19                       &
      9.1           & \obsref{ZN1365}             & Spiral \\
    \ngc{1365}~p     & $03^h33^m35\fs6$  & $-36\degr08\arcmin23\arcsec$   &
      $40\arcsec$ & \isocam                   & 19                       &
      9.1           & \refobs{ZN1365}             & Spiral \\
    \ngc{1365}~e     & $03^h33^m35\fs6$  & $-36\degr08\arcmin23\arcsec$   &
      $100\arcsec-40\arcsec$ & \isocam              & 19                       &
      9.1           & \refobs{ZN1365}             & Spiral \\
    \IC{342}       & $03^h46^m49\fs7$  & $+68\degr05\arcmin45\arcsec$   &
      $40\arcsec$ & \isocam                   & 3.8                      & 
      8.9           & \obsref{Zic}                & SB/AGN \\
    \IC{342}~p       & $03^h46^m49\fs7$  & $+68\degr05\arcmin45\arcsec$   &
      $12\arcsec$ & \isocam                   & 3.8                      & 
      8.9           & \refobs{Zic}                & SB/AGN \\
    \IC{342}~e       & $03^h46^m49\fs7$  & $+68\degr05\arcmin45\arcsec$   &
      $40\arcsec-12\arcsec$ & \isocam               & 3.8                      & 
      8.9           & \refobs{Zic}                & SB/AGN \\
    \IC{342}~map     & $03^h46^m49\fs7$  & $+68\degr05\arcmin45\arcsec$   &
      $3\arcsec$       & \isocam               & 3.8                      & 
      8.9           & \refobs{Zic}                & SB/AGN \\
    \ngc{1569}     & $04^h30^m49\fs1$  & $+64\degr50\arcmin52\arcsec$   &
      $120\arcsec$ & \isocam                  & 2.2                      & 
      8.2           & \obsref{ZN1569}             & Dwarf \\
    \ngc{1569}~e     & $04^h30^m49\fs1$  & $+64\degr50\arcmin52\arcsec$   &
      $120\arcsec-12\arcsec$ & \isocam              & 2.2                      & 
      8.2           & \refobs{ZN1569}             & Dwarf \\
    \ngc{1808}     & $05^h07^m42\fs3$  & $-37\degr30\arcmin47\arcsec$   &
      $50\arcsec$ & \isocam                   & 11                       &
      9.1           & \obsref{ZN1808}             & SB/AGN \\
    \orb~D8      & $05^h35^m18\fs2$  & $-05\degr24\arcmin40\arcsec$   &
      $14\arcsec\times20\arcsec$ & \isosws            & 475 pc                 &
      \nodata       &                             & \hii\ region \\ 
    \orb~D5      & $05^h35^m19\fs8$  & $-05\degr25\arcmin10\arcsec$   &
      $14\arcsec\times20\arcsec$ & \isosws            & 475 pc                &
      \nodata       &                             & \hii\ region \\ 
    \orb~D2      & $05^h35^m21\fs4$  & $-05\degr25\arcmin40\arcsec$   &
      $14\arcsec\times20\arcsec$ & \isosws            & 475 pc                &
      \nodata       &                             & \hii\ region \\ 
    \orb           & $5^h35^m20\fs0$  & $-05\degr25\arcmin20\arcsec$   &
      $6\arcsec$ & \isocam            & 475 pc                  &
      \nodata       &                             & PDR \\ 
    \xxxdor        & $05^h38^m34\fs0$  & $-69\degr05\arcmin57\arcsec$   &
      $120\arcsec$ & \isocam                  & 0.05                     &
      8.4           & \refobs{ZMC}             & Magellanic \\
    \xxxdor~p        & $05^h38^m34\fs0$  & $-69\degr05\arcmin57\arcsec$   &
      $40\arcsec$ & \isocam                   & 0.05                     &
      8.4           & \refobs{ZMC}             & Magellanic \\
    \xxxdor~e        & $05^h38^m34\fs0$  & $-69\degr05\arcmin57\arcsec$   &
      $120\arcsec-40\arcsec$ & \isocam              & 0.05                     &
      8.4           & \refobs{ZMC}             & Magellanic \\
    \xxxdor~map      & $05^h38^m34\fs0$  & $-69\degr05\arcmin57\arcsec$   &
      $6\arcsec$ & \isocam              & 0.05                     &
      8.4           & \refobs{ZMC}             & Magellanic \\
    \ngc{2023}      & $05^h41^m38\fs3$  & $-02\degr16\arcmin33\arcsec$   &
      $14\arcsec\times20\arcsec$ & \isosws            & 475 pc                 &
      \nodata       &                             & PDR \\ 
    \iizw          & $05^h55^m42\fs7$  & $+03\degr23\arcmin29\arcsec$   &
      $24\arcsec$ & \isocam                   & 10                       & 
      8.1           & \obsref{Ziizw}              & Dwarf \\
    \hen           & $08^h36^m15\fs2$  & $-26\degr24\arcmin34\arcsec$   &
      $10\arcsec$ & \irs                        & 8.7                      & 
      8.9           & \refobs{ZN1569}             & Dwarf \\
    \M{82}         & $09^h55^m51\fs8$  & $+69\degr40\arcmin46\arcsec$   &
      $90\arcsec$ & \isocam                     & 3.6                      & 
      9.0           & \obsref{ZM82}               & SB/AGN \\
    \M{82}~p         & $09^h55^m51\fs8$  & $+69\degr40\arcmin46\arcsec$   &
      $24\arcsec$ & \isocam                     & 3.6                      & 
      9.0           & \refobs{ZM82}               & SB/AGN \\
    \M{82}~e         & $09^h55^m51\fs8$  & $+69\degr40\arcmin46\arcsec$   &
      $90\arcsec-24\arcsec$ & \isocam                 & 3.6                  & 
      9.0           & \refobs{ZM82}               & SB/AGN \\
    \M{82}~map         & $09^h55^m51\fs8$  & $+69\degr40\arcmin46\arcsec$   &
      $3\arcsec$ & \isocam                 & 3.6                      & 
      9.0           & \refobs{ZM82}               & SB/AGN \\
    \ngc{3256}     & $10^h27^m51\fs1$  & $-43\degr54\arcmin17\arcsec$   &
      $24\arcsec$ & \isocam                     & 37                       &
      8.9           & \obsref{ZN3256}             & SB/AGN \\
    \ngc{3256}~p     & $10^h27^m51\fs1$  & $-43\degr54\arcmin17\arcsec$   &
      $10\arcsec$ & \isocam                     & 37                       &
      8.9           & \refobs{ZN3256}             & SB/AGN \\
    \ngc{3256}~e     & $10^h27^m51\fs1$  & $-43\degr54\arcmin17\arcsec$   &
      $24\arcsec-10\arcsec$ & \isocam                 & 37                    &
      8.9           & \refobs{ZN3256}             & SB/AGN \\
    \mrk{33}       & $10^h32^m31\fs9$  & $+54\degr24\arcmin04\arcsec$   &
      $10\arcsec$ & \irs                        & 20                        & 
      8.4           & \refobs{ZN3256}             & Dwarf \\
    \arp{299}      & $11^h28^m31\fs0$  & $+58\degr33\arcmin39\arcsec$   &
      $24\arcsec$ & \isocam                     & 41                        &
      \nodata         &                             & SB/AGN \\ 
    \um{448}       & $11^h42^m12\fs4$  & $+00\degr20\arcmin03\arcsec$   &
      $10\arcsec$ & \irs                        & 70                       &
      8.0           & \obsref{ZUM448}             & Dwarf \\ 
    \IR{12331}     & $12^h36^m01\fs9$  & $-61\degr51\arcmin04\arcsec$   &
      $14\arcsec\times20\arcsec$ & \isosws            & 4.5 kpc               &
      \nodata       &                             & \hii\ region \\ 
    \ngc{4945}     & $13^h05^m26\fs2$  & $-49\degr28\arcmin15\arcsec$   &
      $60\arcsec$ & \isocam                     & 3.9                      &
      \nodata       &                             & SB/AGN \\
    \ngc{4945}~p     & $13^h05^m26\fs2$  & $-49\degr28\arcmin15\arcsec$   &
      $20\arcsec$ & \isocam                     & 3.9                      &
      \nodata       &                             & SB/AGN \\
    \ngc{4945}~e     & $13^h05^m26\fs2$  & $-49\degr28\arcmin15\arcsec$   &
      $60\arcsec-20\arcsec$ & \isocam                 & 3.9                   &
      \nodata       &                             & SB/AGN \\
    \cenA          & $13^h25^m28\fs0$  & $-43\degr01\arcmin06\arcsec$   &
      $30\arcsec$ & \isocam                     & 3.8                      &
      $\sim9$       & \obsref{ZcenA}              & SB/AGN \\
    \cenA~e          & $13^h25^m28\fs0$  & $-43\degr01\arcmin06\arcsec$   &
      $10\arcsec$ & \isocam                     & 3.8                      &
      $\sim9$       & \refobs{ZcenA}              & SB/AGN \\
    \M{51}         & $13^h29^m52\fs7$  & $+47\degr11\arcmin43\arcsec$   &
      $140\arcsec$ & \isocam                    & 8.4                      &
      8.7           & \obsref{ZM51}               & Spiral \\
    \M{51}~p         & $13^h29^m52\fs7$  & $+47\degr11\arcmin43\arcsec$   &
      $60\arcsec$ & \isocam                     & 8.4                      &
      8.7           & \refobs{ZM51}               & Spiral \\
    \M{51}~e         & $13^h29^m52\fs7$  & $+47\degr11\arcmin43\arcsec$   &
      $140\arcsec-60\arcsec$ & \isocam                & 8.4                   &
      8.7           & \refobs{ZM51}               & Spiral \\
    \M{51}~map         & $13^h29^m52\fs7$  & $+47\degr11\arcmin43\arcsec$   &
      $6\arcsec$ & \isocam                & 8.4                      &
      8.7           & \refobs{ZM51}               & Spiral \\
    \M{83}         & $13^h37^m00\fs7$  & $-29\degr51\arcmin58\arcsec$   &
      $200\arcsec$ & \isocam                    & 4.5                      &
      9.2           & \obsref{ZM83}               & Spiral \\
    \M{83}~p         & $13^h37^m00\fs7$  & $-29\degr51\arcmin58\arcsec$   &
      $50\arcsec$ & \isocam                     & 4.5                      &
      9.2           & \refobs{ZM83}               & Spiral \\
    \M{83}~e         & $13^h37^m00\fs7$  & $-29\degr51\arcmin58\arcsec$   &
      $200\arcsec-50\arcsec$ & \isocam                & 4.5                   &
      9.2           & \refobs{ZM83}               & Spiral \\
    \M{83}~map         & $13^h37^m00\fs7$  & $-29\degr51\arcmin58\arcsec$   &
      $6\arcsec$ & \isocam                & 4.5                      &
      9.2           & \refobs{ZM83}               & Spiral \\
    Circinus       & $14^h13^m09\fs6$  & $-65\degr20\arcmin21\arcsec$   &
      $40\arcsec$ & \isocam                     & 4.0                      & 
      \nodata       &                             & SB/AGN \\
    Circinus~e     & $14^h13^m09\fs6$  & $-65\degr20\arcmin21\arcsec$   &
      $40\arcsec-10\arcsec$ & \isocam                 & 4.0                   & 
      \nodata       &                             & SB/AGN \\
    \arp{220}      & $15^h34^m57\fs2$  & $+23\degr30\arcmin11\arcsec$   &
      $20\arcsec$ & \isocam                     & 73                       & 
      \nodata       &                             & ULIRG \\
    \arp{220}~p    & $15^h34^m57\fs2$  & $+23\degr30\arcmin11\arcsec$   &
      $10\arcsec$ & \isocam                     & 73                       & 
      \nodata       &                             & ULIRG \\
    \IR{15384}     & $15^h42^m17\fs1$  & $-53\degr58\arcmin31\arcsec$   &
      $14\arcsec\times20\arcsec$ & \isosws            & 2.7 kpc               &
      \nodata       &                             & \hii\ region \\ 
    \ngc{6240}     & $16^h52^m58\fs8$  & $+02\degr24\arcmin06\arcsec$   &
      $20\arcsec$ & \isocam                     & 98                       &
      \nodata       &                             & SB/AGN \\
    \ngc{6240}~p   & $16^h52^m58\fs8$  & $+02\degr24\arcmin06\arcsec$   &
      $10\arcsec$ & \isocam                     & 98                       &
      \nodata       &                             & SB/AGN \\
    \M{17}~map     & $18^h20^m22\fs0$  & $-16\degr12\arcmin40\arcsec$   &
      $6\arcsec$ & \isocam            & 1.5 kpc                       &
      \nodata       &                             & PDR \\ 
    \IR{18317}     & $18^h34^m24\fs9$  & $-07\degr54\arcmin47\arcsec$   &
      $14\arcsec\times20\arcsec$ & \isosws            & 4.9 kpc             &
      \nodata       &                             & \hii\ region \\ 
    \ngc{6946}     & $20^h34^m51\fs2$  & $+60\degr09\arcmin17\arcsec$   &
      $140\arcsec$ & \isocam                    & 5.5                      &
      9.1           & \obsref{ZN6946}             & Spiral \\
    \ngc{6946}~p     & $20^h34^m51\fs2$  & $+60\degr09\arcmin17\arcsec$   &
      $40\arcsec$ & \isocam                     & 5.5                      &
      9.1           & \refobs{ZN6946}             & Spiral \\
    \ngc{6946}~e     & $20^h34^m51\fs2$  & $+60\degr09\arcmin17\arcsec$   &
      $140\arcsec-40\arcsec$ & \isocam                & 5.5                   &
      9.1           & \refobs{ZN6946}             & Spiral \\
    \ngc{7027}      & $21^h07^m01\fs7$  & $+42\degr14\arcmin09\arcsec$   &
      $14\arcsec\times20\arcsec$ & \isosws            & 700 pc               &
      \nodata       &                             & PDR \\ 
    \IR{22308}     & $22^h32^m45\fs9$  & $+58\degr28\arcmin21\arcsec$   &
      $14\arcsec\times20\arcsec$ & \isosws            & 5.5 kpc               &
      \nodata       &                             & \hii\ region \\ 
    \IR{23030}     & $23^h05^m10\fs6$  & $+60\degr14\arcmin41\arcsec$   &
      $14\arcsec\times20\arcsec$ & \isosws            & 5.2 kpc             &
      \nodata       &                             & \hii\ region \\ 
    \IR{23133}     & $23^h15^m31\fs4$  & $+61\degr07\arcmin08\arcsec$   &
      $14\arcsec\times20\arcsec$ & \isosws            &  5.5 kpc         &
      \nodata       &                             & \hii\ region \\ 
    \IR{23128}     & $23^h15^m46\fs0$  & $-59\degr03\arcmin17\arcsec$   &
      $20\arcsec$ & \isocam            &  180                        &
      \nodata       &                             & SB/AGN \\ 
    \IR{23128}~p     & $23^h15^m46\fs0$  & $-59\degr03\arcmin17\arcsec$   &
      $10\arcsec$ & \isocam            &  180                        &
      \nodata       &                             & SB/AGN \\ 
    \ngc{7714}     & $23^h36^m14\fs1$  & $+02\degr09\arcmin19\arcsec$   &
      $10\arcsec$ & \irs                        & 37                       & 
      8.5           & \obsref{ZN7714}             & Dwarf \\ 
  \enddata
  \label{tab:source}
  \tablerefs{\refobs{ZHaro11}~\citet{bergvall00};
             \refobs{ZMC}~\citet{dufour82};
             \refobs{ZN253}~\citet{zaritsky94};
             \refobs{ZN613}~\citet{alloin79};
             \refobs{ZN891}~\citet{otte01};
             \refobs{ZN1068}~\citet{dutil99};
             \refobs{ZN1097}~\citet{storchi-bergmann95};
             \refobs{heckman}~\citet{heckman98};
             \refobs{ZN1365}~\citet{roy97};
             \refobs{Zic}~\citet{pilyugin04};
             \refobs{ZN1569}~\citet{kobulnicky97};
             \refobs{ZN1808}~\citet{ravindranath01};
             \refobs{Ziizw}~\citet{perez-montero03};
             \refobs{ZM82}~\citet{boselli02};
             \refobs{ZN3256}~\citet{mas-hesse99};
             \refobs{ZUM448}~\citet{izotov98};
             \refobs{ZcenA}~\citet{schaerer00};
             \refobs{ZM51}~\citet{bresolin04};
             \refobs{ZM83}~\citet{webster83};
             \refobs{ZN6946}~\citet{kobulnicky99b};
             \refobs{ZN7714}~\citet{gonzalez-delgado95}.}
  \tablecomments{The sources are ordered according to their right ascension.
                 For \isocam\ sources, the apertures are circular, centered 
                 on the coordinates listed in the second and third columns.
                 For \isosws\ sources, the beam is $14\arcsec\times20\arcsec$ 
                 (bands 1 and 2).
                 The diameter of the aperture is given in the fourth column.
                 When the diameter has the form 
                 $\theta_1\arcsec-\theta_2\arcsec$, it means that
                 the we subtracted from the flux in the circular
                 aperture 
                 of diameter $\theta_1\arcsec$,
                 the flux in a concentric circular aperture of diameter 
                 $\theta_2\arcsec<\theta_1\arcsec$.
                 The sources which are followed by ``map'' are those for which
                 the signal-to-noise of the spectral-image is good; 
                 in this case, the aperture refers to the pixel field of view.
                 The letter after the name of the source designates the type
                 of aperture:
                 when there is no letter, most of the emission of the object
                 is encompassed in the aperture;
                 ``p'' is for peak emission; ``e'' for extended emission.}
\end{deluxetable}
\clearpage

  \subsection{\isocam\ Data Reduction}
  \label{sec:cam}

Most of the sources in \reftab{tab:source} were observed with \isocam\ 
\citep{cesarsky96cam} on board the \iso\ satellite \citep{kessler96}.
We refer to the work of \citet{madden06} for a detailed description of the 
data reduction applied to all of the \isocam\ data used in this study.
The CVF performed spectral imaging using a $32\times 32$ detector 
array, with a sampling of $3\upfov$ or $6\upfov$ in our cases, from 
$\lambda=5\mic$ to $16.5\mic$ with one pointing of two CVFs, from $\lambda=5$ 
to $9.5\mic$ and from $\lambda=9.0$ to $16.5\mic$.
The spectral resolution increases from $\lambda/\Delta\lambda=35$ to 51 across 
the full spectra.

For the \iso\ data treatment, we used the CAM Interactive Reduction
\citep[CIR, version AUG01;][]{chanial03}. 
The subtraction of the dark currents was performed using the \citet{biviano98}
model which predicts the time evolution for each row of the detector, taking 
into account drifts along each orbit and each revolution. 
We masked the glitches using multi-resolution median filtering 
\citep{starck99} on each block of data after slicing the cube. 
Additional deglitching was performed manually, examining the temporal cut for 
each pixel. 
We corrected the systematic memory effects using the Fouks-Schubert method 
\citep{coulais00}. 
We computed a hybrid flat-field image placing a mask on the source and
computing a flat field outside this mask from the median of the temporal cut 
for each pixel. 
For the pixels which were on-source, the flat-field response was set to the 
corresponding calibration flat-field.
The conversion from Analog to Digital Units to mJy/pixel was performed using the 
standard in-flight calibration data base.
To remove the sky contribution, sources smaller than the array were masked
and, for a given wavelength, the median of the pixels which are off-source 
were subtracted from each pixel. 
For the more extended sources, we subtracted an independently observed 
zodiacal spectrum.
The contribution of this spectrum was a free parameter varied in order to match
the properly sky subtracted fluxes in the LW2 ($6.7\mic$) and LW3 ($14.3\mic$)
broadbands.
The final product is a 3D spectral-image.
We integrated the spectrum using an aperture encompassing the entire 
galaxy, and obtained a 1D spectrum.
When the angular size of the source was larger than the array, we 
scaled the spectrum by matching \irasi\ broadband flux with that derived 
from convolving the spectrum with the \irasi\ bandpass.

  \subsection{\irs\ Spectrum Extraction}
  \label{sec:irs}

Several of the low-metallicity sources in \reftab{tab:source} were not 
observed by \isocam, thus we complemented our database with mid-IR spectra 
from the \irs\ spectrometer \citep{houck04irs} on board the \spitzST\ 
\citep{werner04}, when these data were released.
Among these galaxies, the spectrum of \sbs\ has been published by 
\citet{houck04}, \ngc{7714} by \citet{brandl04},
\ngc{5253} by \citet{beirao06}, \haro{11} and \izw\ by \citet{wu06}.
We refer to these studies for a detailed presentation of their mid-IR 
properties.
We considered only low-resolution data, taken with
the SL (Short-Low) module, from $\lambda=5.2\mic$ to $14.5\mic$, 
and the LL (Long-Low) module, from $\lambda=14.0\mic$ to $38.0\mic$, both 
with a spectral resolution of $\lambda/\Delta\lambda\simeq64-128$.

We retrieved the Basic Calibrated Data (BCD) that have been preprocessed by
the Spitzer Science Center (SSC) data reduction pipeline.
They have been converted to flux density, corrected for stray light and
flatfielded.
The extraction of the spectra from the 2D space/wavelength images was 
performed with the Spectral Modeling, Analysis and Reduction Tool 
\citep[SMART, version 5.5.6;][]{higdon04}.
We first inspected the BCD images and identified the hot pixels which had not 
been masked out by the SSC.
We replaced them by the median of their neighbors.
For each module, one order is on-source and the other is off-source, at 
a time.
Then the positions are switched.
We subtracted the off-source spectrum from that of the source, in order to 
remove the sky emission.
The extraction of the 1D spectrum was performed inside a column whose width 
varies with wavelength. 
We have excluded the bonus order.
Then, the various frames, for each nod position, were coadded.
Since the long wavelength end of the SL module and the short wavelength end 
of the LL module were not systematically overlapping, we finally scaled the SL 
module, in order to obtain a continuous spectrum.
This scaling factor can be as large as $50\%$.
At the time of this publication, the \irs\ data handbook recommends
that the signal to noise ratio not be derived from the uncertainties generated
by the pipeline.
Instead, we adopted the recommended systematic error of $20\%$.
To take into account these statistical variations, we smoothed the spectra
into a $\Delta\lambda\simeq 0.2\mic$ window (4 points), and took the standard 
deviation inside this window as the error.
As we did with the \isocam\ spectra (\refS{sec:cam}), in order
to compensate for the fact that we may be overlooking some extended emission,
we scaled the spectrum to match the \irasi\ and \irasii.
In the case of \iizw, the \irs\ spectrum does not exhibit the PAH 
features that \citet{madden06} detected in the extended emission, since
the \irs\ observations were conducted in staring mode.
Indeed, this \irs\ observation does not encompass the region where the 
tail was detected.
Thus we prefer to use the \isocam\ data for this galaxy.


\section{THE SPECTRAL DECOMPOSITION METHODS}
\label{sec:decomp}

    \subsection{Inventory of the Physical Components}
    \label{sec:descMIR}

We restrain our study to the spectral range $5-16\mic$, since it 
is the overlap between the various data sets we have compiled.
The main physical components contributing at these wavelengths are the 
following.
\begin{itemize}
  \item The ionic gas emits strong fine-structure ionic lines.
    The \neiiiline, \neiiline, \sivline, \ariiiline\ and \ariiline\ are 
    the brightest lines.
  \item Several ubiquitous broad features, originating from the molecular 
    modes of stochastically heated PAH molecules.
    The most prominent of them are centered at 6.2, 7.7, 8.6, 11.3 and 
    $12.7\mic$.
    The \neiiline\ line is blended with the $12.7\mic$ feature.
  \item Very small grains (VSG), fluctuating in temperature
    around a few hundred degrees Kelvin, produce a continuum emission.
    In addition, hot grains in \hii\ regions, at thermal equilibrium with 
    the radiation field, may contribute to this continuum.
  \item The extinction feature at $9.7\mic$, attributed to amorphous
    silicates, can be seen toward the most embedded sources.
  \item Molecular lines are also present, especially the \hmolooiii\ line.
    However, they are much weaker than the other components.
\end{itemize}

\begin{figure}[htbp]
  \centering
  \includegraphics[width=0.48\textwidth]{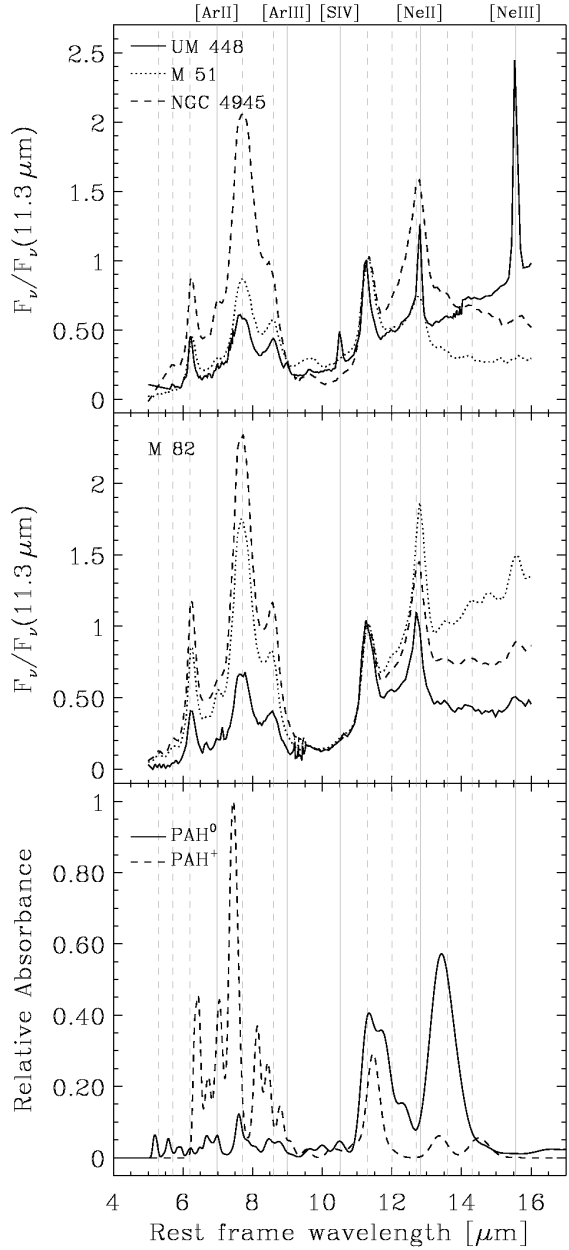}
  \caption{Various spectra normalized to the $11.3\mic$ feature.
           The top panel shows the total integrated spectra of a low-metallicity
           galaxy (\um{448}), a normal spiral (\M{51}), and a LIRG
           (\ngc{4945}).
           The middle panel shows the spectra of three different regions inside
           the starburst galaxy \M{82}.
           The bottom panel shows, for comparison, the absorption 
           coefficient of neutral and cationic PAHs measured in laboratory by 
           \citet{allamandola99}.
           The solid vertical lines mark the wavelengths of the brightest lines,
           and the dashed vertical lines mark the wavelengths of the major 
           bands.
           In this figure and in what follows, $F_\nu(\lambda)$ is the 
           monochromatic flux density at wavelength $\lambda\mic$.}
  \label{fig:spec}
\end{figure}
\reffig{fig:spec} shows the variations of the aromatic feature spectrum among
galaxies (top panel), inside one galaxy (middle panel) and from a theoretical
point of view (bottom panel).
This figure shows that the 6.2, 7.7 and $8.6\mic$ features are
qualitatively tied together, and that most of the variations in the
mid-IR aromatic spectrum is a variation of these three features relative
to the $11.3\mic$ band.
The lower panel of \reffig{fig:spec} suggests that most of these variations
could be explained by a variation of the neutral-to-cationic-PAH ratio.
In what follows, we quantify these variations in our sample, in order to
explore the validity of this explanation.

    \subsection{Measuring the Intensity of the Aromatic Features}
    \label{sec:method}

Measuring the intensity of the aromatic features is uncertain, due to the
intrinsic width ($\Delta\lambda\simeq1\mic$) and the complexity of the band 
profiles.
Indeed, contrary to gas lines, a large fraction of the energy of the PAH bands is radiated in the {\it wings}.
These wings extend far outside of the central wavelength and can be difficult
to reliably disentangle from the underlying continuum emission.
Therefore, a proper extraction of the feature requires an assumed band profile.
However, the actual profile of each band is not known.
It is asymmetric and varies as a function of environment 
\citep[{e.g.}][]{peeters02b,van-diedenhoven04}.
Consequently, several profiles have been used in the literature, motivated by 
different physical arguments.
For example, \citet{boulanger98} proposed that the width of the band is a 
consequence of the continuous redistribution of intramolecular vibrational 
energy, between different excitation levels, at high temperature.
They estimated the transition timescales to be $\lesssim 10^{-13}$~s, and 
showed that the PAH bands were well represented by Lorentzian profiles.
Conversely, a Drude profile was used by \citet{li01}.
Such a profile describes the electric permeability in a solid or a large 
molecule.
Finally, numerous studies simply measure the tip of the aromatic band, and 
assumes that the total flux radiated in the band scales with this quantity.

\begin{figure*}[htbp]
  \centering
  \begin{tabular}{cc}
    \includegraphics[width=0.48\textwidth]{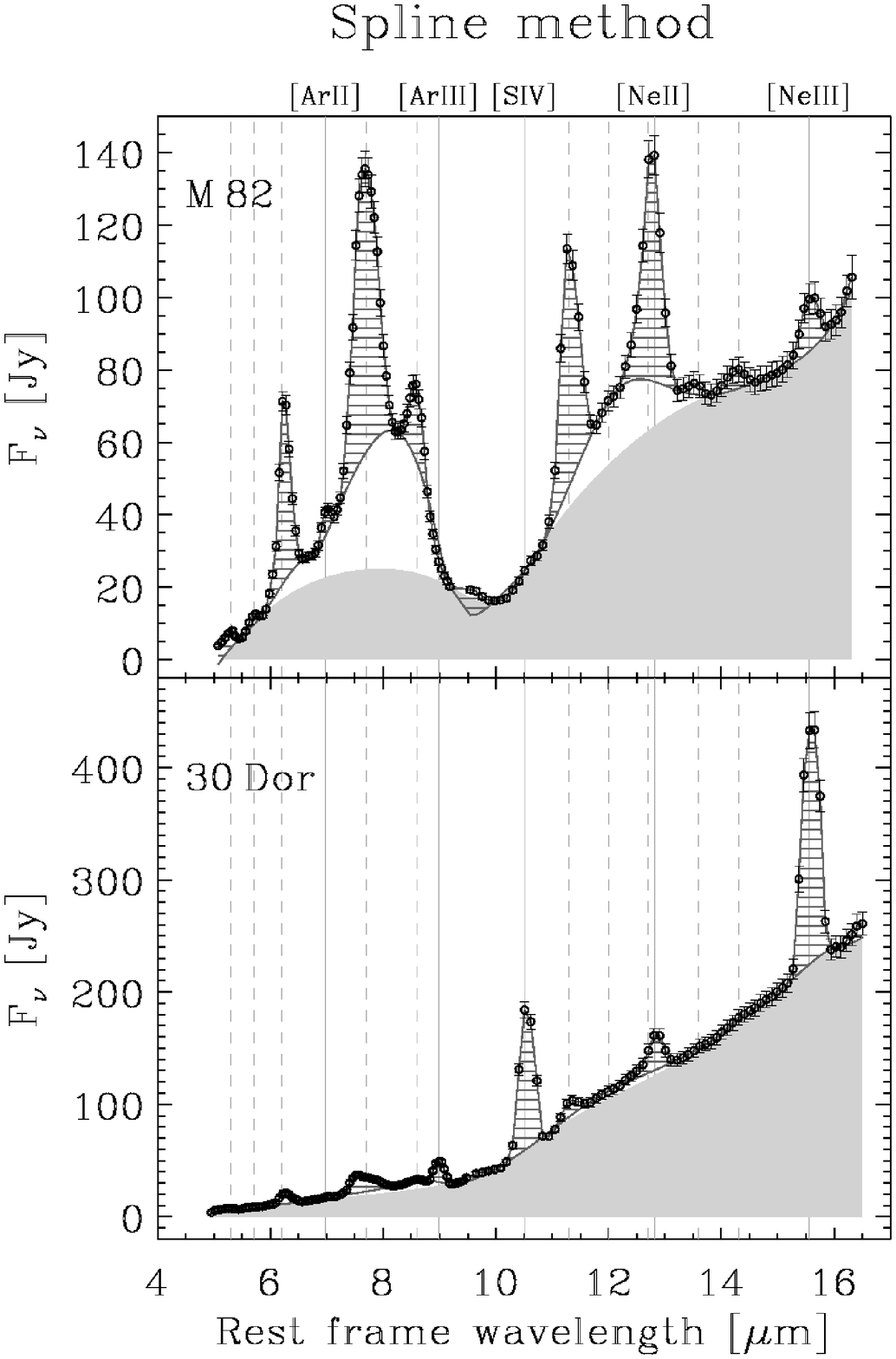} &
    \includegraphics[width=0.48\textwidth]{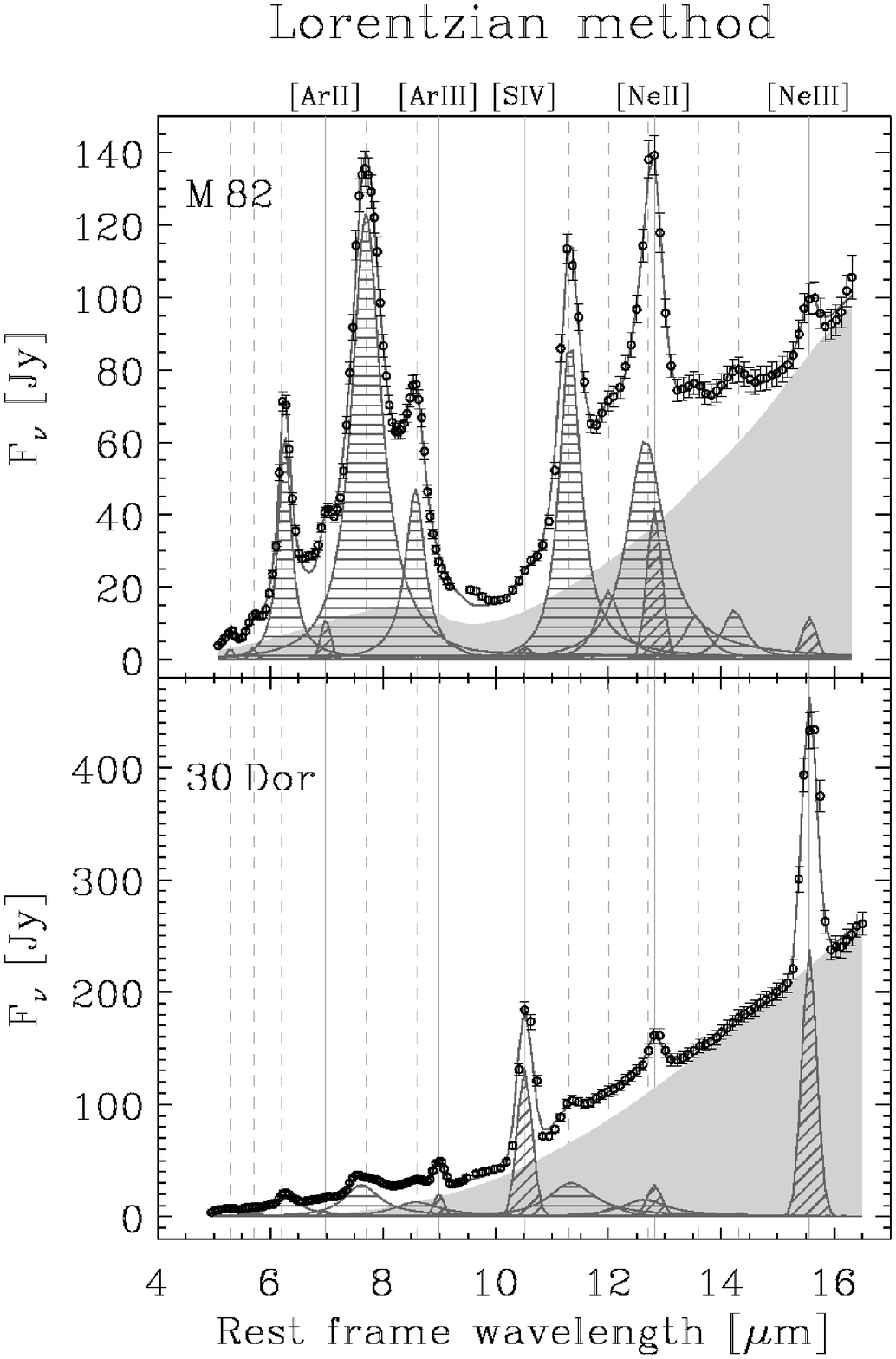} \\
  \end{tabular}
  \caption{Demonstration of the two spectral decomposition methods for 
           the total spectrum of \M{82} and \xxxdor.}
  \label{fig:methods}
\end{figure*}
In order to take into account the uncertainty of the actual shape of the
aromatic bands, we systematically analysed our spectra using two different 
spectral decomposition methods.
These two methods differ in the way the continuum and the aromatic bands are fitted.
They are demonstrated in \reffig{fig:methods}, on a PAH-rich spectrum (\M{82}), and PAH-poor one (\xxxdor).
\begin{description}
  \item[The \nlmet] (hereafter identified by the letter $\mathcal{S}$), 
    used e.g.\ by \citet{hony01} and \citet{vermeij02}, takes into account 
    only the tip of the aromatic bands when computing their intensities, 
    as shown in left panels of \reffig{fig:methods}.
    The continuum (grey filled curve) is fitted with a spline function, 
    constrained at the wavelengths $\lambda=5.04$, 5.47, 5.84, 9.18, 10.85, 
    13.82, 14.60, 15.08, 16.00, $16.15\mic$.
    A second spline function is fitted to the spectrum, corresponding to the
    previous continuum, plus the ``{\it plateau}'' under the 6.2, 7.7 and 
    $8.6\mic$ bands, and the plateau under the 11.3 and $12.7\mic$ bands.
    In addition to the previous wavelengths, this spline function is also 
    constrained at $\lambda=6.61$, 6.80, 7.13, 8.24, 8.80, 11.85, 12.20, 
    $13.22\mic$.
    The difference between the total spectrum and this continuum+plateau 
    component defines the lines and features (line filled areas of the left
    panels of \reffig{fig:methods}).
    This method is relatively robust.
    However, it does not allow us to separate the \neiiline\ from the 
    $12.7\mic$ band (most of our spectra have a low spectral resolution).
  \item[The \frmet] (hereafter identified by the letter $\mathcal{L}$), used 
    e.g.\ by \citet{boulanger98}, \citet{laurent00} and \citet{verstraete01},
    assumes that the aromatic bands have lorentz profiles.
    The wings of these lorentzians account for the underlying plateau.
    A variation of this method, implemented by \citet{smith07}, consists 
    of modeling the PAH features with Drude profiles.
    We proceed as following.
    \begin{enumerate}
    \item The \neiiiline, \neiiline, \sivline, \ariiiline, and \ariiline\ 
      ionic lines are fitted with gaussian functions (diagonal line filling 
      in the right panels of \reffig{fig:methods}).
      For each (number $i$) of the $N_\sms{line}$ lines, the central 
      frequency, $\nu_i^\sms{line}$, is fixed, 
      the width is fixed by the spectral resolution of the instrument, 
      $\Delta\nu_\sms{spectro}$, and the total flux of each line, 
      $F_i^\sms{line}$, is the free parameter.
    \item The PAH bands are fitted with lorentzian functions (horizontal line 
      filling in the right panels of \reffig{fig:methods}).
      The considered features are centered at: 
      $\lambda=5.3$, 5.7, 6.2, 7.7,
      8.6, 11.3, 12.0, 12.7, 13.6 and $14.3\mic$.
      For each (number $i$) of the $N_\sms{band}$ band, 
      the central frequency, $\nu_i^\sms{band}$, 
      the width, $\Delta\nu_i^\sms{band}$, and the total flux, $F_i^\sms{band}$,
      are the free parameters.
      We emphasize the fact that we fit only one band around $7.7\mic$, where
      there are actually two bands centered at $7.6\mic$ and $7.8\mic$.
      We proceed this way, in order to keep the number of parameters reasonably
      low.
      We will use the variation of the centroid of the $7.7\mic$ feature, 
      noted $\lambda_{7.7}$, as an indicator of the $7.6\mic$ to 
      $7.8\mic$ ratio.
    \item The VSG continuum is fitted with the sum of $N_\sms{cont}$ 
      modified black-bodies having the absorption efficiency, 
      $Q_\sms{abs}(\nu)$, of graphite in the Rayleigh approximation
      \citep[grey filled area on the right panels of \reffig{fig:methods};][]
            {laor93}.
      The temperatures, $T_i^\sms{cont}$, and the total fluxes, 
      $F_i^\sms{cont}$, are the free parameters.
      We adopt $N_\sms{cont}=2$.
    \item The sum of the previous components is multiplied
      by $\exp (-\tau)$, where $\tau(\nu)$ is the \citet{mathis90}
      extinction law, with the \citet{dudley97} silicate features.
      The column density is the only free parameter.
  \end{enumerate}
  All these components are fitted simultaneously as part of one mathematical
  function:
  \begin{equation}
    F_\nu (\nu) =
      \left(F_\nu^\sms{line}(\nu)+F_\nu^\sms{band}(\nu)+F_\nu^\sms{cont}(\nu)
            \right)\times\exp\left(-\tau(\nu)\right),
  \end{equation}
  with:
  \begin{equation}
    \left\{
    \begin{array}{rcl}
      F_\nu^\sms{line}(\nu) & = & \displaystyle\sum_{i=1}^{N_{\rm line}} 
        F_i^\sms{line} \,\sqrt{\frac{2}{\pi\Delta\nu_\sms{spectro}^2}}
        \,\exp\left(-2\frac{(\nu-\nu_i^\sms{line})^2}
                         {\Delta\nu_\sms{spectro}^2}\right) \\
      &&\\
      F_\nu^\sms{band}(\nu) & = & \displaystyle\sum_{i=1}^{N_{\rm band}} 
        F_i^\sms{band} \,\frac{1}{2\pi}
        \,\frac{\Delta\nu_i^\sms{band}}{(\nu-\nu_i^\sms{band})^2
                                       +(\Delta\nu_i^\sms{band}/2)^2} \\
      &&\\
      F_\nu^\sms{cont}(\nu) & = & \displaystyle\sum_{i=1}^{N_{\rm cont}}
        F_i^\sms{cont} 
        \,\frac{B_\nu (T_i^\sms{cont},\nu)\,Q_\sms{abs}(\nu)}
               {\displaystyle\int_{c/(16\mic)}^{c/(10\mic)} 
                B_\nu (T_i^\sms{cont},\nu)\,Q_\sms{abs}(\nu)\ddiff\nu},
    \end{array}
    \right.
  \end{equation}
  the various $F_\nu$ being the monochromatic flux densities.
  An advantage of the \frmet\ over the \nlmet\ is that it allows us to 
  separate the \neiiline\ line and the $12.7\mic$ feature, and to study the 
  variations of the centroids of the bands.
  However, when the PAH-to-VSG ratio is very low, as in the case of \xxxdor\
  (\reffig{fig:methods}), the width of the features is more uncertain.
\end{description}
In both cases, $\icont$ is defined as the integrated intensity of the continuum
between 10 and $16\mic$, and $\iPah$, the sum of the intensities of all the 
bands between 5 and $16\mic$.
We emphasize that our methods work automatically, without any {\it by eye}
adjustment.
This condition prevents systematic effects that could result from arbitrary
choices of parameters.
These two methods are orthogonal but give similar trends \citep[see 
\refS{sec:correl} and][]{smith07}. 
The \nlmet\ is clearly continuum-biased in its assumptions whereas the
\frmet\ is a line-biased assumption method.

Our two methods can be applied to an integrated spectrum, or to each pixel
of a spectro-image, provided that the signal-to-noise ratio is sufficient.
We have systematically applied the two methods to all the spectra presented in
\reftab{tab:source}.
In our analysis, we will systematically consider the results of the two methods
before drawing conclusions on the PAH properties.


\section{SYSTEMATIC ANALYSIS OF THE MAJOR BAND RATIOS}
\label{sec:correl}

  \subsection{Correlations Exhibited Among Integrated Spectra}
  \label{sec:corglo}

\begin{figure*}[htbp]
  \centering
  \includegraphics[width=\textwidth]{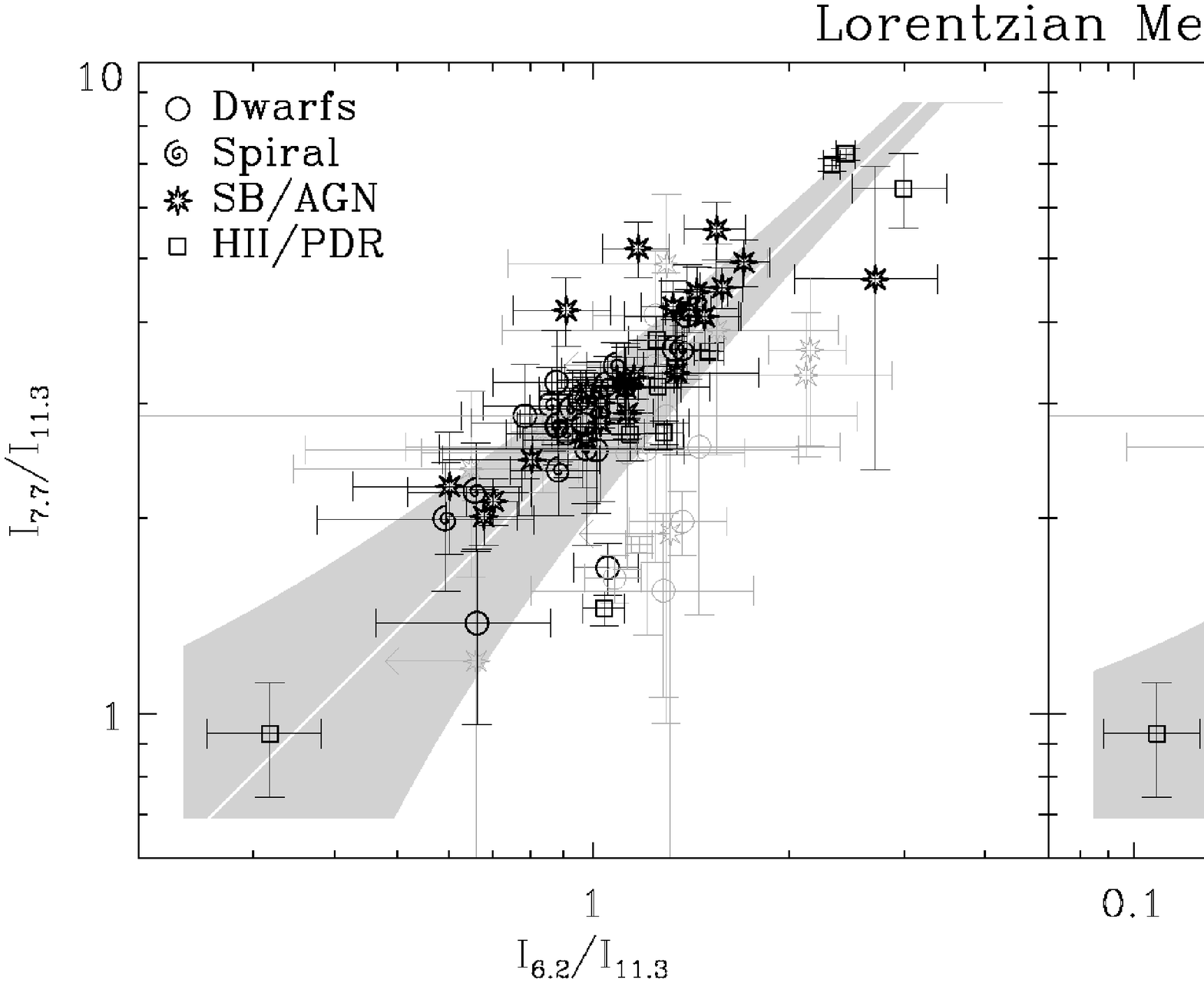} \\
  \includegraphics[width=\textwidth]{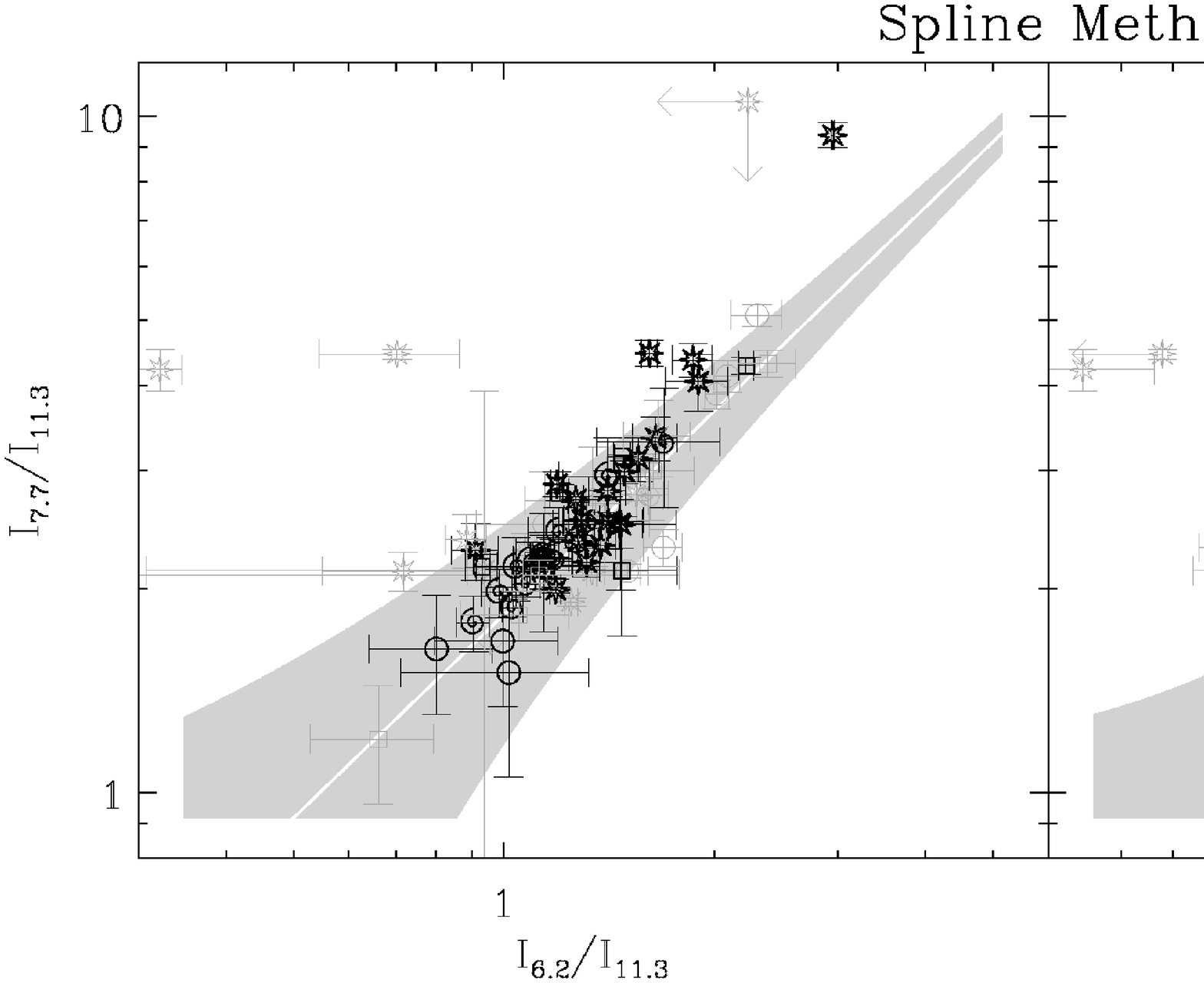}
  \caption{PAH properties for integrated spectra.
           The black error bars are for the measurements that we consider to be
           reliable, while the grey are more uncertain.
           In each panel representing a relation 
           $Y$ versus $X$, the grey filled area is the domain between 
           $Y = a_{Y/X}\times X - 1\sigma_{Y/X}$, and 
           $Y = a_{Y/X}\times X + 1\sigma_{Y/X}$, 
           $Y = a_{Y/X}\times X$ being the linear correlation of the data, and 
           $\sigma_{Y/X}$, the dispersion of the data around this correlation.}
  \label{fig:corglo1}
\end{figure*}

\begin{figure*}[htbp]
  \centering
  \begin{tabular}{cc}
    \includegraphics[width=0.48\textwidth]{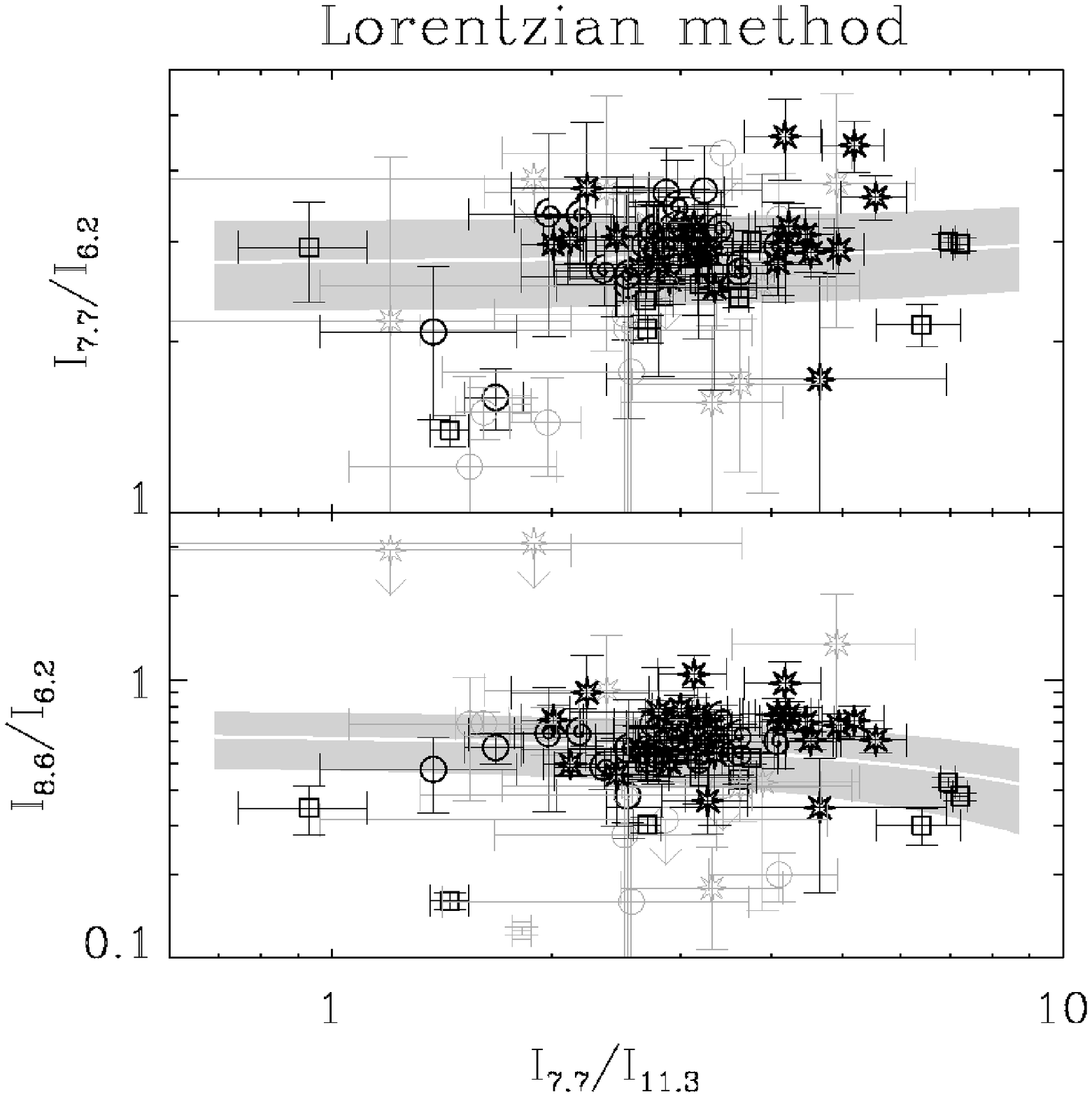} &
    \includegraphics[width=0.48\textwidth]{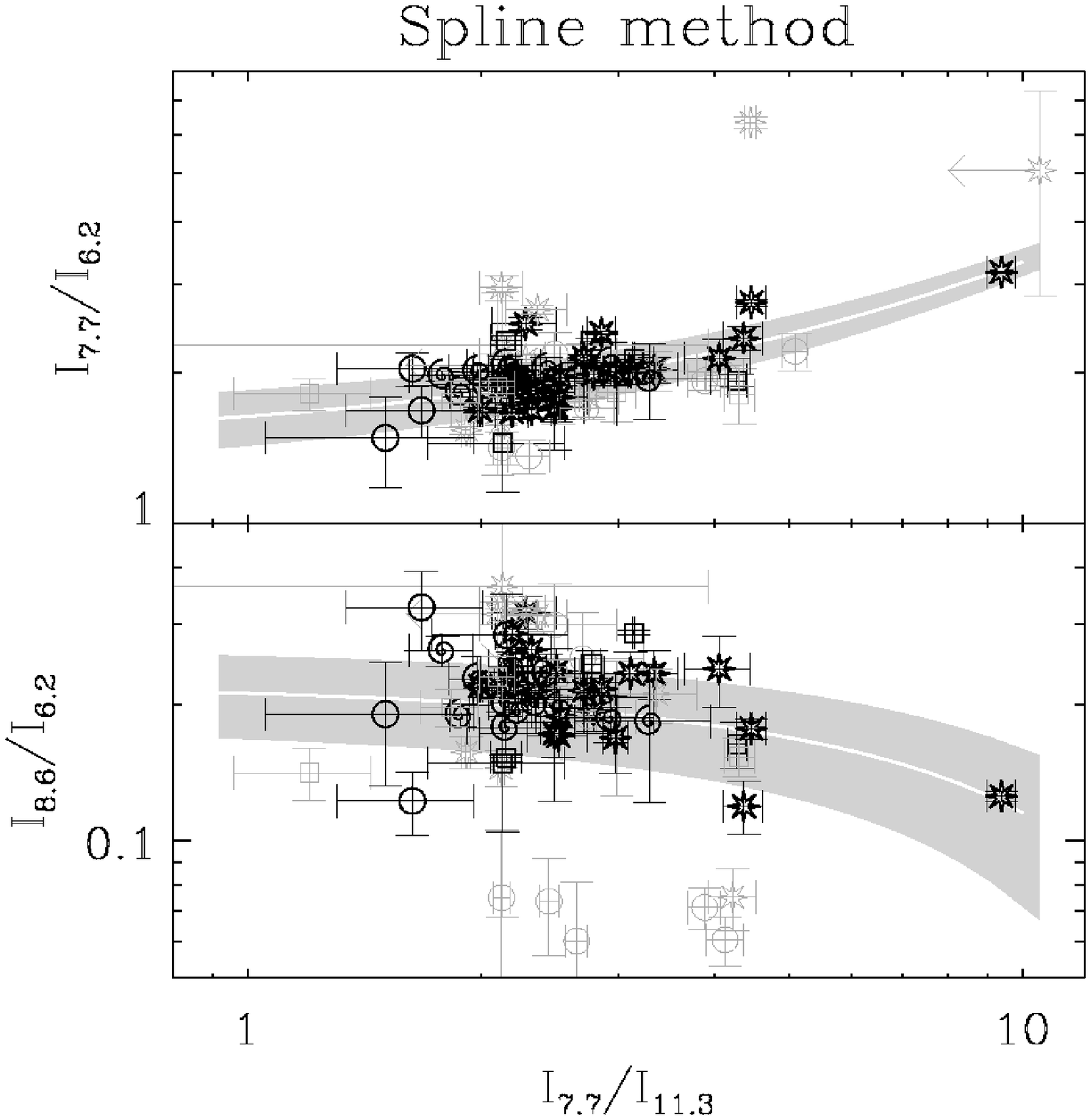} \\
  \end{tabular}
  \caption{PAH properties for integrated spectra (continued).
           The same symbol conventions are adopted as in \reffig{fig:corglo1}.}
  \label{fig:corglo2}
\end{figure*}

\reffigs{fig:corglo1}-\ref{fig:corglo2} show select correlations between 
band ratios of the 
integrated spectra of our sample (\reftab{tab:source}), obtained with the 
two methods presented in \refS{sec:method}.
We focus on the four brightest bands at 6.2, 7.7, 8.6 and $11.3\mic$, and 
study the correlations between the various ratios. 
The intensities of the features are reported in \refapp{sec:tabfig} 
(\reftab{tab:intens}), and 
the parameters of the various correlations are given in \reftab{tab:correl}.
As mentioned in \refS{sec:method}, the two methods are uncertain for
very low values of the PAH-to-VSG ratio.
Thus we define two subgroups of data.
\begin{enumerate}
  \item The data that we consider to be reliable are marked with black symbols 
        in \reffigs{fig:corglo1}-\ref{fig:corglo2}.
        We define them as the measurements which have a signal-to-noise ratio,
        at $\lambda=7.7\mic$, larger than 6, and 
        $\iPah/\icont \geq 0.5$ with the \frmet, and $\iPah/\icont \geq 0.3$ 
        with the \nlmet.
  \item The fits that we consider to be less certain are the complementary data 
        (grey symbols in \reffigs{fig:corglo1}-\ref{fig:corglo2}), having lower
        signal-to-noise and PAH-to-VSG ratios.
\end{enumerate}
  
\reffig{fig:corglo1} shows that there is an excellent linear correlation between the ratios $\ipah{6.2}/\ipah{11.3}$, $\ipah{7.7}/\ipah{11.3}$ and 
$\ipah{8.6}/\ipah{11.3}$, using either of the two methods.
The variations of these ratios spread roughly over one order of magnitude, while 
the ratios $\ipah{6.2}/\ipah{7.7}$, and $\ipah{7.7}/\ipah{8.6}$ are roughly 
constant within the error bars (\reffig{fig:corglo2}).
In general, the measure of the intensity of the $8.6\mic$ band is
less accurate than for the other main bands.
Indeed, it is less intense than the 6.2 and $7.7\mic$ features.
In addition, it is merged with the long wavelength wing of the $7.7\mic$ band.
Finally, this particular feature is significantly affected by the silicate 
extinction feature around $9.7\mic$, when the source is deeply embedded.
This extinction is not corrected in the case of the \nlmet, and is
corrected very simply in the case of the \frmet.
That is the reason why correlations involving the $8.6\mic$ feature are
always more dispersed than the others.
Quantitatively, the correlation coefficients are around $0.8$ for the 
four panels of \reffig{fig:corglo1}, except for that involving the 
$\ipah{8.6}$ feature, with the \nlmet.
Most of the outsiders are based on fits that we consider to be 
uncertain.
  
\begin{figure}[htbp]
  \centering
  \includegraphics[width=0.48\textwidth]{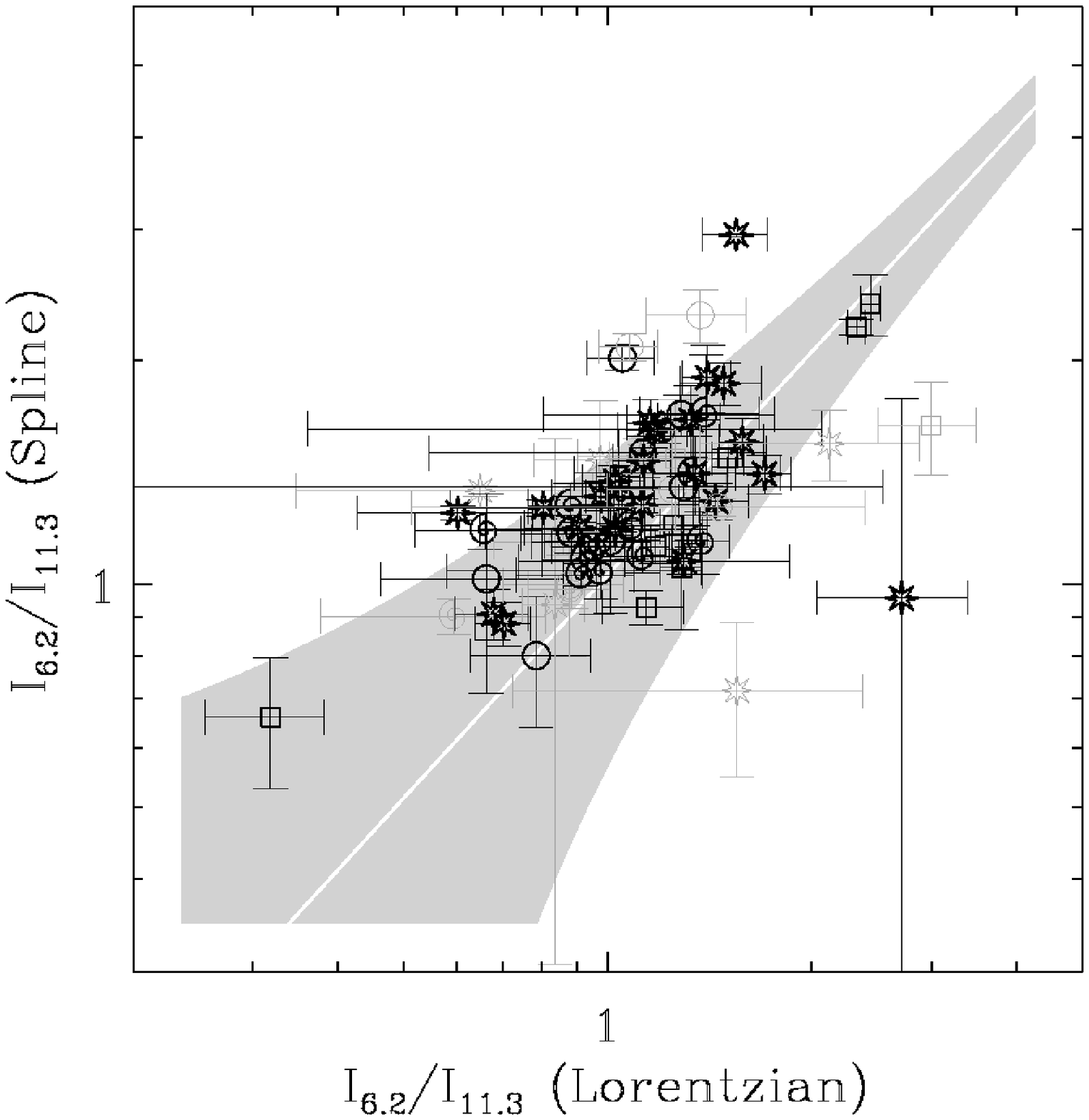}
  \caption{Comparison between the two methods.
           The same symbol conventions are adopted as in \reffig{fig:corglo1}.}
  \label{fig:compglo}
\end{figure}

\reffig{fig:compglo} shows the consistency between the two methods.
It demonstrates that, whatever method we use to measure the band ratios,
the order of the various measures is conserved; i.e.\ a spectrum 
$\mathcal{A}$ having a lower $\ipah{6.2}/\ipah{11.3}$ ratio than a spectrum
$\mathcal{B}$, using the \nlmet, will also have a lower ratio than 
$\mathcal{B}$ using the \frmet.
Moreover, the fact that both methods give similar trends means that
we did not artificially incorporate part of the continuum flux within the
aromatic band intensities.
Indeed, it is possible to accidentally account for a fraction of the continuum 
intensity within the wings of the PAH profile, with the \frmet,
especially within the $11.3\mic$ feature.
However, this bias is not possible with the \nlmet, since it integrates only
the tip of the band.
Therefore, the agreement between the two methods allow us to rule out this bias
and claim that the trends of \reffig{fig:corglo1} are not induced by the fitting
methods.

These first relations, established on integrated spectra, indicate that the 
properties of the PAHs throughout different types of galaxies and Galactic 
regions are remarkably homogeneous.
They are consistent with a significant variation of the 6.2, 7.7 and $8.6\mic$
features relative to the $11.3\mic$ band, coupled with an absence of significant variations among the 6.2, 7.7 and $8.6\mic$ bands.

\begin{figure}[htbp]
  \centering
  \includegraphics[width=0.48\textwidth]{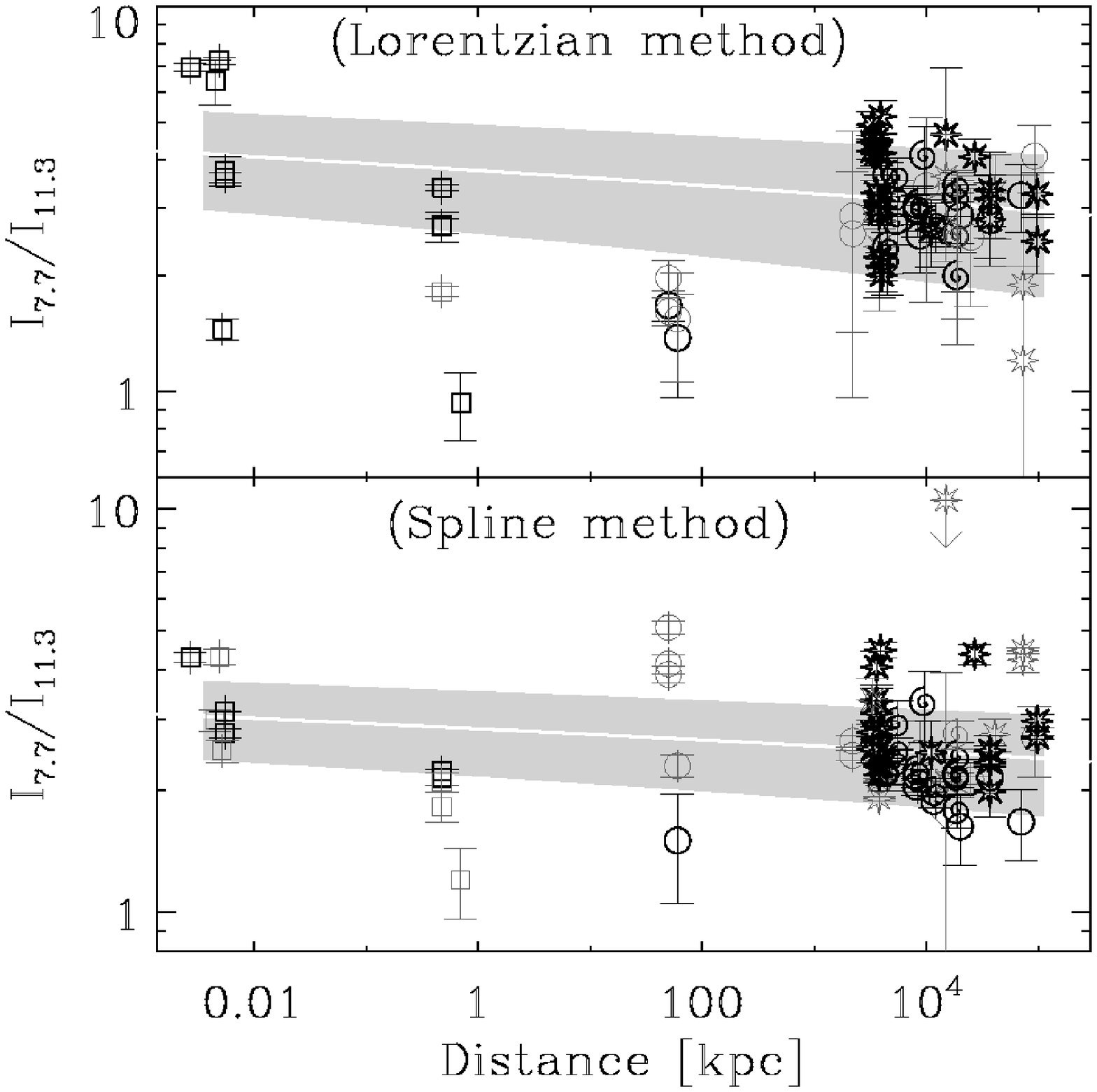}
  \caption{Correlation of the band ratio with the distance.
           The absence of correlation shows the uniformity of the sample.
           The same symbol conventions are adopted as in \reffig{fig:corglo1}.}
  \label{fig:cordist}
\end{figure}

Finally, as a consistency check, \reffig{fig:cordist} shows the 
variations of the band ratios as a function of distance.
The correlation coefficient is 0.23 for the \frmet, and 0.20 for the \nlmet.
This absence of correlation shows the uniformity of the sample and rules out
the possibility of variations induced by aperture effects.

  \subsection{Spatial Variations of the Band Properties}
  \label{sec:cormap}

\begin{figure*}[htbp]
  \centering
  \includegraphics[width=\textwidth]{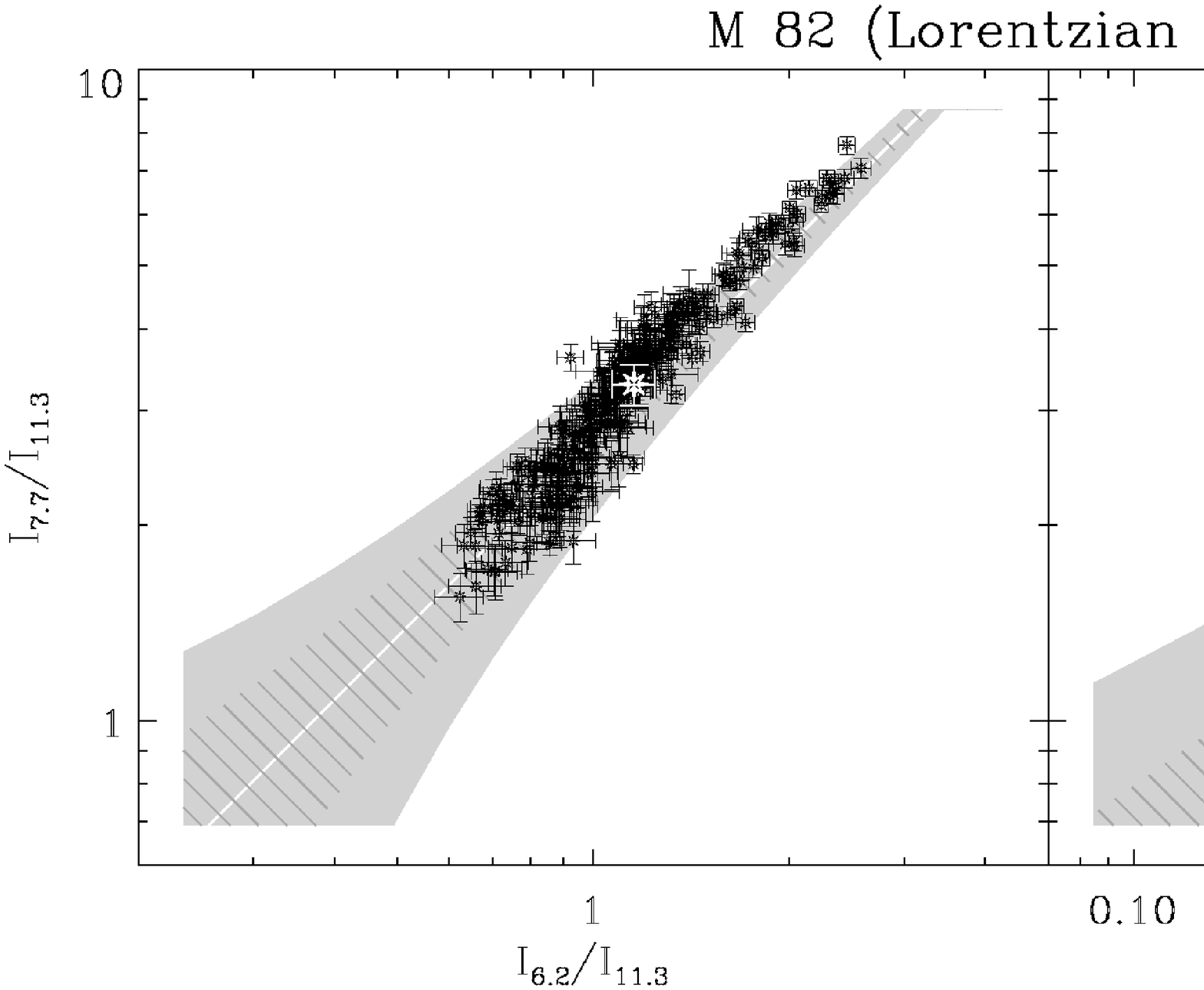} \\
  \includegraphics[width=\textwidth]{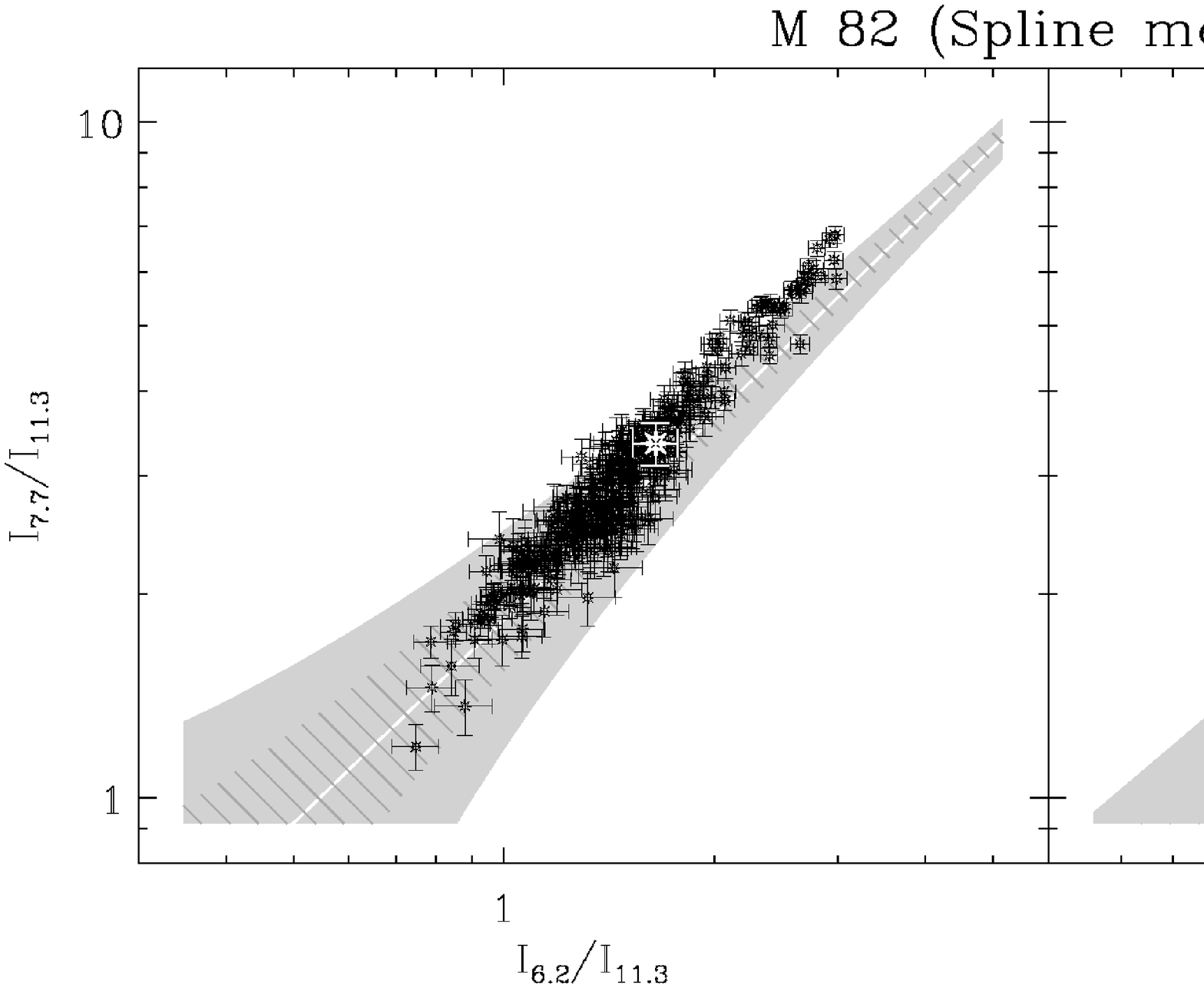}
  \caption{PAH band ratios within \M{82}.
           The grey filled areas are the correlations obtained for the
           integrated spectra (\reffig{fig:corglo1}).
           The correlations obtained inside \M{82} are represented by the
           hatched region.
           The white symbol is the value of the global measurement over 
           the entire galaxy.}
  \label{fig:cormap1m82}
\end{figure*}

\begin{figure*}[htbp]
  \centering
  \begin{tabular}{cc}
    \includegraphics[width=0.48\textwidth]{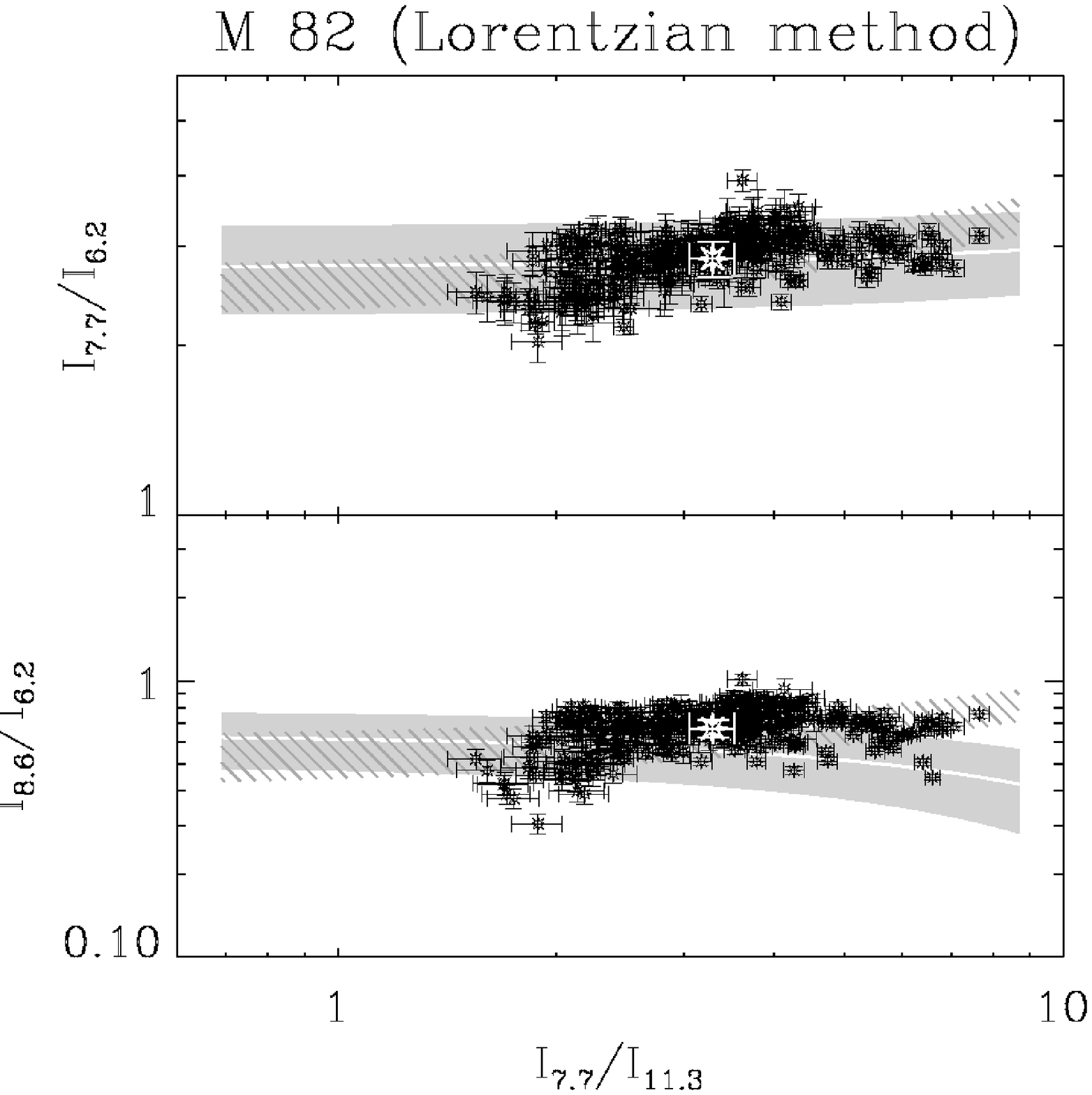} &
    \includegraphics[width=0.48\textwidth]{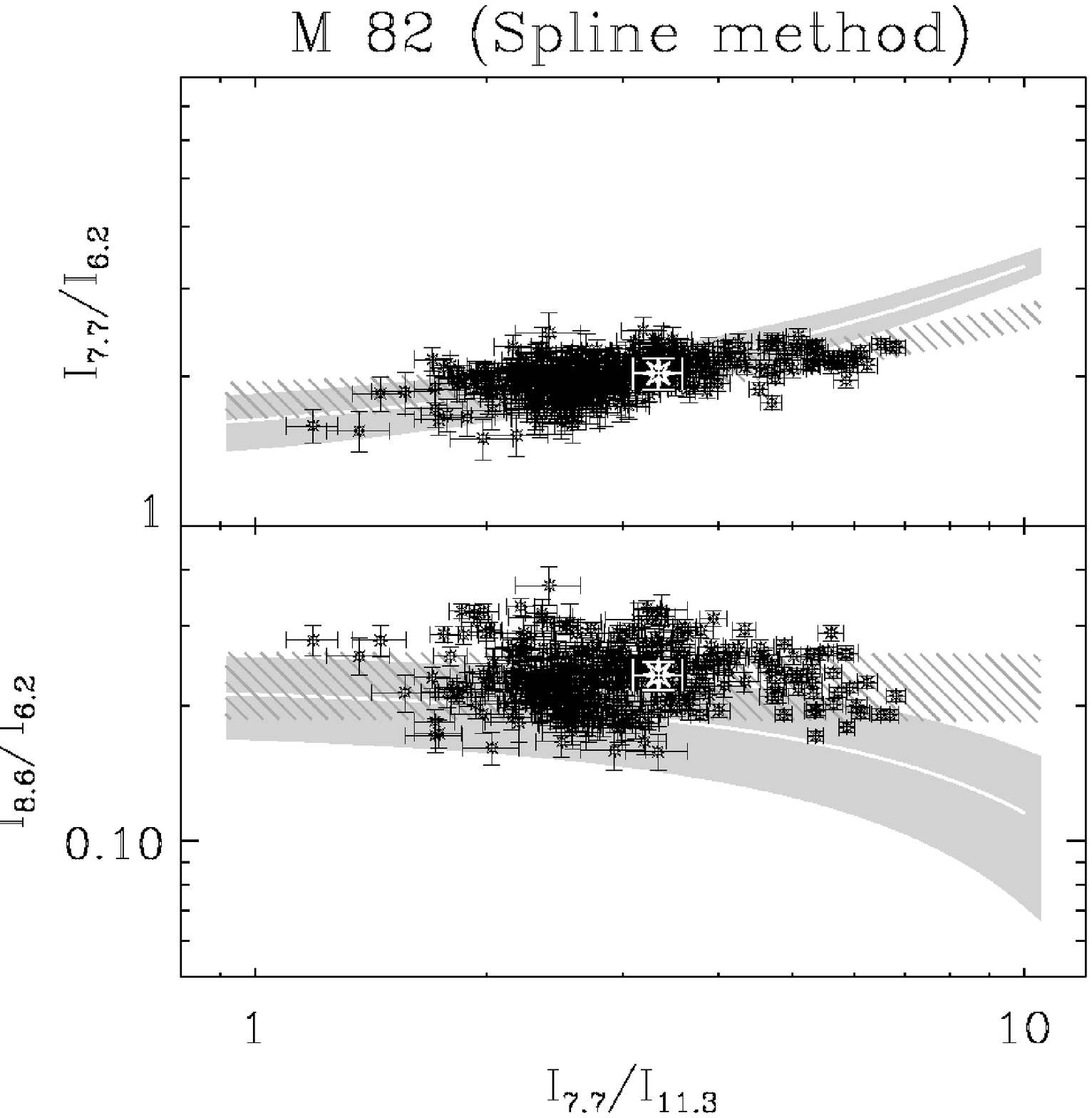} \\
  \end{tabular}
  \caption{PAH band ratios within \M{82} (continued).}
  \label{fig:cormap2m82}
\end{figure*}

\begin{figure}[htbp]
  \centering
  \includegraphics[width=0.48\textwidth]{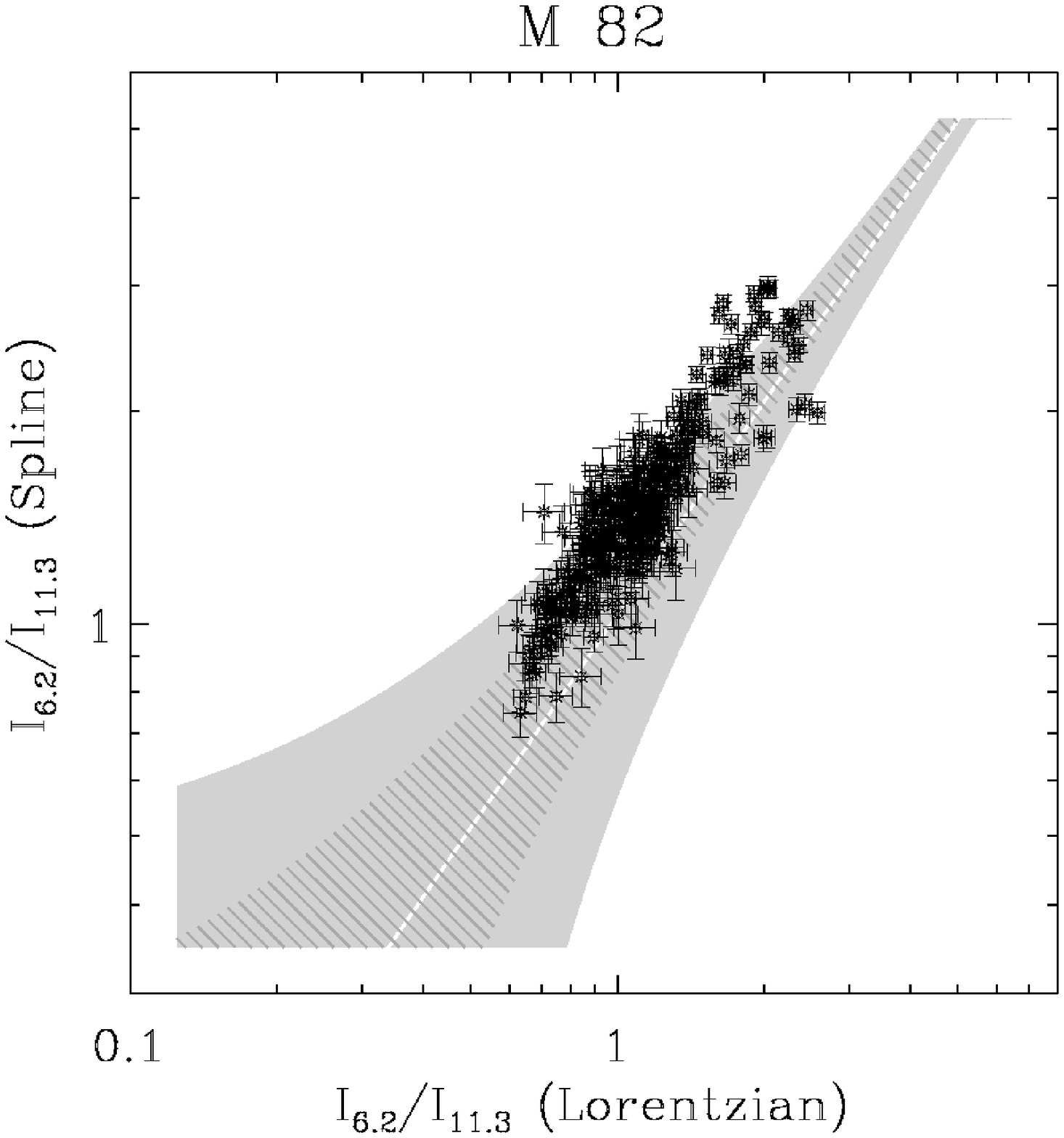}
  \caption{Comparison between the two methods in \M{82}.
           The same symbol conventions are adopted as in 
           \reffig{fig:cormap1m82}.}
  \label{fig:compmapm82}
\end{figure}

\begin{figure*}[htbp]
  \centering
  \includegraphics[width=\textwidth]{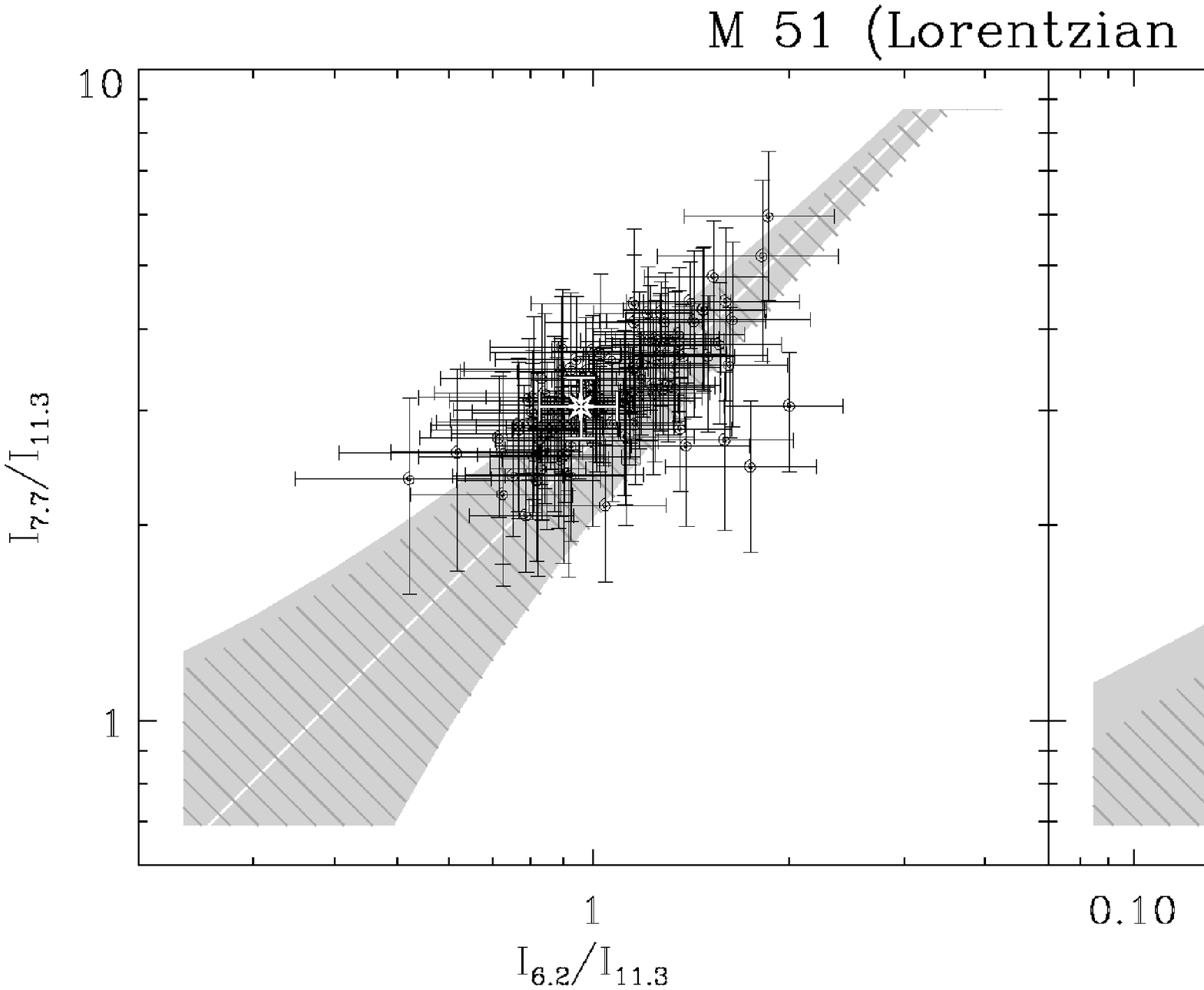} \\
  \includegraphics[width=\textwidth]{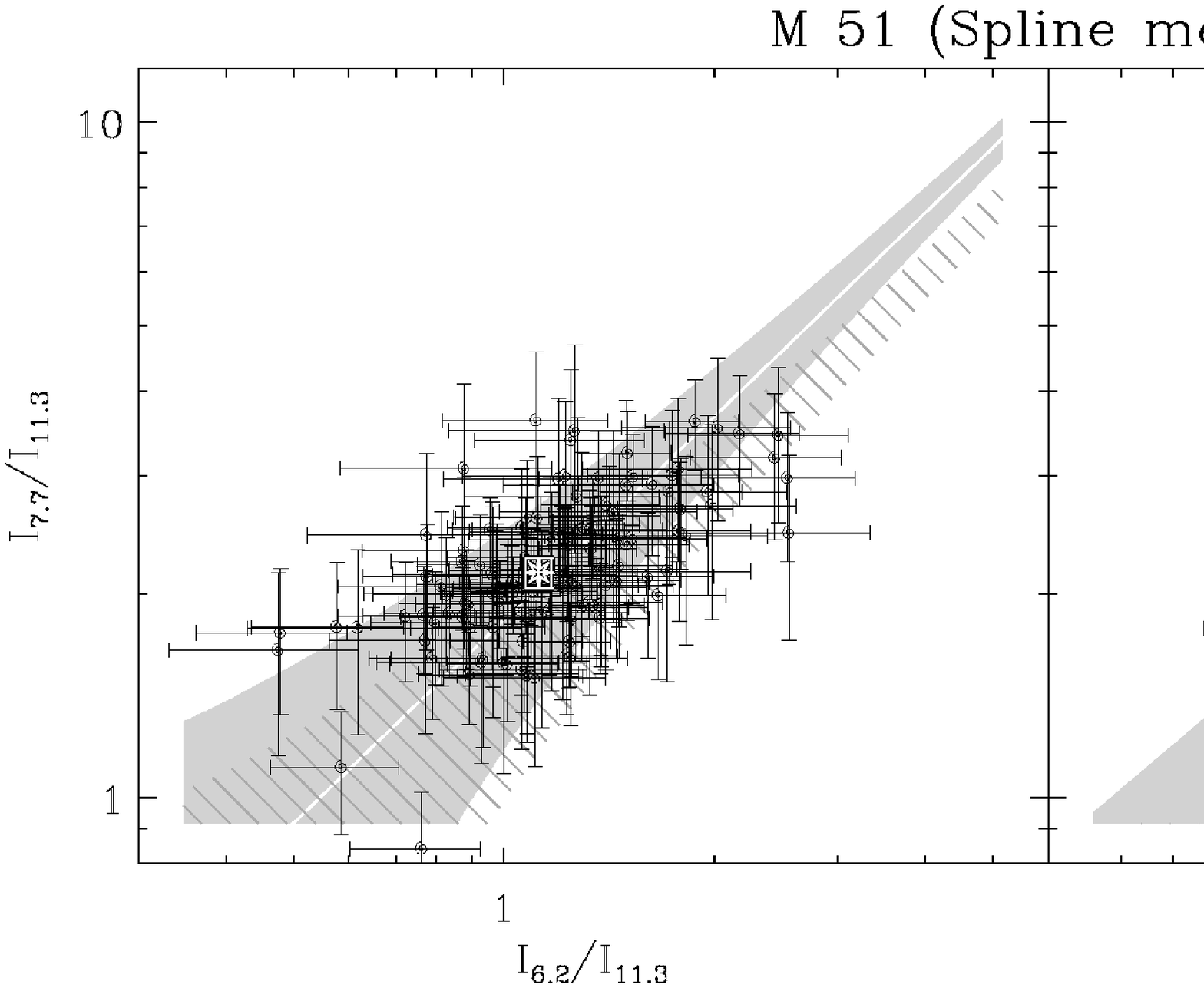}
  \caption{PAH band ratios within \M{51}.
           The same symbol conventions are adopted as in 
           \reffig{fig:cormap1m82}.}
  \label{fig:cormap1m51}
\end{figure*}

\begin{figure*}[htbp]
  \centering
  \begin{tabular}{cc}
    \includegraphics[width=0.48\textwidth]{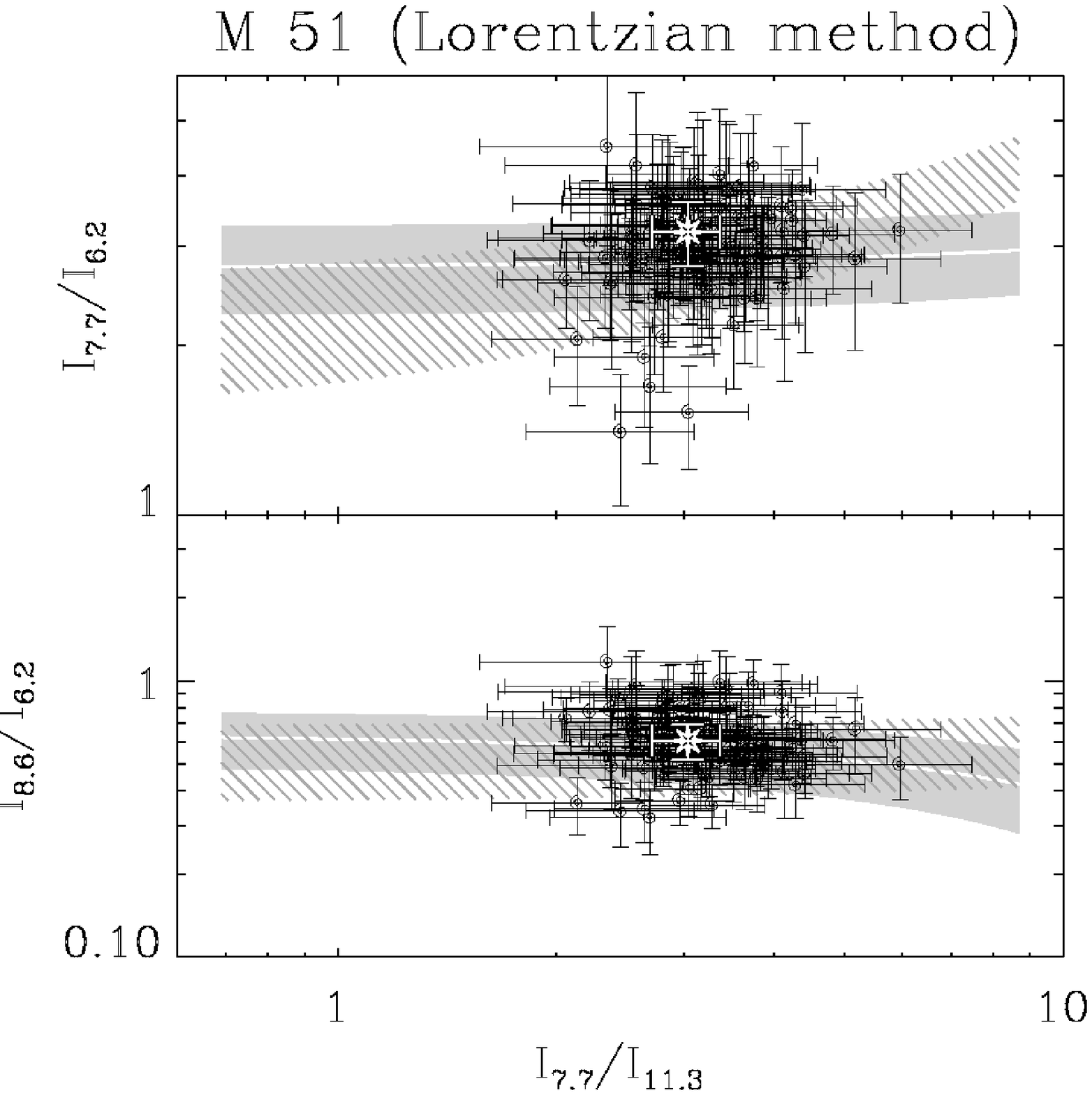} &
    \includegraphics[width=0.48\textwidth]{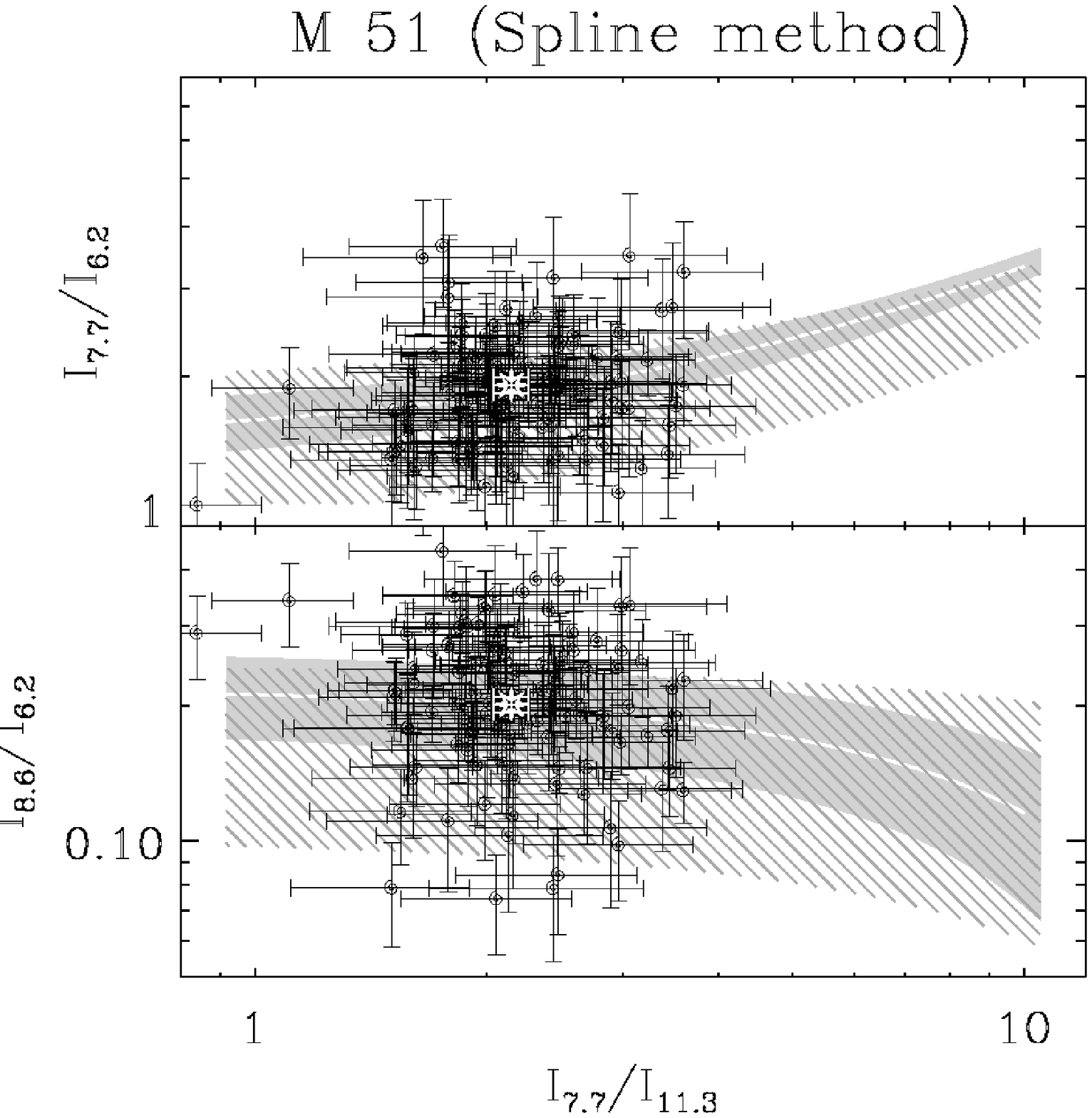} \\
  \end{tabular}
  \caption{PAH band ratios within \M{51} (continued).}
  \label{fig:cormap2m51}
\end{figure*}

\begin{figure}[htbp]
  \centering
  \includegraphics[width=0.48\textwidth]{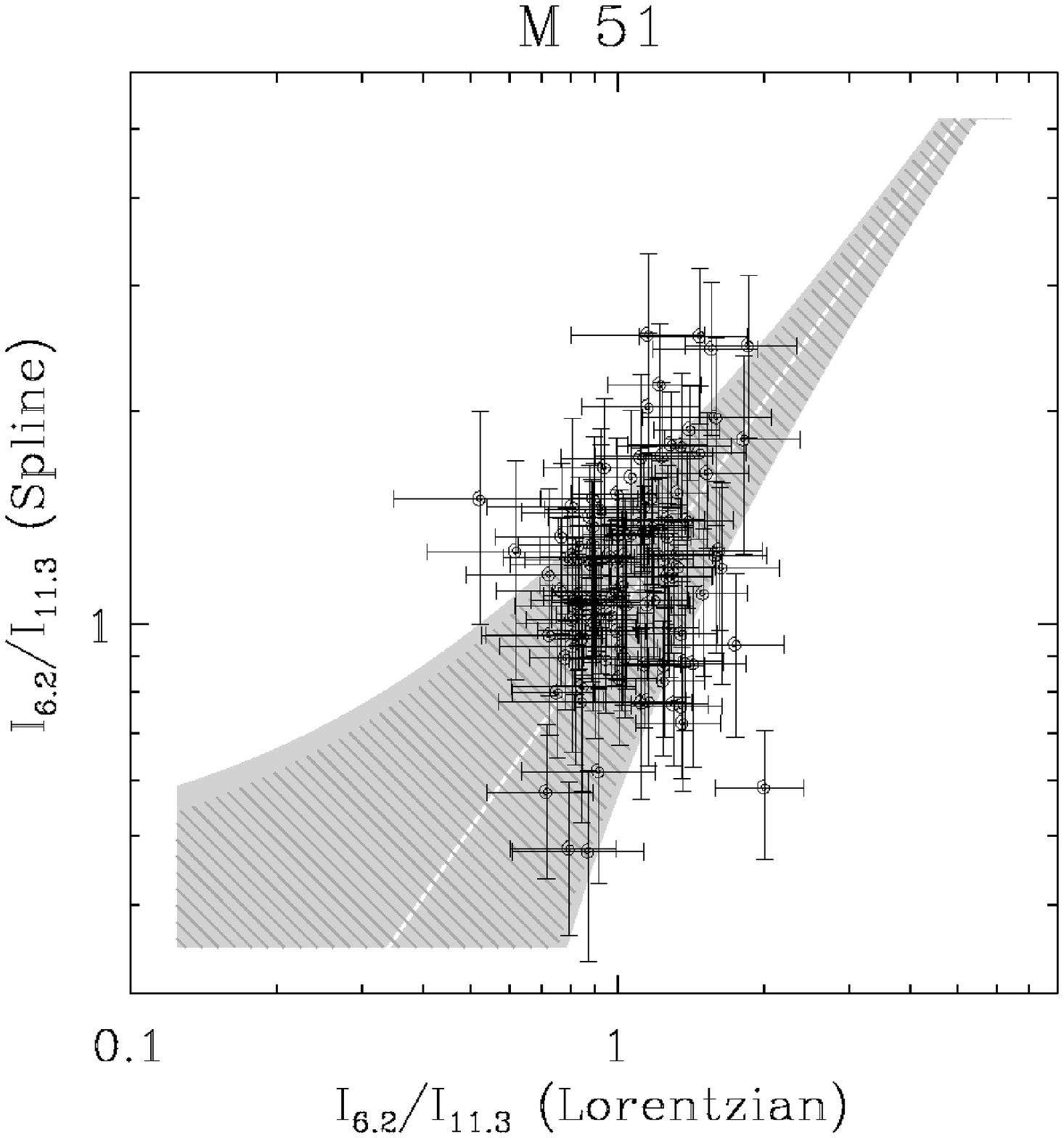}
  \caption{Comparison between the two methods in \M{51}.
           The same symbol conventions are adopted as in 
           \reffig{fig:cormap1m82}.}
  \label{fig:compmapm51}
\end{figure}

\begin{figure*}[htbp]
  \centering
  \plottwo{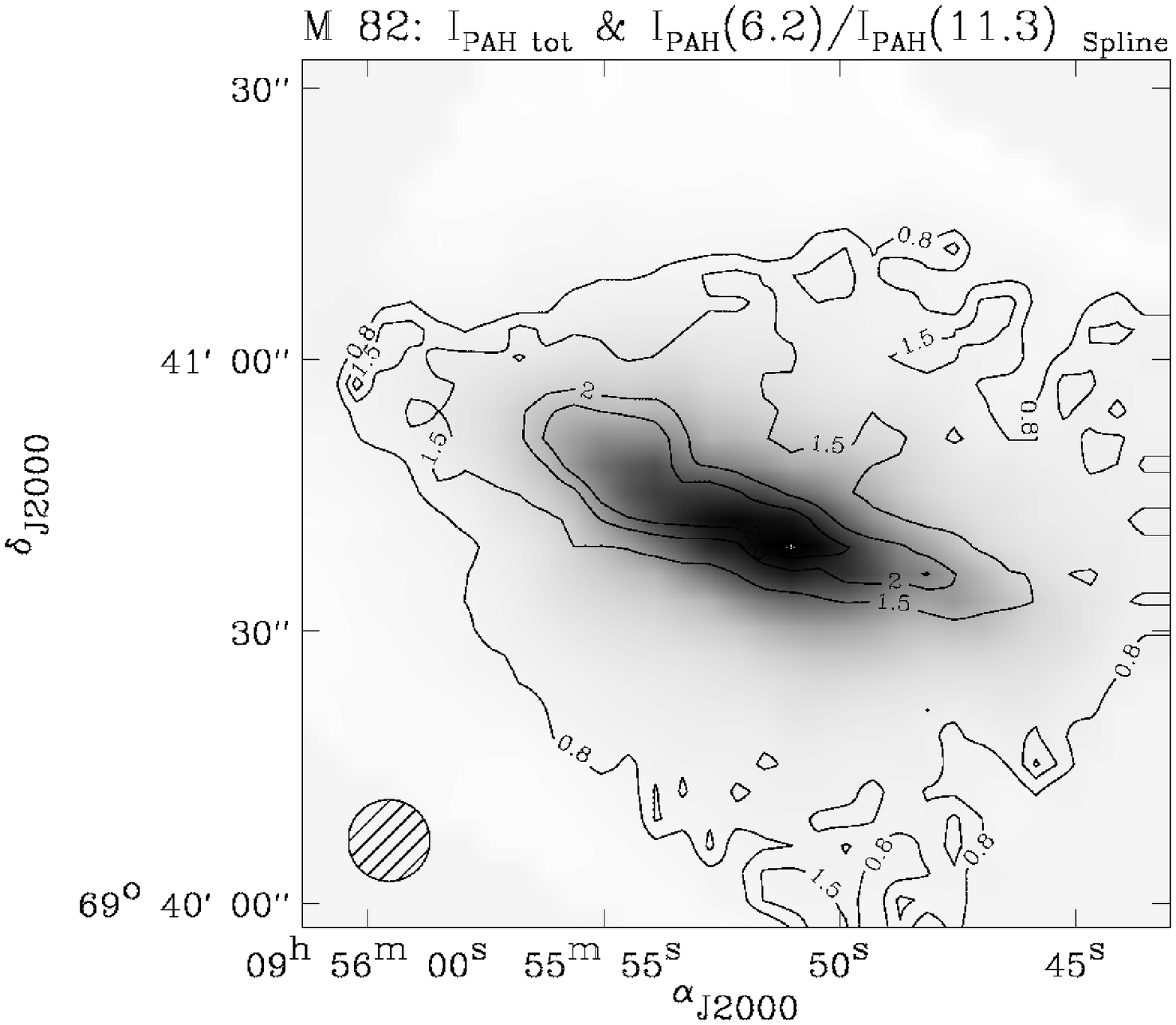}{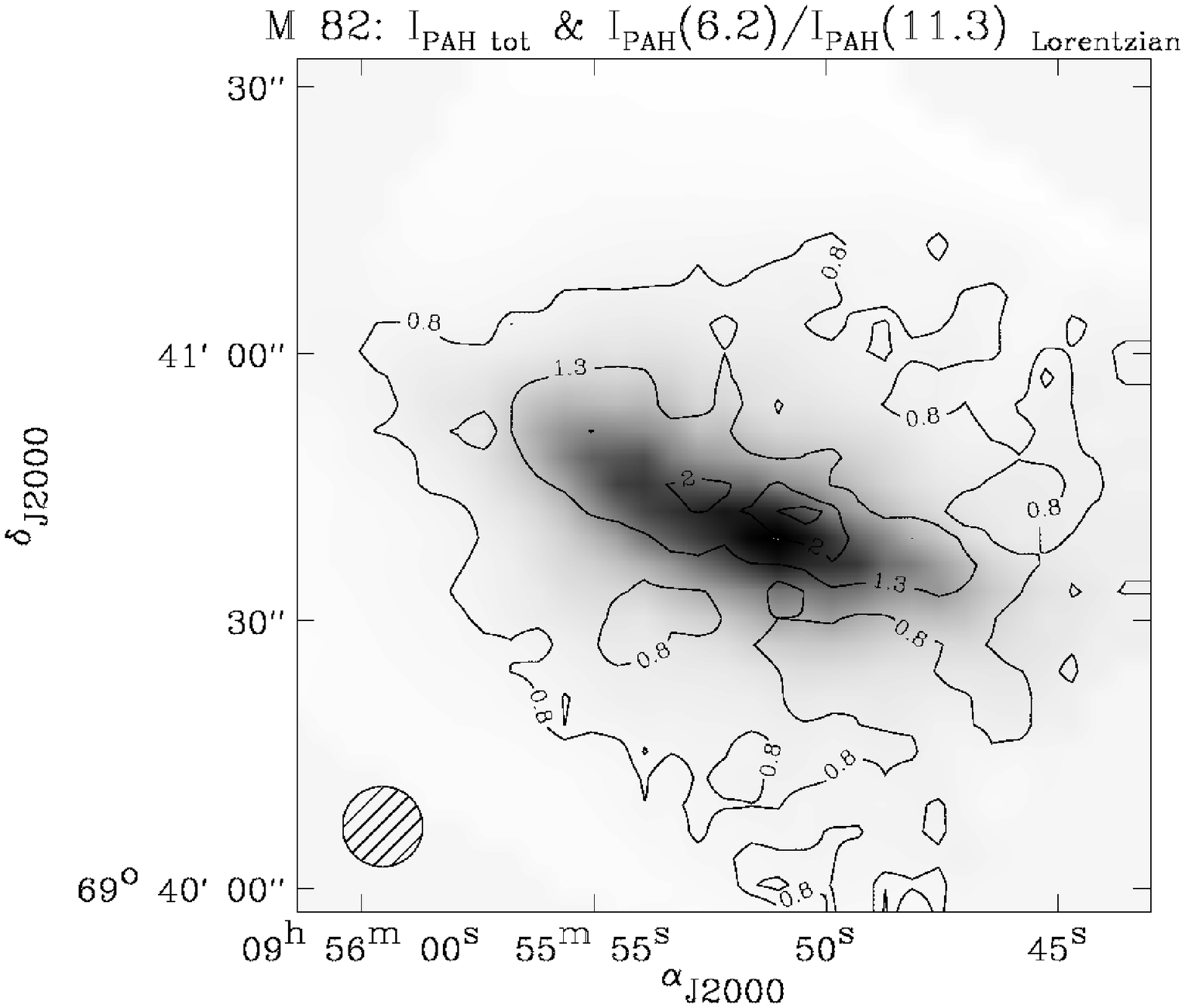}
  \caption{Spatial distribution of the PAHs in \M{82}.
           For each method, the image is the total PAH intensity, and the 
           contours are the $\ipah{6.2}/\ipah{11.3}$ ratio.
           The shaded circle indicates the beam size.}
  \label{fig:imm82}
  \centering
  \plottwo{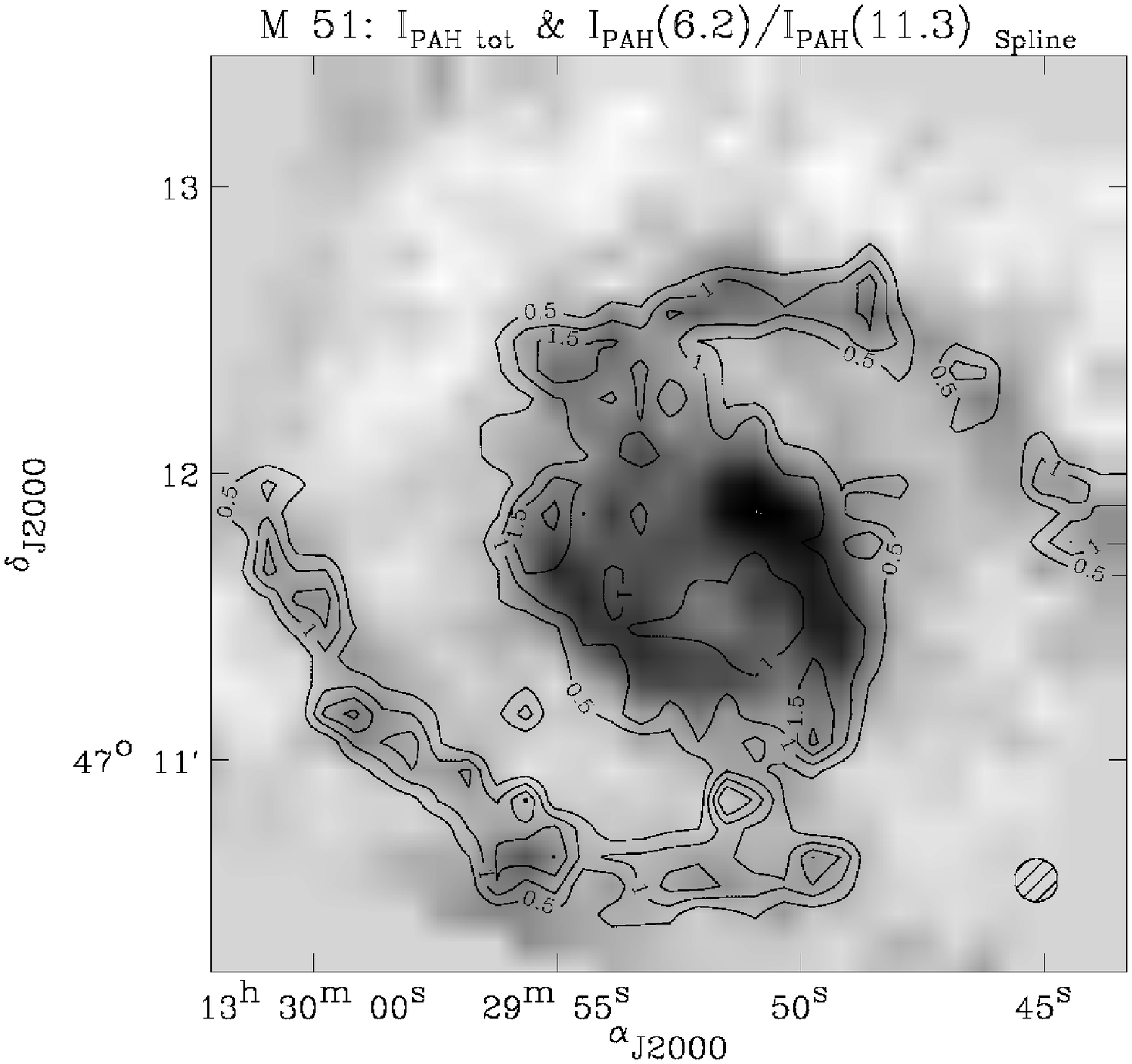}{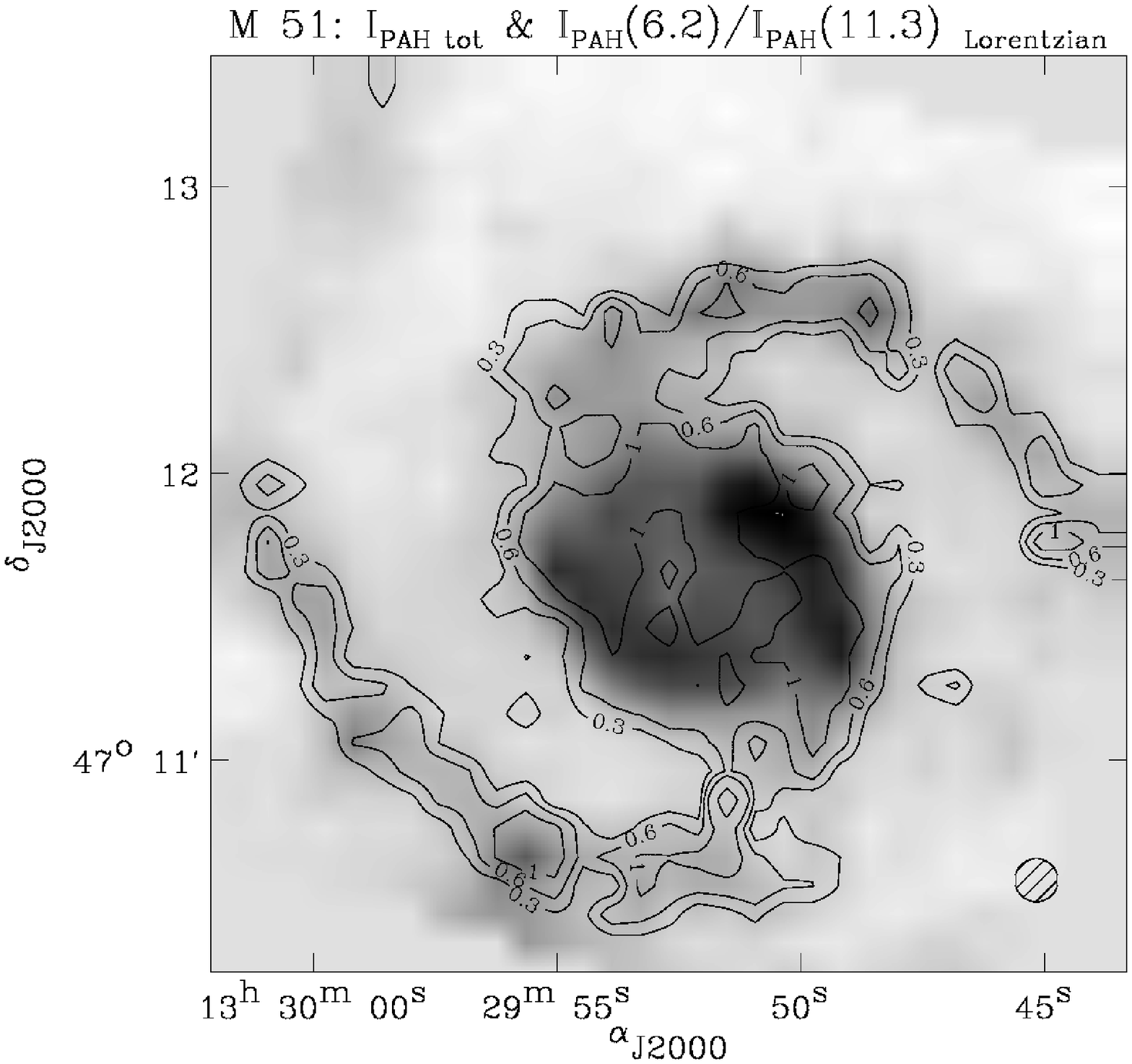}
  \caption{Spatial distribution of the PAHs in \M{51}.
           For each method, the image is the total PAH intensity, and the 
           contours are the $\ipah{6.2}/\ipah{11.3}$ ratio.
           The shaded circle indicates the beam size.}
  \label{fig:imm51}
\end{figure*}

We now study the spatial variations of the properties of the mid-IR features, 
by analyzing the spectral maps of a sub-sample of sources:
\IC{342}, \M{17}, \M{51}, \M{82}, \M{83}, \xxxdor, and the \orb.
These objects are those which are spatially resolved, satisfy the two
PAH-to-continuum and signal-to-noise ratio thresholds listed in 
\refS{sec:corglo}, and show significant variations of the band ratios.
Similar to \refS{sec:corglo}, we systematically compare the results of the 
two methods, but instead of applying them to integrated spectra, we fit the
spectrum of each pixel of the spectral maps.
We then degrade the images of each spectral component to the spatial resolution 
of the longest wavelength of the spectrum (${\rm FWHM} \simeq 9''$).
We achieve this by convolving the image of a given component
(e.g.\ $\ipah{6.2}$), with the PSF at the longest wavelength 
(i.e.\ $\lambda=16\mic$) deconvolved beforehand by the PSF of the component
(i.e.\ $\lambda=6.2\mic$).
\reffigs{fig:cormap1m82} to \ref{fig:compmapm51} (as well as
\reffigs{fig:cormap1ic342} to \ref{fig:compmapm17} in \refapp{sec:tabfig}) show 
the same correlations as in \reffigs{fig:corglo1}-\ref{fig:compglo},
obtained inside resolved sources, instead of integrated spectra.
Each data point represents the fit of the spectrum of one pixel.
We selected the points, according to the criterion defined in 
\refS{sec:corglo}: 
the signal-to-noise ratio at $\lambda=7.7\mic$ must be larger than 6, and 
$\iPah/\icont \geq 0.5$ with the \frmet, and $\iPah/\icont \geq 0.3$ 
with the \nlmet.
The pixels which do not fall into this category have not been considered here.
The parameters and statistical quantities relative to these correlations
are given in \reftab{tab:correl}.
\reffigs{fig:imm82} to \ref{fig:imm51} (as well as \reffigs{fig:imic342}
to \ref{fig:imm17} in \refapp{sec:tabfig}) 
show the spatial distributions of the components.

In general, the correlations presented in 
\reffigs{fig:corglo1}-\ref{fig:compglo} still hold inside individual 
objects as well.
Therefore, the variations of the PAH band ratios are independent of the 
spatial resolution.
For example, the pixel size is $\simeq0.1$~pc in \M{17} and the \orb, while 
it is $\simeq1$~kpc in external galaxies.
From the images, we notice that the ratio $\ipah{6.2}/\ipah{11.3}$ 
(as well as the ratios $\ipah{7.7}/\ipah{11.3}$ and $\ipah{8.6}/\ipah{11.3}$)
is roughly correlated with the intensity of the PAH emission, the highest 
ratios being found in the brightest regions (spiral arms, starburst region, 
etc.), while the lowest ratios are generally found in the extended, 
low-luminosity regions.
We do not detect any significant variations of the centroid of the $7.7\mic$
feature inside each source.
However, the average centroid varies from one source to another.

\refapp{sec:tabfig} comments on each individual source.
Here we will illustrate the systematic variations within the individual sources 
by focusing on two extragalactic sources: the edge-on irregular
starburst galaxy \M{82} and the face-on spiral galaxy \M{51}.
For \M{82} (\reffig{fig:imm82}), the maximum values of the 
$\ipah{7.7}/\ipah{11.3}$ ratio are found along the disc, in the star forming 
region, where the infrared emission is the highest. 
The value of the band ratio drops by a factor of $\simeq3$ at $\simeq200$~pc
above and below the disc, in the halo.
For \M{51} (\reffig{fig:imm51}), the $\ipah{7.7}/\ipah{11.3}$ ratio is maximum 
along the circumnuclear ring.
It is somewhat lower in the $10''$ nuclear region.
The band ratio is also high along the spiral arms, coinciding with the infrared
bright blobs, but can exhibit variations by a factor of $\simeq2$.
Finally, the ratio drops in the interarm region.

The highest values of the $\ipah{7.7}/\ipah{11.3}\simeq7$ ratio
(\frmet), found in the nuclear star forming regions of \M{82} and \M{51}, are similar to the values found in the compact Galactic \hii\ regions \IR{15384} and
\IR{18317} (\reftab{tab:intens}).
Intermediate values of $\ipah{7.7}/\ipah{11.3}\simeq3$ (\frmet), similar to the
Galactic reflection nebula \ngc{2023} are found in the halo of \M{82} and along
spiral arms of \M{51}.
The lowest values of $\ipah{7.7}/\ipah{11.3}\lesssim 1$ (\frmet), seen in the planetary nebula \ngc{7027}, correspond to the outermost regions of the two 
galaxies.

\clearpage
\clearpage

\clearpage

\begin{deluxetable}{llllrrrrrrrl}
  \tabletypesize{\footnotesize}
  \rotate
  \tablecolumns{8}
  \tablewidth{0pc}
  \tablecaption{Parameters of the various correlations.}
  \tablehead{
    \colhead{Source} & \colhead{$X$} & \colhead{$Y$} & \colhead{Method}
    & \colhead{$a_{Y/X}$} & \colhead{$b_{Y/X}$} & \colhead{$\sigma_{Y/X}$}
    & \colhead{$\rho_{Y/X}$} }
  \startdata
    Integrated spectra    & $\ipah{6.2}/\ipah{11.3}$ 
                          & $\ipah{7.7}/\ipah{11.3}$ & $\mathcal{S}$ 
    & $1.78$              & $0$                 & $0.69$
    & $0.89$ \\           
    (\reffig{fig:corglo1}) 
                          &              &              & $\mathcal{L}$ 
    & $2.70$              & $0$                 & $0.64$
    & $0.85$ \\           
                          & $\ipah{8.6}/\ipah{11.3}$ 
                          & $\ipah{7.7}/\ipah{11.3}$ & $\mathcal{S}$ 
    & $8.21$              & $0$                 & $1.09$
    & $0.44$ \\           
                          &              &              & $\mathcal{L}$ 
    & $4.43$              & $0$                 & $0.79$
    & $0.78$ \\           
                          & $\ipah{6.2}/\ipah{11.3}$ 
                          & $\ipah{6.2}/\ipah{11.3}$ & $\mathcal{S}/\mathcal{L}$ 
    & $1.03$              & $0$                 & $0.46$
    & $0.45$ \\           
    %
  \hline
    \M{82}                & $\ipah{6.2}/\ipah{11.3}$ 
                          & $\ipah{7.7}/\ipah{11.3}$ & $\mathcal{S}$ 
    & $1.86$              & $0$                 & $0.32$
    & $0.97$ \\           
    (\reffig{fig:cormap1m82}) 
                          &              &              & $\mathcal{L}$ 
    & $2.70$              & $0$                 & $0.32$
    & $0.97$ \\           
                          & $\ipah{8.6}/\ipah{11.3}$ 
                          & $\ipah{7.7}/\ipah{11.3}$ & $\mathcal{S}$ 
    & $8.06$              & $0$                 & $0.53$
    & $0.87$ \\            
                          &              &              & $\mathcal{L}$ 
    & $4.04$              & $0$                 & $0.39$
    & $0.95$ \\           
                          & $\ipah{6.2}/\ipah{11.3}$ 
                          & $\ipah{6.2}/\ipah{11.3}$ & $\mathcal{S}/\mathcal{L}$ 
    & $1.07$              & $0$                 & $0.23$
    & $0.88$ \\            
                          & $\ipah{7.7}/\ipah{11.3}$ 
                          & $\lambda_{7.7}\;[\mu m]$ & $\mathcal{L}$ 
    & $0$                 & $7.68$                 & $0.012$
    & \nodata \\            
  \hline
    \IC{342}              & $\ipah{6.2}/\ipah{11.3}$ 
                          & $\ipah{7.7}/\ipah{11.3}$ & $\mathcal{S}$ 
    & $1.50$              & $0$                 & $0.63$
    & $0.82$ \\           
    (\reffig{fig:cormap1ic342}) 
                          &              &              & $\mathcal{L}$ 
    & $2.33$              & $0$                 & $1.35$
    & $0.79$ \\           
                          & $\ipah{8.6}/\ipah{11.3}$ 
                          & $\ipah{7.7}/\ipah{11.3}$ & $\mathcal{S}$ 
    & $5.71$              & $0$                 & $0.69$
    & $0.81$ \\            
                          &              &              & $\mathcal{L}$ 
    & $1.89$              & $0$                 & $1.41$
    & $0.66$ \\           
                          & $\ipah{6.2}/\ipah{11.3}$ 
                          & $\ipah{6.2}/\ipah{11.3}$ & $\mathcal{S}/\mathcal{L}$ 
    & $1.12$              & $0$                 & $0.40$
    & $0.74$ \\            
                          & $\ipah{7.7}/\ipah{11.3}$ 
                          & $\lambda_{7.7}\;[\mu m]$ & $\mathcal{L}$ 
    & $0$                 & $7.68$                 & $0.012$
    & \nodata \\            
  \hline
    \M{51}                & $\ipah{6.2}/\ipah{11.3}$ 
                          & $\ipah{7.7}/\ipah{11.3}$ & $\mathcal{S}$ 
    & $1.49$              & $0$                 & $0.50$
    & $0.62$ \\           
    (\reffig{fig:cormap1m51}) 
                          &              &              & $\mathcal{L}$ 
    & $2.59$              & $0$                 & $0.58$
    & $0.65$ \\           
                          & $\ipah{8.6}/\ipah{11.3}$ 
                          & $\ipah{7.7}/\ipah{11.3}$ & $\mathcal{S}$ 
    & $7.19$              & $0$                 & $0.62$
    & $0.41$ \\            
                          &              &              & $\mathcal{L}$ 
    & $4.55$              & $0$                 & $0.56$
    & $0.65$ \\           
                          & $\ipah{6.2}/\ipah{11.3}$ 
                          & $\ipah{6.2}/\ipah{11.3}$ & $\mathcal{S}/\mathcal{L}$ 
    & $1.01$              & $0$                 & $0.42$
    & $0.33$ \\            
                          & $\ipah{7.7}/\ipah{11.3}$ 
                          & $\lambda_{7.7}\;[\mu m]$ & $\mathcal{L}$ 
    & $0$                 & $7.72$                 & $0.012$
    & \nodata \\            
  \hline
    \M{83}                & $\ipah{6.2}/\ipah{11.3}$ 
                          & $\ipah{7.7}/\ipah{11.3}$ & $\mathcal{S}$ 
    & $1.30$              & $0$                 & $0.52$
    & $0.62$ \\           
    (\reffig{fig:cormap1m83}) 
                          &              &              & $\mathcal{L}$ 
    & $2.19$              & $0$                 & $0.54$
    & $0.67$ \\           
                          & $\ipah{8.6}/\ipah{11.3}$ 
                          & $\ipah{7.7}/\ipah{11.3}$ & $\mathcal{S}$ 
    & $7.18$              & $0$                 & $0.69$
    & $0.43$ \\            
                          &              &              & $\mathcal{L}$ 
    & $3.52$              & $0$                 & $0.70$
    & $0.66$ \\           
                          & $\ipah{6.2}/\ipah{11.3}$ 
                          & $\ipah{6.2}/\ipah{11.3}$ & $\mathcal{S}/\mathcal{L}$ 
    & $1.07$              & $0$                 & $0.41$
    & $0.39$ \\            
                          & $\ipah{7.7}/\ipah{11.3}$ 
                          & $\lambda_{7.7}\;[\mu m]$ & $\mathcal{L}$ 
    & $0$                 & $7.71$                 & $0.016$
    & \nodata \\            
  \hline
    \xxxdor               & $\ipah{6.2}/\ipah{11.3}$ 
                          & $\ipah{7.7}/\ipah{11.3}$ & $\mathcal{S}$ 
    & $1.43$              & $0$                 & $0.79$
    & $0.78$ \\           
    (\reffig{fig:cormap130dor}) 
                          &              &              & $\mathcal{L}$ 
    & $2.31$              & $0$                 & $0.67$
    & $0.68$ \\           
                          & $\ipah{8.6}/\ipah{11.3}$ 
                          & $\ipah{7.7}/\ipah{11.3}$ & $\mathcal{S}$ 
    & $11.48$             & $0$                 & $1.61$
    & $0.35$ \\            
                          &              &              & $\mathcal{L}$ 
    & $2.10$              & $0$                 & $0.62$
    & $0.49$ \\           
                          & $\ipah{6.2}/\ipah{11.3}$ 
                          & $\ipah{6.2}/\ipah{11.3}$ & $\mathcal{S}/\mathcal{L}$ 
    & $2.01$              & $0$                 & $0.86$
    & $0.44$ \\            
                          & $\ipah{7.7}/\ipah{11.3}$ 
                          & $\lambda_{7.7}\;[\mu m]$ & $\mathcal{L}$ 
    & $0$                 & $7.66$                 & $0.016$
    & \nodata \\            
  \hline
    \M{17}                & $\ipah{6.2}/\ipah{11.3}$ 
                          & $\ipah{7.7}/\ipah{11.3}$ & $\mathcal{S}$ 
    & $1.96$              & $0$                 & $0.33$
    & $0.81$ \\           
    (\reffig{fig:cormap1m17}) 
                          &              &              & $\mathcal{L}$ 
    & $2.41$              & $0$                 & $0.45$
    & $0.93$ \\           
                          & $\ipah{8.6}/\ipah{11.3}$ 
                          & $\ipah{7.7}/\ipah{11.3}$ & $\mathcal{S}$ 
    & $11.91$             & $0$                 & $0.81$
    & $0.33$ \\            
                          &              &              & $\mathcal{L}$ 
    & $4.04$              & $0$                 & $0.84$
    & $0.80$ \\           
                          & $\ipah{6.2}/\ipah{11.3}$ 
                          & $\ipah{6.2}/\ipah{11.3}$ & $\mathcal{S}/\mathcal{L}$ 
    & $1.01$              & $0$                 & $0.25$
    & $0.61$ \\            
                          & $\ipah{7.7}/\ipah{11.3}$ 
                          & $\lambda_{7.7}\;[\mu m]$ & $\mathcal{L}$ 
    & $0$                 & $7.69$                 & $0.012$
    & \nodata \\            
  \hline
    \orb                  & $\ipah{6.2}/\ipah{11.3}$ 
                          & $\ipah{7.7}/\ipah{11.3}$ & $\mathcal{S}$ 
    & $1.77$              & $0$                 & $0.32$
    & $0.83$ \\           
    (\reffig{fig:cormap1orionBar}) 
                          &              &              & $\mathcal{L}$ 
    & $2.32$              & $0$                 & $0.21$
    & $0.95$ \\           
                          & $\ipah{8.6}/\ipah{11.3}$ 
                          & $\ipah{7.7}/\ipah{11.3}$ & $\mathcal{S}$ 
    & $8.38$              & $0$                 & $0.38$
    & $0.91$ \\            
                          &              &              & $\mathcal{L}$ 
    & $3.72$              & $0$                 & $0.52$
    & $0.86$ \\           
                          & $\ipah{6.2}/\ipah{11.3}$ 
                          & $\ipah{6.2}/\ipah{11.3}$ & $\mathcal{S}/\mathcal{L}$ 
    & $1.01$              & $0$                 & $0.29$
    & $0.40$ \\            
                          & $\ipah{7.7}/\ipah{11.3}$ 
                          & $\lambda_{7.7}\;[\mu m]$ & $\mathcal{L}$ 
    & $0$                 & $7.68$                 & $0.011$
    & \nodata \\            
  \enddata
  \label{tab:correl}
  \tablecomments{This table gives, for two given measured quantities $X$ and
                 $Y$, the parameters corresponding to the fit of $Y$ by
                 $a_{Y/X}\times X + b_{Y/X}$.
                 The dispersion of the measures around this correlation is 
                 $\sigma_{Y/X}
                  =\sqrt{\langle(Y-a_{Y/X}\times X+ b_{Y/X})^2\rangle}$.
                 In the case where $b_{Y/X}=0$, $\rho_{Y/X}$ is the linear
                 correlation coefficient.
                 }
\end{deluxetable}
\clearpage


\section{INTERPRETATION OF THE AROMATIC FEATURE VARIATIONS}
\label{sec:explanation}

  \subsection{Origin of the Band Ratio Variations}
  \label{sec:pahU}
  
\begin{figure*}[htbp]
  \centering
  \includegraphics[width=\textwidth]{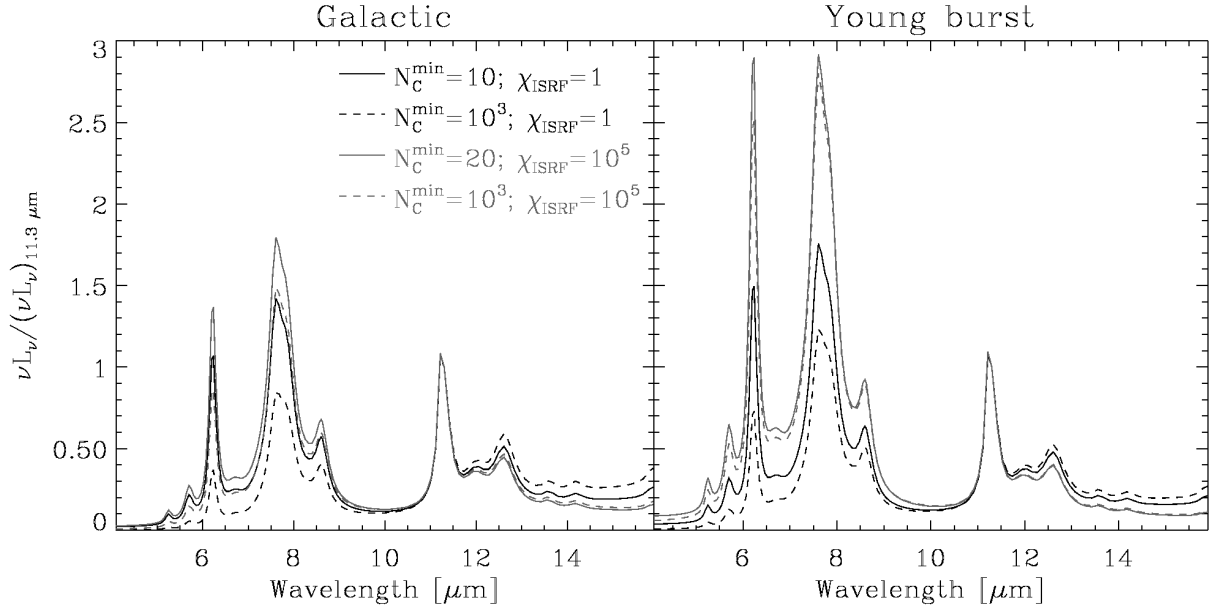}
  \caption{Effects of the temperature distribution on the PAH spectra.
           The various curves are the theoretical emission spectra, varying the
           shape (Galactic or young starburst) and 
           the intensity of the ISRF ($\chi_\sms{ISRF}$), as well as the lower 
           cut-off of the size distribution ($N_\sms{C}^\sms{min}$).
           The size distribution is the BARE-GR-S of \citet{zubko04}.
           The absorption efficiencies are taken from \citet{draine07}.}
  \label{fig:pah_th}
\end{figure*}

\begin{figure*}[htbp]
  \centering
  \includegraphics[width=0.48\textwidth]{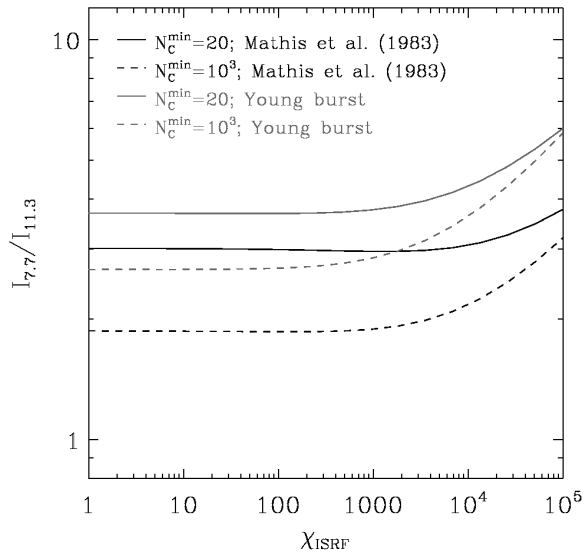}
  \caption{Theoretical modeling of the effect of the PAH temperature 
           distribution on their band ratios, with a fixed
           ionization fraction ($50\%$).
           No extinction is applied.}
  \label{fig:pahU1}
\end{figure*}

\begin{figure*}[htbp]
  \centering
  \includegraphics[width=0.48\textwidth]{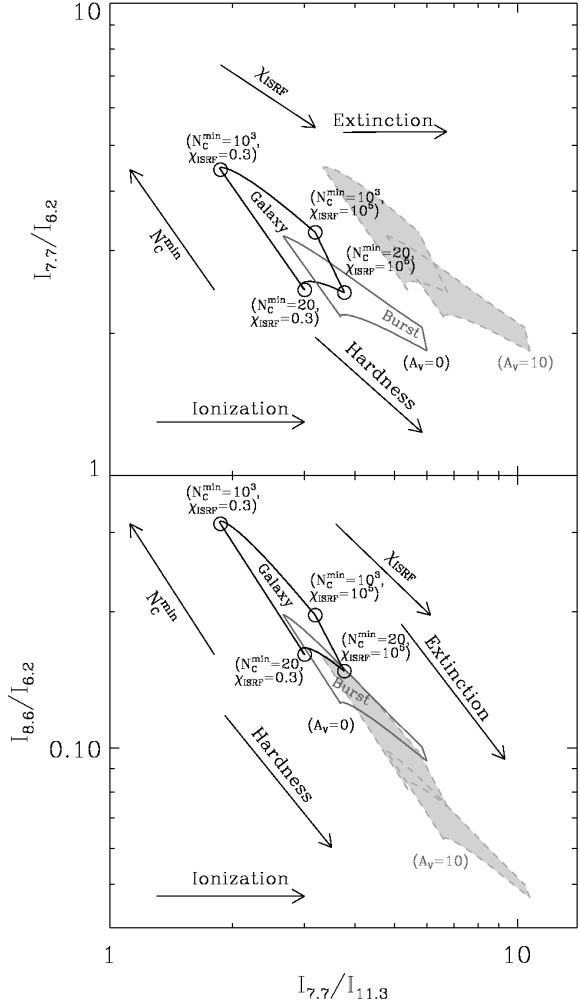}
  \caption{The two panels present the results of the same modeling as in 
           \reffig{fig:pahU1}, the only difference being that the top panel 
           shows the resulting 
           $\ipah{7.7}/\ipah{6.2}$, while the bottom panel shows the 
           $\ipah{8.6}/\ipah{6.2}$.
           To read this figure, first look at the black empty
           polygon with the circled corners.
           This polygon shows the effect of varying both the minimum cut-off 
           PAH size, $N_\sms{C}^\sms{min}$, and the ISRF intensity,
           $\chi_\sms{ISRF}$, with the Galactic ISRF (labeled {\it Galaxy})
           and with no extinction.
           The values between parenthesis, close to each corners are the values
           of these two parameters at the corner.
           Second, the grey empty polygon is the analog of the previous one,
           except that the ISRF is now the Starburst one (labeled {\it Burst}).
           Finally, the two grey filled polygons are the analog of the two 
           previous ones, except that we applied a screen extinction with 
           $A_V=10$.
           As a summary, the various arrows show the sense of variation of the 
           band ratios with each parameter, including the ionization degree of 
           the PAHs.}
  \label{fig:pahU2}
\end{figure*}

The main conclusions of \refS{sec:correl} are 
that\textlist{\thetextlist~the ratios $\ipah{6.2}/\ipah{11.3}$, 
      $\ipah{7.7}/\ipah{11.3}$ and $\ipah{8.6}/\ipah{11.3}$
      are correlated and span one order of magnitude throughout our sample, and
    \thetextlist~the ratios $\ipah{7.7}/\ipah{6.2}$ and $\ipah{8.6}/\ipah{6.2}$
      are not significantly anticorrelated with the 
      $\ipah{7.7}/\ipah{11.3}$ ratio, and do not show significant variations
      within the sources which have the best signal-to-noise ratios (especially
      \M{82}).}As
explained in \refS{sec:intro} and demonstrated in \reffig{fig:spec}, 
the intensity of the features between 6 and $9\mic$ relative to the 
$11.3\mic$ band are one order of magnitude higher for PAH$^+$ than for PAH$^0$.
Therefore, the universal correlations displayed from \reffig{fig:corglo1} to
\reffig{fig:compmaporionBar} can be attributed to variations of the charge of
the carriers of the aromatic features.
However, before concluding, we first need to explore the ability of other
physical processes to reproduce these trends.

First, dehydrogenation of the PAHs have an effect similar to ionization,
on the mid-IR spectrum, as shown by several laboratory studies 
\citep{allamandola89,allain96,jochims99}.
However, as reported by \citet{allain96}, only PAHs containing less than 
$N_\sms{C}\simeq 50$ carbon atoms can be considerably dehydrogenated.
At the same time, this threshold corresponds to the minimum size of PAHs
that can survive in most PDRs \citep{allain96}.
For comparison, the PAH size distribution derived by \citet{zubko04}, for the
diffuse Galactic ISM, has a lower cut-off of $N_\sms{C}^\sms{min}=20$.
Their emission is dominated by PAHs of $N_\sms{C}\simeq50-500$, depending 
on their excitation rate.
This conclusion is supported by the observed linear correlation of the 
3.3$\mic$ band with the 11.3$\mic$ band in Galactic sources \citep{hony01}.
If dehydrogenation were important then a non-linear relation would be expected as duo and trio groups (with bands longwards of 11.3$\mic$) were converted into 
solo groups \citep[with a 11.3$\mic$ band;][]{schutte93,hony01}. 
Likewise, if dehydrogenation were important, then the decrease in the 
$\ipah{11.3}/\ipah{6.2}$ ratio (decreasing H coverage) would be expected to 
be accompanied by a decreasing $\ipah{12.7}/\ipah{11.3}$ ratio (conversion of
duo/trio's into solo's) in contrast to the observations \citep{hony01}.
Therefore, we can neglect the effect of dehydrogenation on our mid-IR spectra.

Second, deep extinction by the silicate feature at $9.7\mic$ can
cause variations of the band ratios, as proposed by \citet{peeters02}, 
\citet{spoon02} or \citet{brandl06}.
Indeed, the $11.3\mic$ feature is more absorbed than the $7.7\mic$ band.
This effect would also cause the $8.6\mic$ feature to suffer from the 
same amount of extinction as the $11.3\mic$ band, and these two bands should
be correlated.
However, this is not what \reffigs{fig:corglo1} to \reffig{fig:cormap2m17}
show.
Moreover, the \frmet\ corrects the bands for extinction (\refS{sec:decomp}).
The good agreement between the \frmet\ and the \nlmet\ tells us that extinction
can not explain the majority of the observed band ratio variations.
Finally, there is no evidence for silicate absorption in the spectra of
the \orb. 
Indeed, the total gas and dust column along the line of sight for this source is 
insufficient to produce noticeable 10$\mic$ absorption.

Third, a modification of the temperature distribution of the PAHs 
can affect the relative weight of the aromatic features.
For example, truncating the highest temperatures would have the effect of 
decreasing the intensity of the short wavelength bands relative to the long 
wavelength ones.
There are essentially two ways of modifying the temperature distribution of the
molecules:\textlist{\thetextlist~by varying the interstellar 
            radiation field (ISRF) intensity or shape, and
          \thetextlist~by changing their size 
distribution.}The latter effect could happen in the vicinity of a very hard 
radiation source, like an AGN, where the smallest PAHs could undergo
photosublimation, as proposed by \citet{smith07}.

We performed theoretical modeling of the PAH emission spectra,
in order to study the latter effects.
These spectra, which take into account the stochastic heating of the particles,
are shown in \reffig{fig:pah_th}.
We adopted the PAH size distribution by \citet[][bare grains, solar abundance constraints; BARE-GR-S]{zubko04}, and the absorption efficiencies by 
\citet{draine07}, with an ionization fraction of $50\%$.
We consider successively two classes of ISRF, in order to explore the effect
of the hardness of the radiation on the PAH 
emission:\textlist{\thetextlist~the Galactic ISRF of \citet{mathis83} and
         \thetextlist~a very hard ISRF, corresponding to an instantaneous burst
           of star formation with a Salpeter IMF, synthesized with the stellar
           evolutionary model \peg\ \citep{fioc97}.}The 
latter is normalised to the intensity of the \citet{mathis83} ISRF.
We also vary the intensity of each ISRF, by multiplying them 
by a factor $\chi_\sms{ISRF}=1-10^5$.
In addition, we explore the effect of a possible small PAH segregation, by 
varying the value of the lower cut-off of the PAH size distribution.
This lower cut-off is $N_\sms{C}^\sms{min}=20$ carbon atoms, for the 
\citet{zubko04} model.
We vary it up to $N_\sms{C}^\sms{min}=10^3$.
Finally, we apply a screen extinction to the modeled spectra
by multiplying them by a factor $\exp[-\tau(\lambda,A_V)]$.
We adopt the \citet{zubko04} optical depth, $\tau(\lambda,A_V)$, and 
plot the results using two values of the V band attenuation: $A_V=0$ 
(no extinction) and $A_V=10$.

\reffigs{fig:pahU1}-\ref{fig:pahU2} synthesize the results of this modeling. 
\reffig{fig:pahU1} demonstrates that effect of the 
radiation field intensity on the $\ipah{7.7}/\ipah{11.3}$ band ratio, for 
the two types of radiation fields and the two extreme PAH size cut-offs.
This figure shows that, for a given size distribution and ISRF type, the
$\ipah{7.7}/\ipah{11.3}$ ratio is independent of the intensity of the ISRF,
up to $\chi_\sms{ISRF}\simeq10^4$.
Indeed, the PAHs are stochastically heated below $\chi_\sms{ISRF}\simeq10^4$,
therefore the shape of their emission spectrum is independent of the ISRF.
However, when $\chi_\sms{ISRF}\gtrsim10^4$, the largest PAHs reach thermal 
equilibrium, and the temperature fluctuation range shrinks significantly enough
to affect the ratio between the bands.
\reffig{fig:pahU2} shows how every effect considered here
affects the relation between $\ipah{7.7}/\ipah{11.3}$ and 
$\ipah{7.7}/\ipah{6.2}$ and $\ipah{8.6}/\ipah{6.2}$.
The arrows indicate the direction toward which these parameters affect the 
relation.
This figure shows that the combination of all the effects affecting the
PAH temperature distribution (ISRF intensity, ISRF hardness and size distribution) can not explain
a variation of the $\ipah{7.7}/\ipah{11.3}$ ratio by more than a factor 
$\simeq3$.
Moreover, it shows that if these effects were responsible for the
variation of the $\ipah{7.7}/\ipah{11.3}$ ratio, then the 
$\ipah{7.7}/\ipah{6.2}$ ratio would also be anticorrelated with it.
The ratio $\ipah{7.7}/\ipah{6.2}$ is particularly very sensitive to 
$N_\sms{C}^\sms{min}$.
According to \reffig{fig:pahU2}, an increase of the 
$\ipah{7.7}/\ipah{11.3}$ ratio by a factor $\simeq3$ would imply a 
decrease of the $\ipah{7.7}/\ipah{6.2}$ ratio by a factor of $\simeq2$.

One of the strong results of \refS{sec:corglo} and \refS{sec:cormap}
is that the ratios $\ipah{7.7}/\ipah{6.2}$ do not show any trend with
$\ipah{7.7}/\ipah{11.3}$.
Therefore, the correlations discussed in \refS{sec:correl} can not be attributed
to a modification of the temperature distribution.
In particular, we can rule out an interpretation of the variation of the 
$\ipah{7.7}/\ipah{11.3}$ in terms of destruction of the smallest PAHs.
These results are in good agreement with those by \citet{hony01}, who
studied a wide sample of Galactic sources and showed that the 
$\ipah{3.3}/\ipah{11.3}$ ratio was not varying significantly, while the 
$\ipah{6.2}/\ipah{11.3}$ ratio was varying by a factor of 5.
In addition, \reffig{fig:pahU2} shows that the effect of extinction
do not account for our observed trends.
Indeed, the bottom panel of \reffig{fig:pahU2} shows that the $8.6\mic$ and
$11.3\mic$ bands would be correlated if extinction was the main cause of 
variation of the aromatic bands.
Consequently, our work shows that the variation of the PAH band ratios,
throughout different environments, and at different spatial scales, is 
primarily controled by the ionization fraction of the molecules.
In addition, it appears that the mixture of PAH molecules, in all these 
environments, is remarkably universal.

We emphasize that our sample contains only a few AGNs.
Therefore, our results reflect primarily the properties of star forming 
regions.
Observations of AGN environments by \citet{smith07} and \citet{kaneda07}
show PAH properties that differ from our trends, suggesting that the AGN 
could alter the PAH composition.

  \subsection{Relating the Band Ratios to the Physical Conditions}
  \label{sec:calibration}
  
In \refS{sec:pahU}, we showed that the band ratio variations were 
essentially due to a variation of the fraction of ionized PAHs in the beam.
The mid-IR feature ratios are therefore related to the physical conditions
where the emission is originating.
Indeed, the fraction of ionized PAHs, in a given region, depends on the quantity
$G_0/n_e\times\sqrt{T_\sms{gas}}$ \citep[][for a review]{tielens05}, where
$G_0$, is the UV radiation field density, $n_e$,
the electron density, and $T_\sms{gas}$, the gas temperature.
The UV field density is the integration between $\lambda=0.09\mic$ and 
$\lambda=0.2\mic$ of the monochromatic mean intensity, 
$J_\lambda$, and is normalised to the solar neighborhood value:
\begin{equation}
  G_0 = \frac{\displaystyle
              \int_{0.09\mic}^{0.2\mic} 4\pi\,J_\lambda\ddiff\lambda}
             {1.6\E{-6}\;\rm W\,m^{-2}}.
  \label{eq:G0}
\end{equation}
We emphasize that, although \S5.1 showed that the intensity of the ISRF
was not responsible for the observed variations of the aromatic band ratios, 
the intensity of the ISRF is likely to vary significantly within and among our
sources.
In this section, we use the most well-studied Galactic regions of our sample, in order to derive an empirical relation between the $\ipah{6.2}/\ipah{11.3}$ band
ratio and the quantity $G_0/n_e\times\sqrt{T_\sms{gas}}$.

\begin{table}[htbp]
  \centering
  \begin{tabular}{lrrrr}
    \hline\hline
      Region     & $G_0$ & $T_\sms{gas}$ [K] 
        & $n_\sms{H}\;\rm [cm^{-3}]$ & Reference                 \\
    \hline
      \ngc{2023} & $1.5\E{4}$      & 750
        & $10^5$                     & \citet{steiman-cameron97} \\
      \ngc{7027} & $6\E{5}$        & 2000
        & $10^7$                     & \citet{justtanont97}      \\
      \orb\ (position 4) & $4\E{4}$        & 500
        & $5\E{4}$                   & \citet{tauber94}          \\
    \hline
  \end{tabular}
  \caption{Physical conditions in select regions.
           The last columns refers to the study where these quantities have
           been derived.}
  \label{tab:g0ne}
\end{table}
\citet{bregman05} gave such a calibration, from the spectro-images of three 
reflection nebulae.
In order to derive the physical conditions in their nebulae, they made 
several hypotheses.
They\textlist{\thetextlist~assumed a constant electronic density throughout
      their regions,
    \thetextlist~they neglected dust extinction, and
    \thetextlist~they derived the relative geometry of the star and nebula
      from the profile of the scattered light.}In
our analysis, we have adopted a slightly different strategy.
We have made the conscious decision to include only the sources for which 
the physical conditions have been reliably determined, using PDR models, 
constrained by several gas lines.
The sources of our sample which have been modeled in detail are 
\ngc{2023}, \ngc{7027} and position~4 of the \orb\ \citep{roche89}.
Their properties are summarized in \reftab{tab:g0ne}.

\begin{figure*}[htbp]
  \centering
  \begin{tabular}{cc}
    \includegraphics[width=0.48\textwidth]{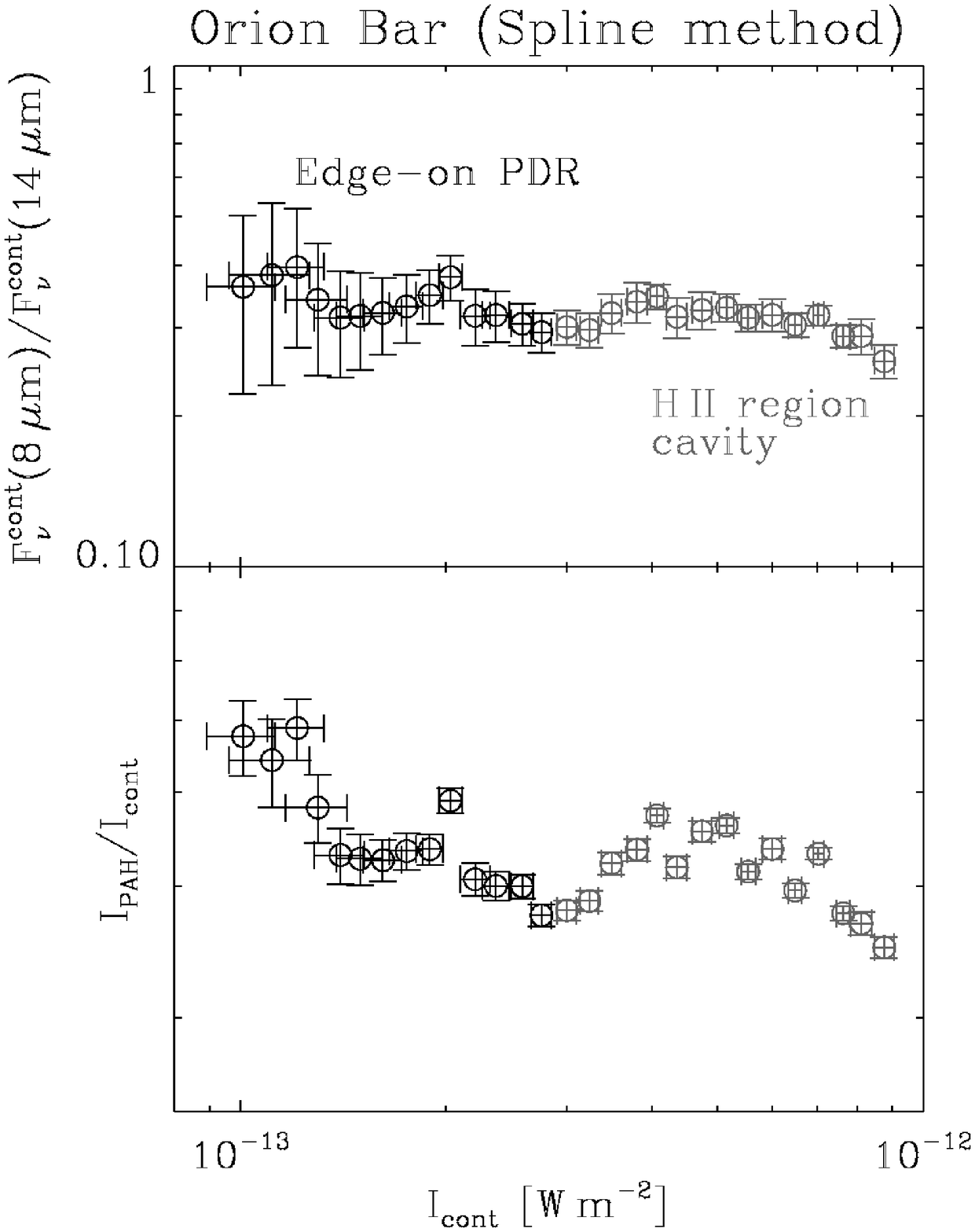} &
    \includegraphics[width=0.48\textwidth]{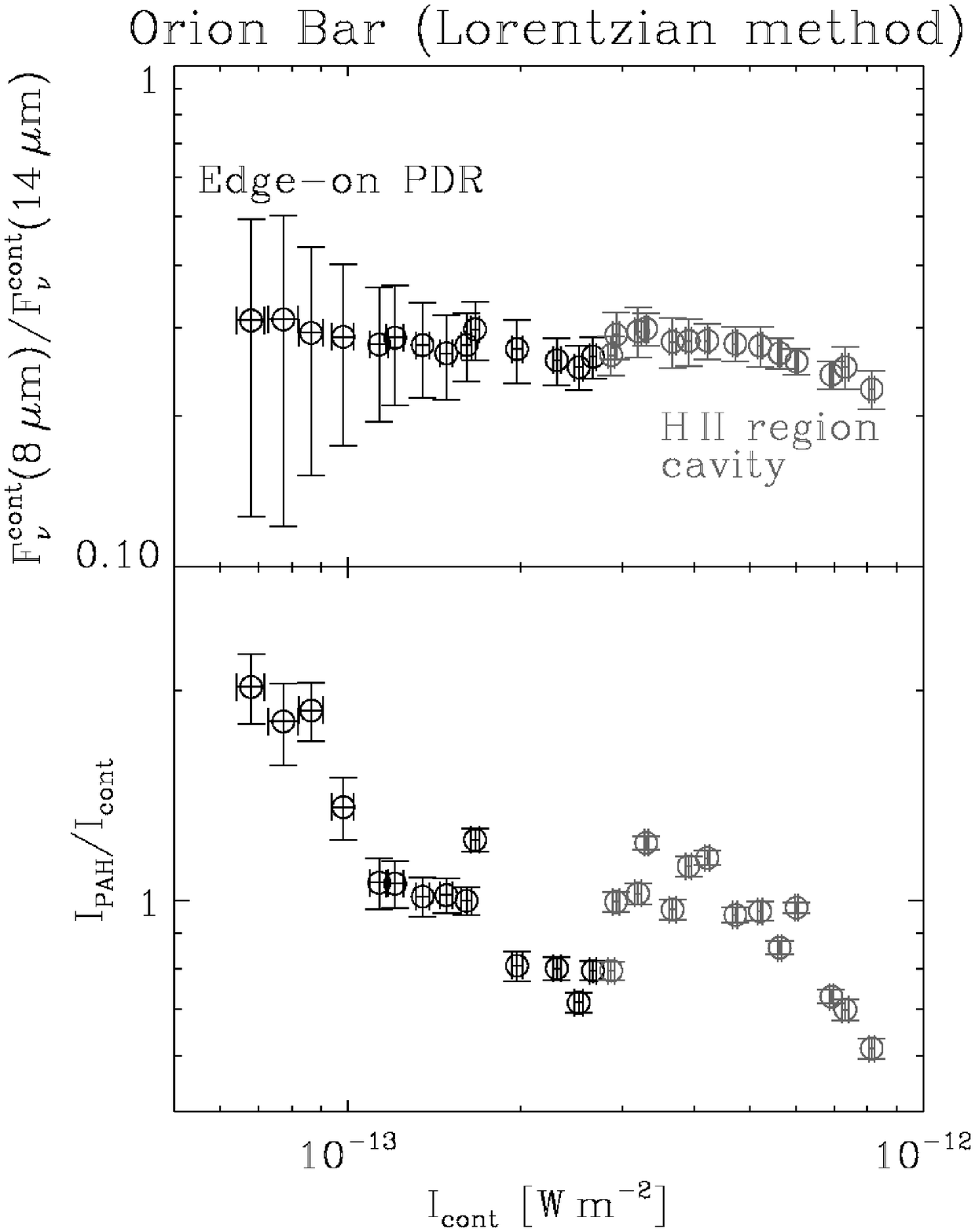} \\
  \end{tabular}
  \caption{{\bf Top panels:} variation of the color of the VSG continuum
           (ratio of the flux at 8 and $14\mic$) as a function of the intensity
           of the VSG continuum, inside the \orb.
           These figures show that the color of the VSG continuum is virtually 
           independent of its intensity, since these grains are stochastically 
           heated.
           {\bf Bottom panels:} variation of the PAH-to-VSG ratio 
           as a function of the intensity.
           These figures show that the relative strength of the PAHs in the PDR
           decreases slightly toward the \hii\ region (when $\icont$ increases),
           as an effect of their photo-depletion.}
  \label{fig:colorVSG}
\end{figure*}
\begin{figure}[htbp]
  \centering
  \includegraphics[width=0.48\textwidth]{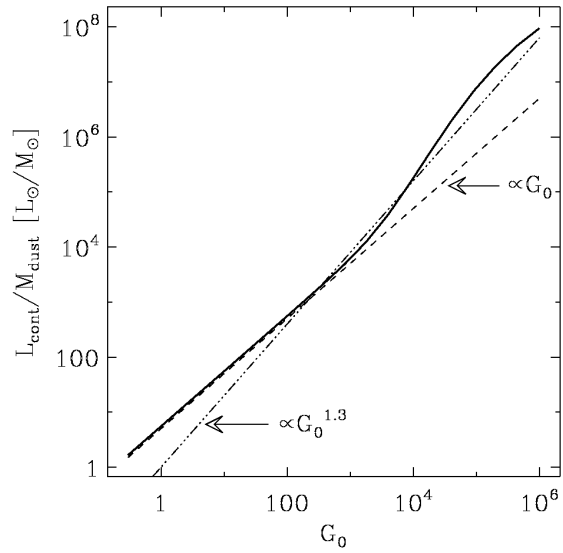}
  \caption{Theoretical modeling of the continuum intensity as a function
           of the UV radiation field, $G_0$, in black.
           $L_\sms{cont}/M_\sms{dust}$ is the luminosity of the
           graphite component of \citet{zubko04}, integrated between 10 and
           $16\mic$, and normalized to the dust mass.
           The dotted line varies as $G_0$.
           The dashed-dotted line is a power-law fit to the black line,
           in the range $10^2<G_0<10^4$.}
  \label{fig:contU}
\end{figure}
\begin{figure*}[htbp]
  \centering
  \begin{tabular}{cc}
    \includegraphics[width=0.48\textwidth]{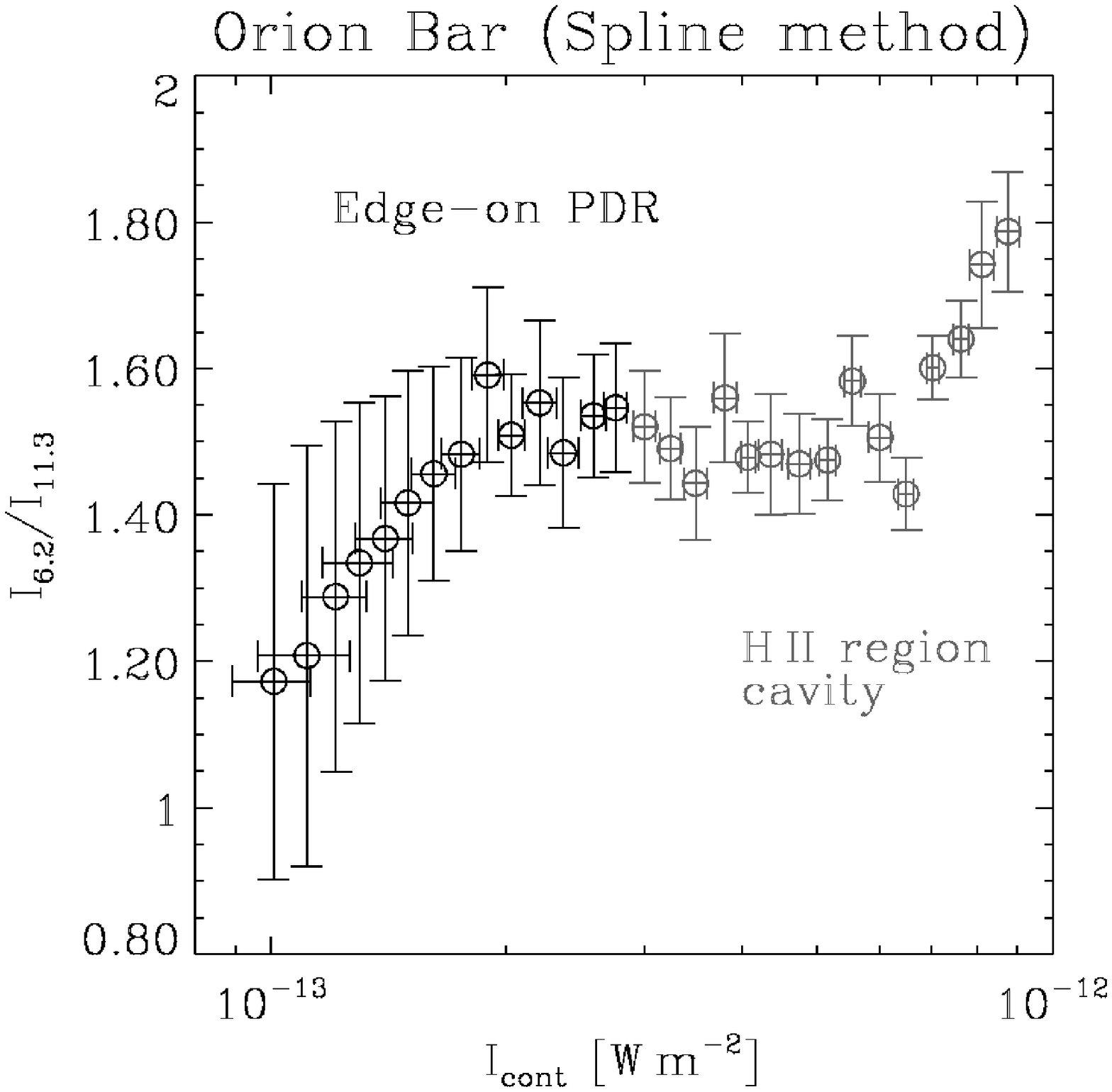} &
    \includegraphics[width=0.48\textwidth]{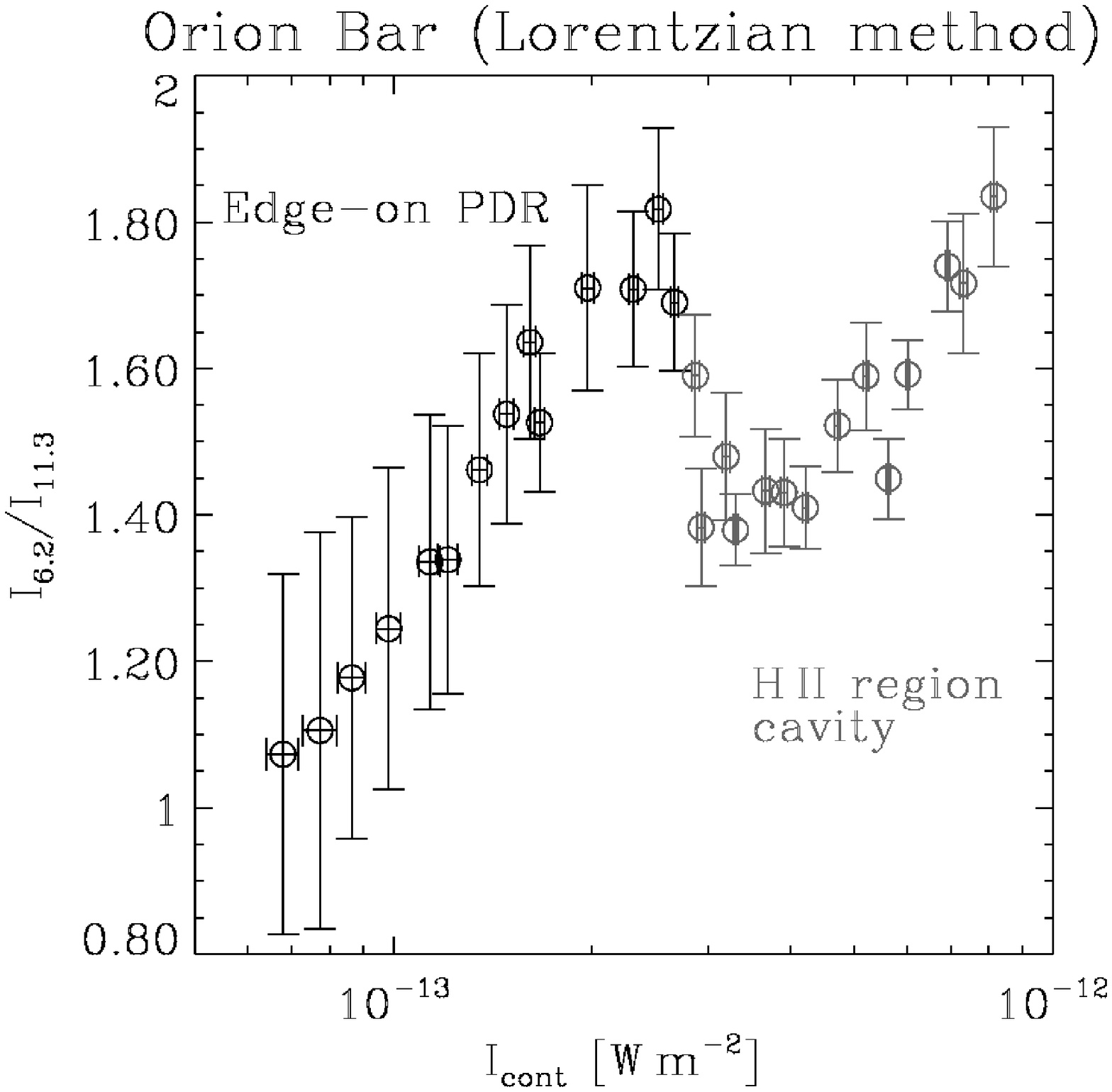} \\
    \includegraphics[width=0.48\textwidth]{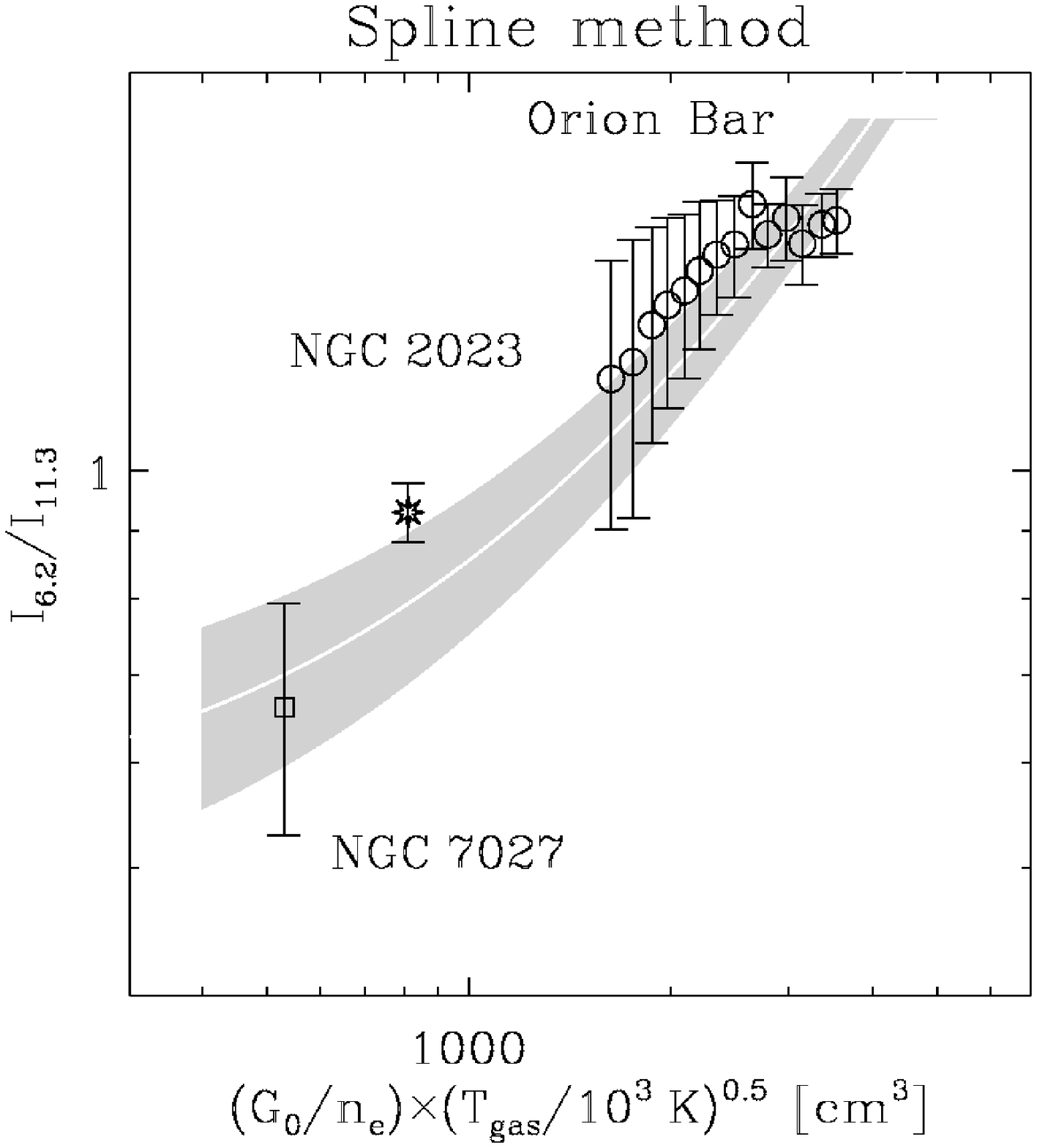} &
    \includegraphics[width=0.48\textwidth]{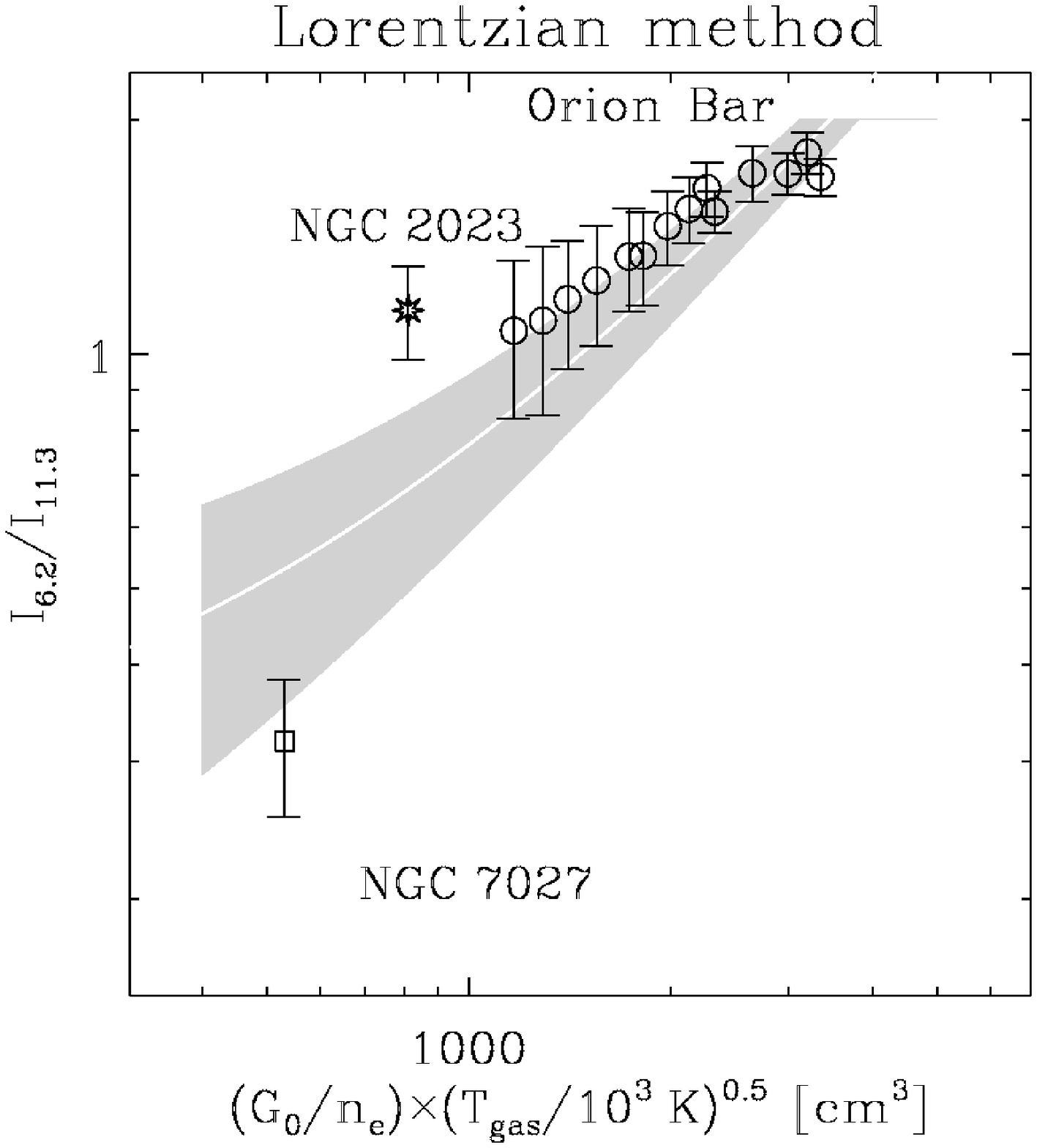} \\
  \end{tabular}
  \caption{Empirical calibration of the band ratio.
           The two upper panels show the variation of the band ratio inside
           the \orb, as a function of the VSG intensity.
           The black error bars correspond to the edge-on PDR,
           and the grey error bars, to the \hii\ region cavity.
           The two lower panels show the final calibration, including 
           \ngc{2023} and \ngc{7027}.
           We kept only the edge-on PDR measurements, in the \orb.}
  \label{fig:g0ne}
\end{figure*}

We adopt the values given in \reftab{tab:g0ne} and
assume that the electrons result from photo-ionization of carbon and that all
the gas phase carbon is ionized in PDRs, so that:
\begin{equation}
 n_e \simeq \displaystyle\left(\frac{\rm C}{\rm H}\right) \times n_\sms{H}
     \simeq 1.6\E{-4}\, n_\sms{H},
\end{equation}
where we have adopted the carbon abundance by \citet{sofia04}.
For \ngc{2023} and \ngc{7027}, we do not have reliable spatial information,
hence we will only use the global values of the ratios.
In the \orb, we study the spatial variations by proceeding as follows.
\begin{enumerate}
  \item We exclude the region north of the bar (see \reffig{fig:imorionBar}), 
    which is the location of the \hii\ region cavity.
    Indeed, the PAH-to-continuum ratio is very low there, thus the fits are
    uncertain.
    Moreover, the spherical geometry of the region causes confusion along the
    lines of sight.
    To the contrary, the region south of the bar, is an edge-on PDR, where
    the conditions are believed to be roughly homogeneous along each line
    of sight.
    Indeed, it is observed to be well stratified in the main PDR tracers 
    (CO data as well as \hmol, \ci, \cii\ and PAH data; 
    \citealt{tielens93,tauber94,wyrowski97}; and HCN 
    data by \citealt{fuente96}).
  \item We assume that the column density is homogeneous in the edge-on PDR,
    so that we can use the intensity of the VSG continuum, in order to scale 
    the intensity of the UV radiation field, $G_0$.
    Indeed, the top panels of \reffig{fig:colorVSG} show that the 
    VSGs are stochastically heated.
    Therefore, their intensity scales with the radiation field density.
    Moreover, the intensity of the VSG continuum is probably a better tracer
    of the radiation density than the PAHs, since the latter are gradually 
    depleted in direction of the \hii\ region (bottom panels of 
    \reffig{fig:colorVSG}).
    \reffig{fig:contU} illustrates this fact by showing the theoretical
    variation of the integrated intensity of the VSG continuum between 10 and 
    $16\mic$, as a function of $G_0$.
    To perform this simulation, we used the radiation field defined in 
    \refeq{eq:G0}, and the Galactic graphite size distribution of 
    \citet[with solar abundance constraints]{zubko04}.
    For $G_0\lesssim10^3$, the mid-IR continuum intensity scales 
    perfectly with $G_0$, since the grains are out of thermal equilibrium.
    At higher $G_0$, the relation between $G_0$ and $I_\sms{cont}$
    becomes non-linear due to temperature effects.
    In the range $10^2\lesssim G_0\lesssim10^4$, which are the conditions in the
    \orb, the UV radiation field is very well approximated by 
    $G_0\propto I_\sms{cont}^{1/1.3}$.
    This method has the advantage of taking into account the actual 
    variations of $G_0$, with both the distance from the star cluster, and
    the dust extinction.
  \item In order to improve the signal-to-noise ratio of our spectra, we 
    average the pixels into 30 bins of energy ($I_\sms{cont}$).
    The two top panels of \reffig{fig:g0ne} show the variation of the
    $\ipah{6.2}/\ipah{11.3}$ ratio, as a function of the continuum intensity,
    with the two methods.
  \item Finally, we assume a constant electron density and temperature
    throughout the PDR.
\end{enumerate}
The two bottom panels of \reffig{fig:g0ne} shows the final calibration of the
band ratio, with the two methods.
As found by \citet{bregman05}, the ratio $\ipah{6.2}/\ipah{11.3}$ 
(or $\ipah{7.7}/\ipah{11.3}$) increases in the range 
$10^2\lesssim G_0/n_e\lesssim 10^3$.
The grey stripes are the following linear fits:
\begin{equation}
  \left\{
  \begin{array}{rcll}
    \displaystyle \frac{\ipah{6.2}}{\ipah{11.3}} & \simeq & 
      \left(\displaystyle\frac{G_0}{n_e\;\rm[cm^{-3}]}
             \sqrt{\displaystyle\frac{T_\sms{gas}}{10^3\;\rm K}}\right)/3040
      + 0.53 \pm 0.10 & \mbox{ (\nlmet),} \\
    & & \\
    \displaystyle \frac{\ipah{6.2}}{\ipah{11.3}} & \simeq &
      \left(\displaystyle\frac{G_0}{n_e\;\rm[cm^{-3}]}
             \sqrt{\displaystyle\frac{T_\sms{gas}}{10^3\;\rm K}}\right)/1990
      + 0.26 \pm 0.16 & \mbox{ (\frmet),} \\
  \end{array} \\
  \right.
  \label{eq:calib}
\end{equation}
valid in the range $400\lesssim
\displaystyle\frac{G_0}{n_e\;\rm[cm^{-3}]}
\sqrt{\displaystyle\frac{T_\sms{gas}}{10^3\;\rm K}} \lesssim 4000$.

  \subsection{Diagnosing Galaxies Using PAH Band Ratios}

In principle, \refeq{eq:calib} can be used to derive the ratio 
$G_0/n_e\times\sqrt{T_\sms{gas}}$ from any mid-IR spectrum.
However, if several regions with different physical conditions are encompassed
within the spectrograph's beam, the situation is more complex.
The band ratio is then averaged over these various environments and its global 
value does not reflect a single $G_0/n_e\times\sqrt{T_\sms{gas}}$.

\begin{figure*}[htbp]
  \centering
  \begin{tabular}{cc}
    \includegraphics[width=0.48\textwidth]{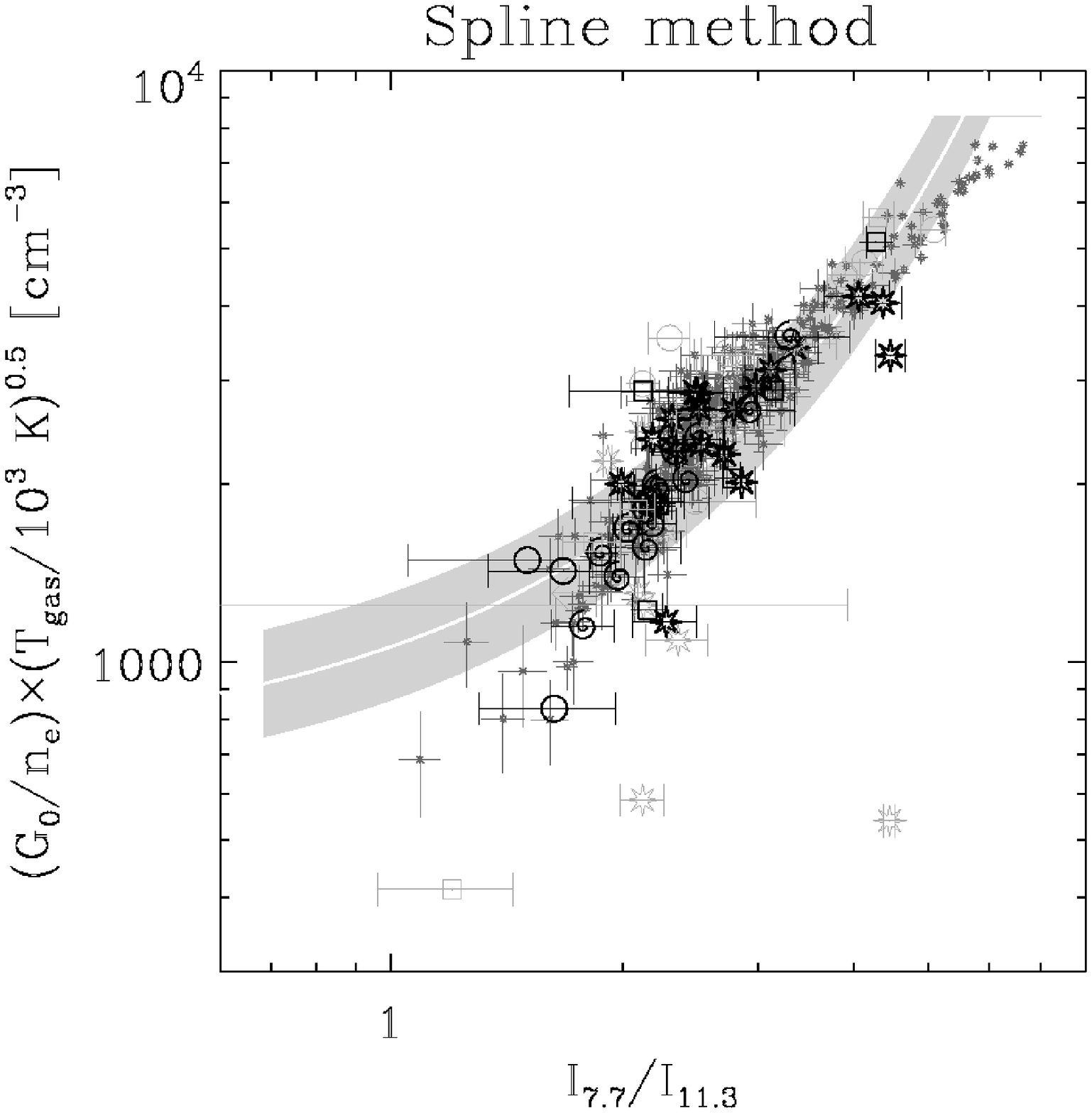} &
    \includegraphics[width=0.48\textwidth]{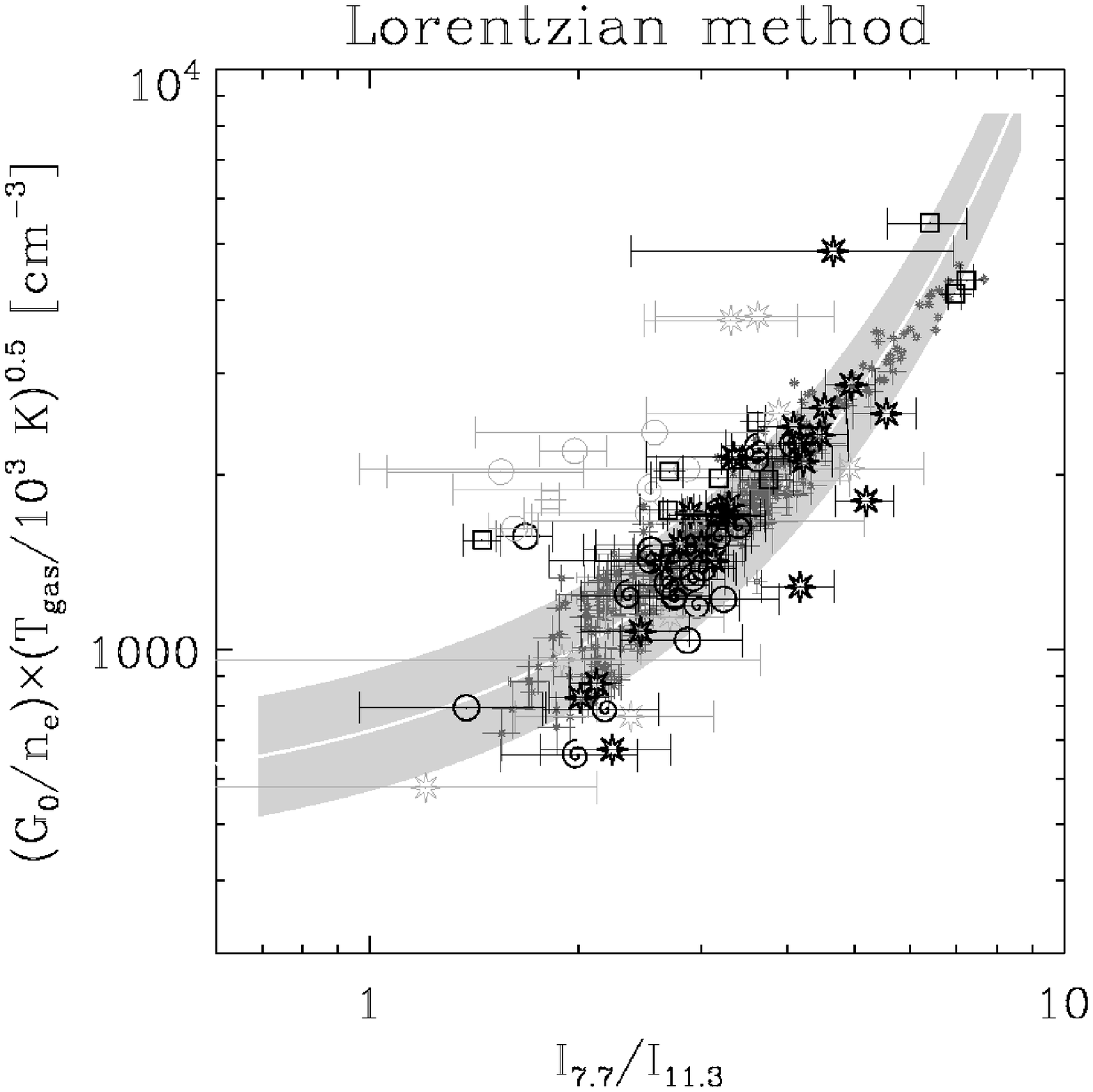} \\
  \end{tabular}
  \caption{Determination of $G_0/n_e\times\sqrt{T_\sms{gas}}$ for the
           sources in our sample.
           These figures are similar to \reffig{fig:corglo1}, where the 
           $\ipah{6.2}/\ipah{11.3}$ has been converted into 
           $G_0/n_e\times\sqrt{T_\sms{gas}}$ using \refeq{eq:calib}.
           The same symbol conventions are adopted as in \reffig{fig:corglo1}. 
           The small additional symbols are the spatial variations within 
           \M{82}.}
  \label{fig:final}
\end{figure*}
In the case of the global spectrum of a star forming region, we can 
expect the $\ipah{7.7}/\ipah{11.3}$ ratio to depend on the age and initial 
mass function of the stellar cluster, that will determine the $G_0$ at the 
edge of the PDR, as well as on the geometry of the ISM (clumpiness, densities),
that will determine the transfer of the UV light into the PDR, and the electron 
density.
If this hypothesis is correct, the value of $\ipah{7.7}/\ipah{11.3}$ can
potentially be used in combination with other PDR tracers, like the 
\ciiline, \oiline, \siIIline\ and \hmol\ mid-IR rotational lines, to 
determine the geometry and physical conditions of an unresolved 
star forming region.
\reffig{fig:final} demonstrates the relation between the observed band 
ratios and the averaged $G_0/n_e\times\sqrt{T_\sms{gas}}$ quantity.

It has long been surmised that the IR emission features provide a clear 
mid-IR signature of the interaction of FUV photons with cloud surfaces and 
hence a probe of the importance of (massive) star formation in a region 
\citep{genzel98,peeters04}.
The present study has extended this by developing the observed 6.2 to 
11.3$\mic$ band ratio as a quantitative tool to probe the physical conditions 
({e.g.} $G_0/n_e\times\sqrt{T_\sms{gas}}$) in the emitting regions.
We expect that this study will be of fundamental value for the interpretation 
of \spitz\ data as well as future \sofia\ and \jwst\ observations of galaxies in 
the nearby and early universe.


\section{SUMMARY AND CONCLUSION}
\label{sec:concl}

We presented the results of a systematic study
aimed at understanding the main properties of the mid-IR features in 
different astrophysical environments.
It is based on observations of Galactic regions and galaxies of various types,
observed by the satellites \iso\ and \spitz.
We have developed two distinct methods of spectral decomposition with
different hypotheses, in order to test the robustness of our trends, and 
overcome eventual biases.
These two methods have shown similar trends between the physical quantities 
that we have studied.
We explored the variations of the different mid-IR features between the
wavelengths 5 and $16\mic$, among integrated spectra of galaxies, and
inside galaxies and Galactic regions.
Our main results are the followings.
\begin{enumerate}
  \item We find that the 6.2, 7.7, $8.6\mic$ features are essentially tied 
    together, 
    while the ratio of these bands to the $11.3\mic$ feature can vary by one
    order of magnitude, in our sample.
    These variations are seen both inside individual sources (like \M{82},
    \M{17}, the \orb, \xxxdor, etc.), as well as among integrated spectra.
    In general, the $\ipah{6.2}/\ipah{11.3}$ ratio is spatially correlated
    with the power radiated by the PAHs.
    It indicates that the ratio $\ipah{6.2}/\ipah{11.3}$ is higher in regions
    of intense star formation.
  \item With the help of a stochastic heating model and realistic absorption
    efficiencies, we show that the variations of the mid-IR features are
    essentially due to the variation of the fraction of ionized PAHs.
    We conclude that the properties of the PAHs, throughout our sample, 
    are remarkably universal.
    In particular, we rule out both the modification of the grain size 
    distribution, and the extinction by the $9.7\mic$ silicate feature,
    as an explanation of these variations.
    Indeed, we show that a modification of the size distribution could explain
    the observed variation of the $\ipah{6.2}/\ipah{11.3}$ ratio.
    However, it would cause the $\ipah{6.2}/\ipah{7.7}$ and 
    $\ipah{6.2}/\ipah{8.6}$ (and $\ipah{3.3}/\ipah{11.3}$) ratios to vary 
    significantly and to be correlated with the $\ipah{6.2}/\ipah{11.3}$, 
    contradicting our observations.
    Similarly, a deep absorption by the silicate feature would vary the 
    $\ipah{6.2}/\ipah{11.3}$ ratio, but it would decouple the $8.6\mic$
    from the 6.2 and $7.7\mic$ features at the same time.
  \item The universality of the properties of the PAHs and the fact that the 
    band ratios are mainly sensitive to the charge of the molecules allow us 
    to use these band ratios as a tracer of the physical conditions inside the 
    emitting region.
    Using a few well-studied Galactic regions (including the spectral image 
    of the \orb), we give an empirical relation between $\ipah{6.2}/\ipah{11.3}$ 
    and the ratio $G_0/n_e\times\sqrt{T_\sms{gas}}$.
  \item In the case where several regions with different physical conditions
    are integrated, the band ratios are dependent on the morphology of the ISM,
    as well as on the evolutionary properties of the illuminating
    stellar cluster.
    Thus, we find that the $\ipah{6.2}/\ipah{11.3}$ band ratios in the
    star forming nuclei of \M{51} and \M{82} are very similar to those 
    in the galactic compact \hii\ regions, \IR{15384} and \IR{18317},
    while the halo of \M{82} and the spiral arms of \M{51} are similar 
    to those in the Galactic reflection nebula \ngc{2023}. 
    These differences in band ratio reflect differences in the 
    ionization over recombination rate and hence trace back to 
    variations in the ratio of the ionizing radiation field to the  
    electron density.
\end{enumerate}


\appendix
\section{ADDITIONAL TABLE AND FIGURES}
\label{sec:tabfig}

  \subsubsection*{Notes on Individual Sources}

\begin{figure*}[htbp]
  \centering
  \includegraphics[width=\textwidth]{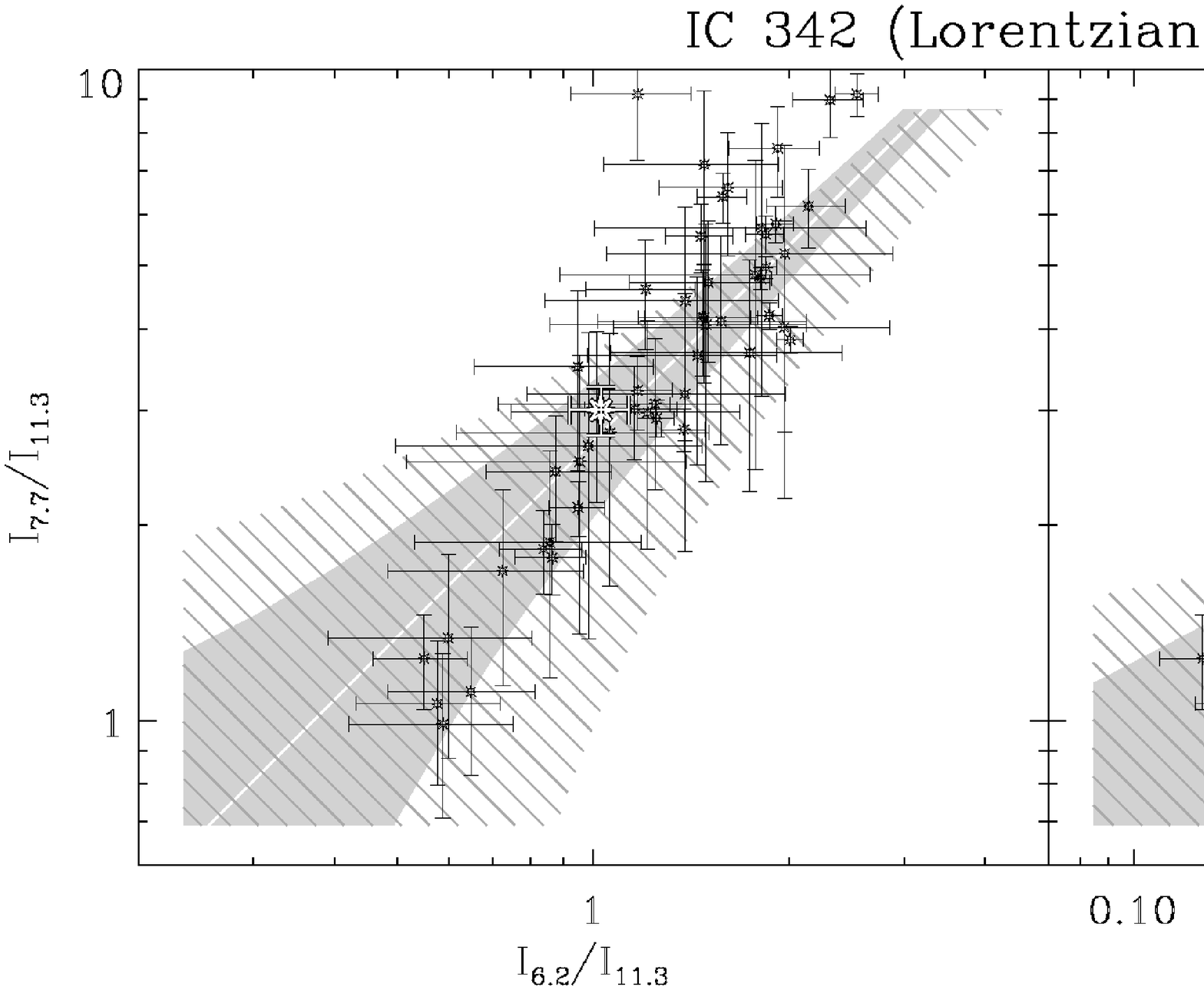} \\
  \includegraphics[width=\textwidth]{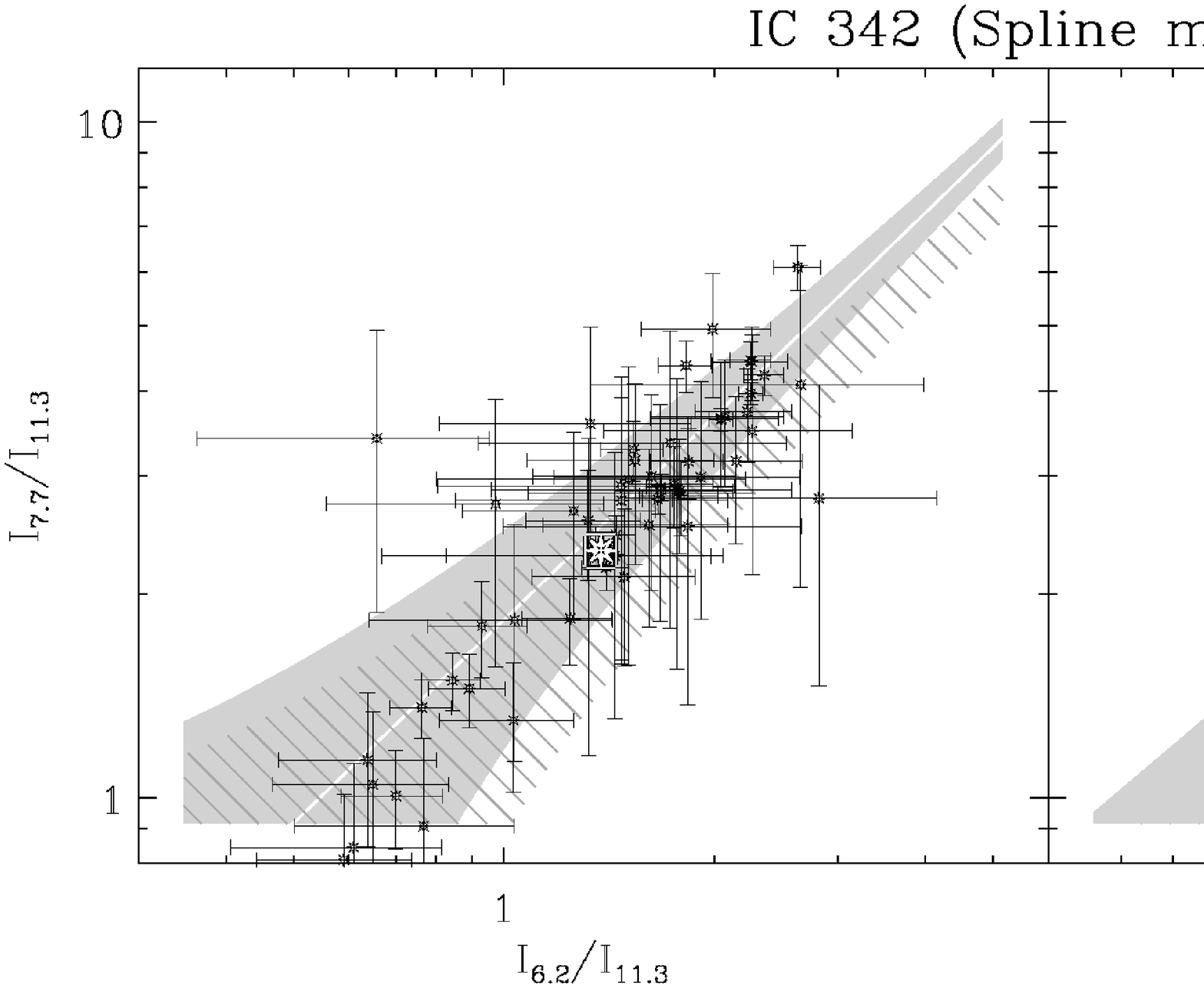}
  \caption{PAH band ratios within \IC{342}.
           The same symbol conventions are adopted as in 
           \reffig{fig:cormap1m82}.}
  \label{fig:cormap1ic342}
\end{figure*}

\begin{figure*}[htbp]
  \centering
  \begin{tabular}{cc}
    \includegraphics[width=0.48\textwidth]{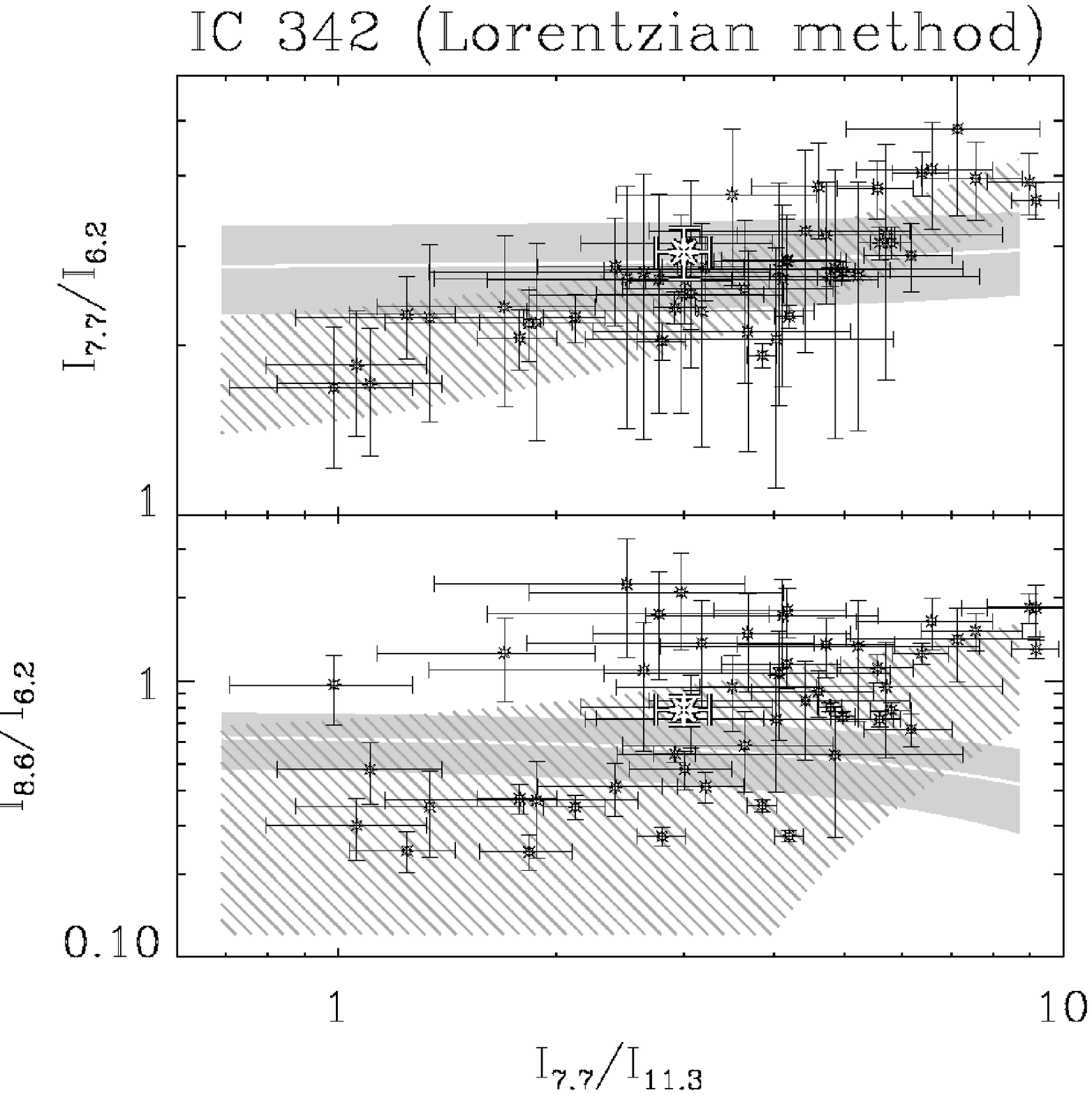} &
    \includegraphics[width=0.48\textwidth]{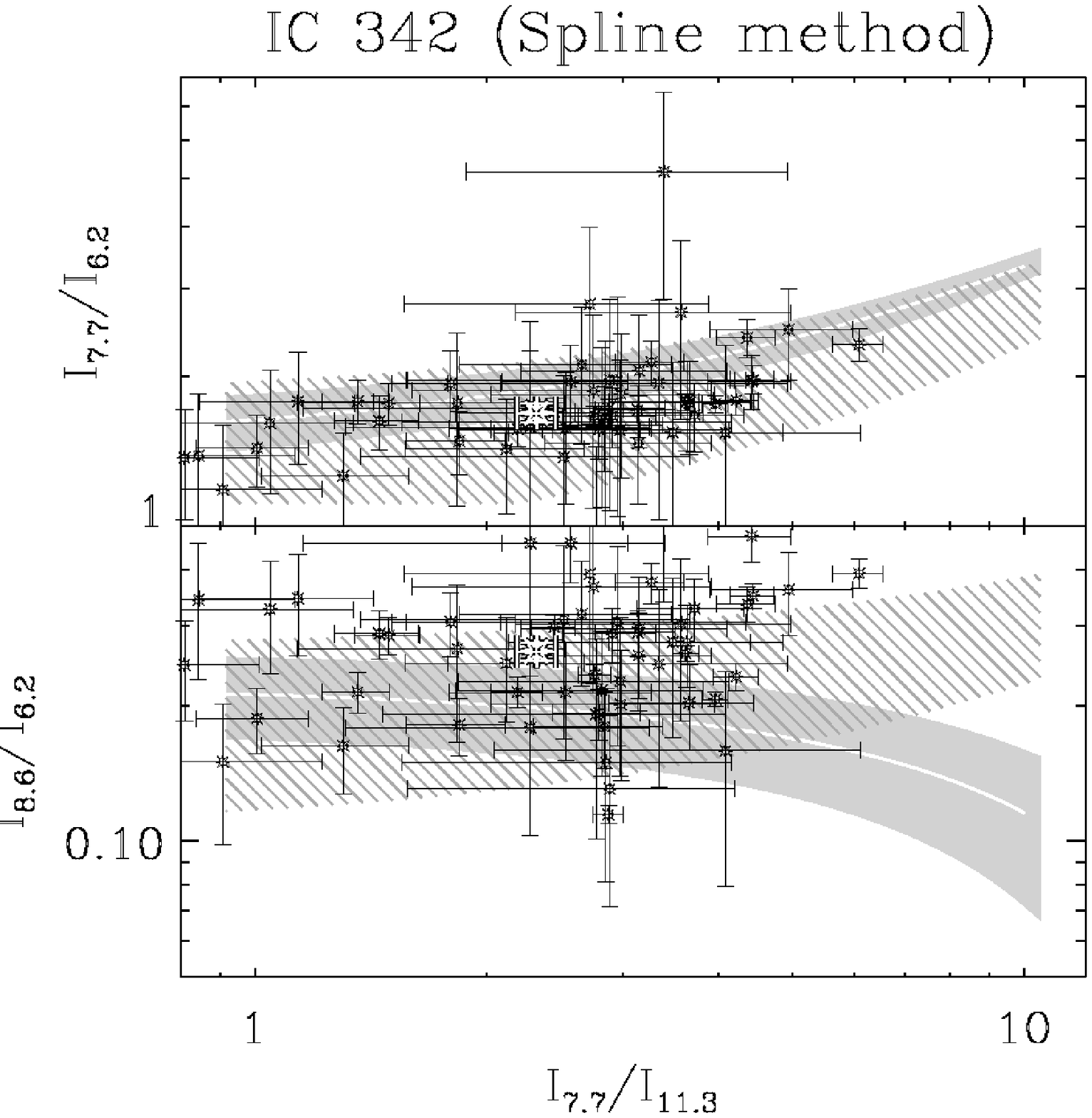} \\
  \end{tabular}
  \caption{PAH band ratios within \IC{342} (continued).}
  \label{fig:cormap2ic342}
\end{figure*}

\begin{figure*}[htbp]
  \centering
  \includegraphics[width=\textwidth]{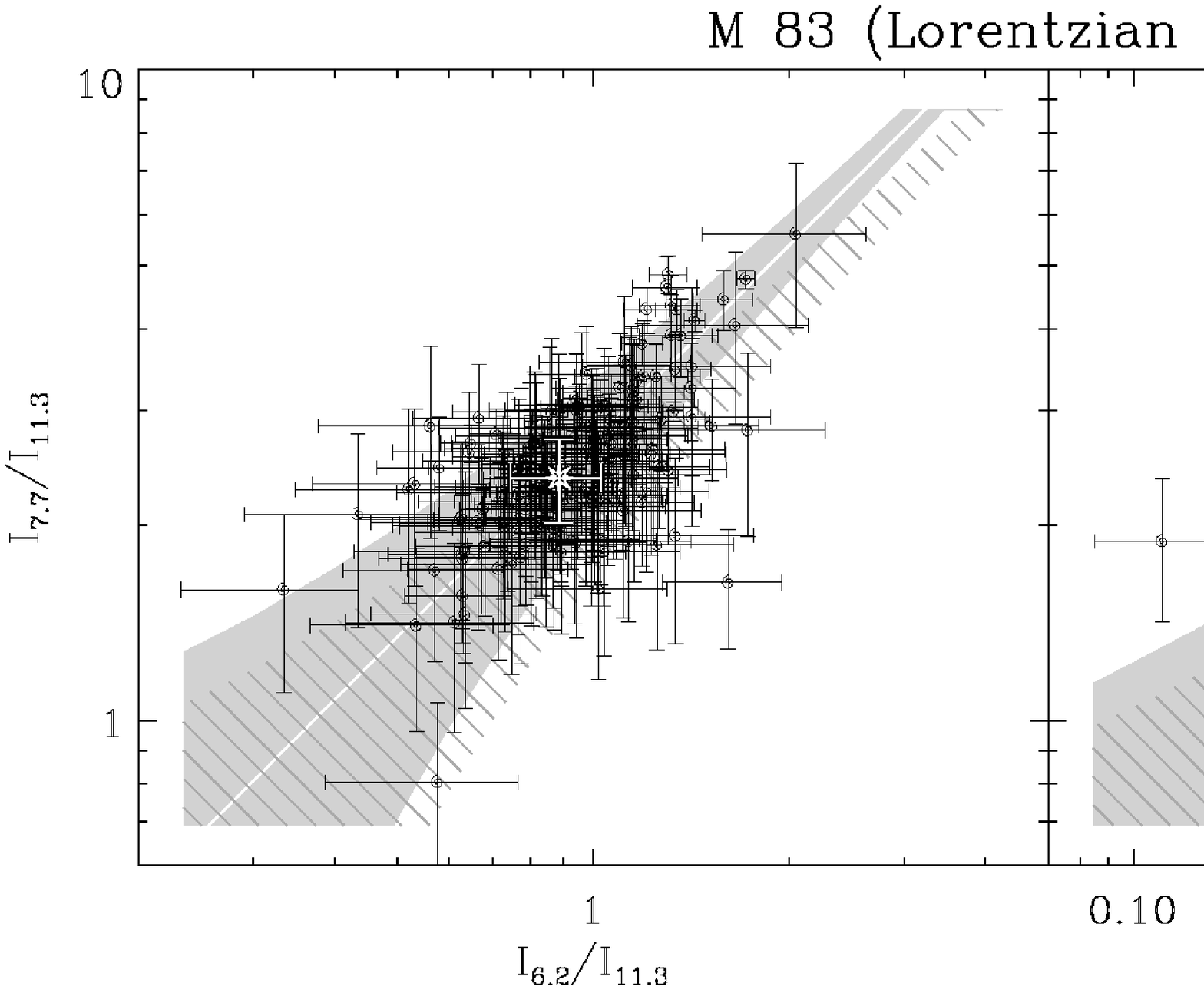} \\
  \includegraphics[width=\textwidth]{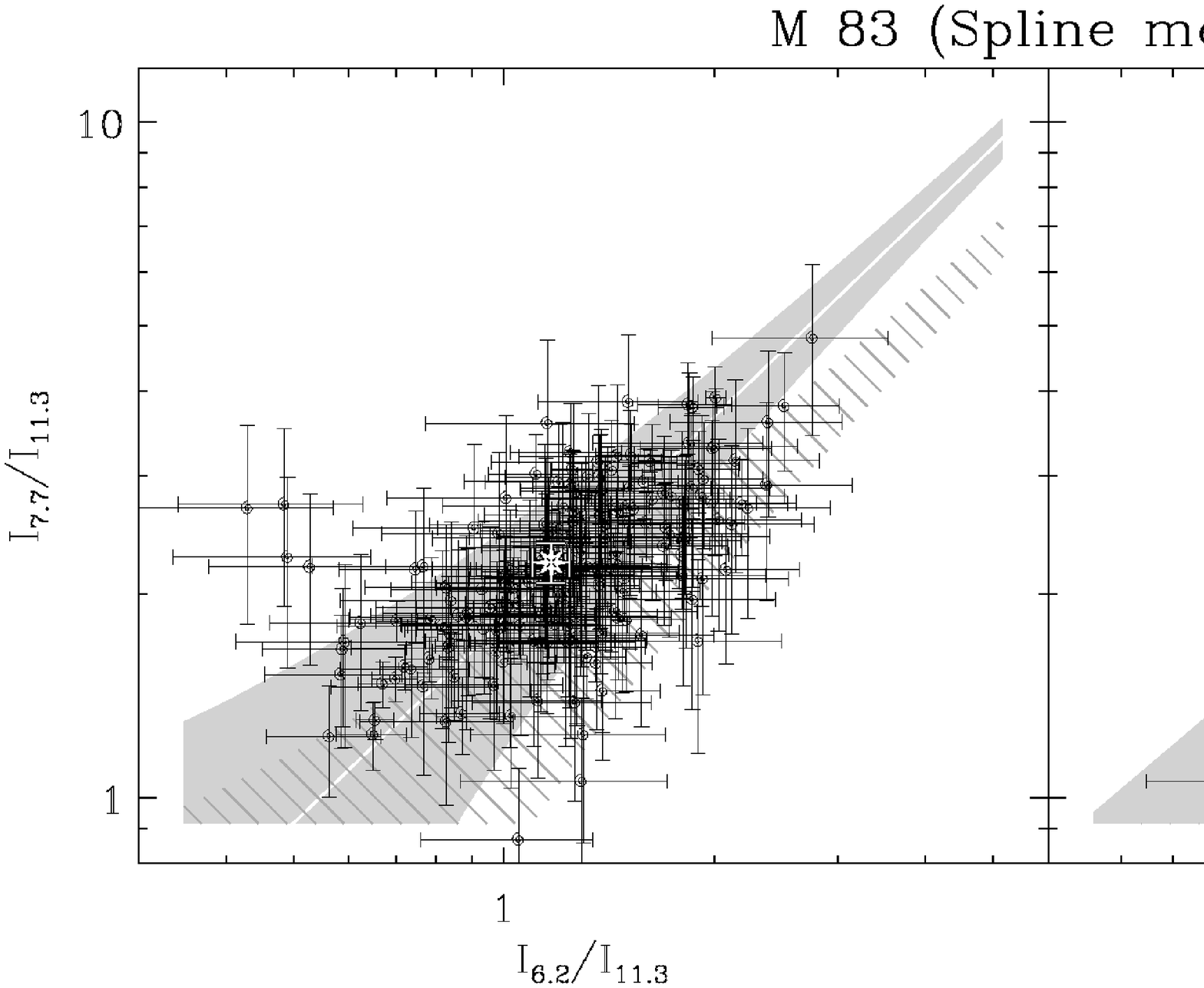}
  \caption{PAH band ratios within \M{83}.
           The same symbol conventions are adopted as in 
           \reffig{fig:cormap1m82}.}
  \label{fig:cormap1m83}
\end{figure*}

\begin{figure*}[htbp]
  \centering
  \begin{tabular}{cc}
    \includegraphics[width=0.48\textwidth]{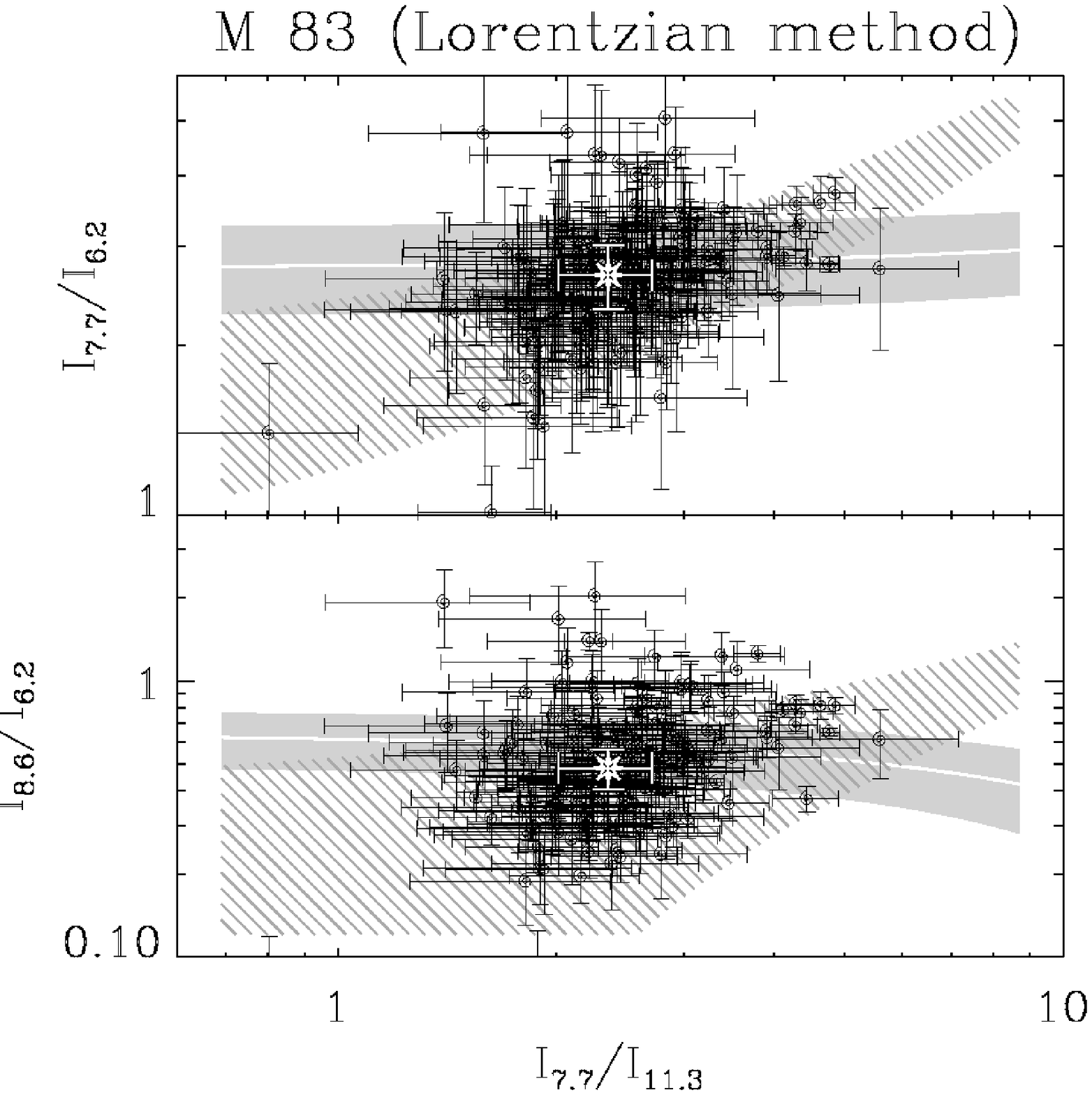} &
    \includegraphics[width=0.48\textwidth]{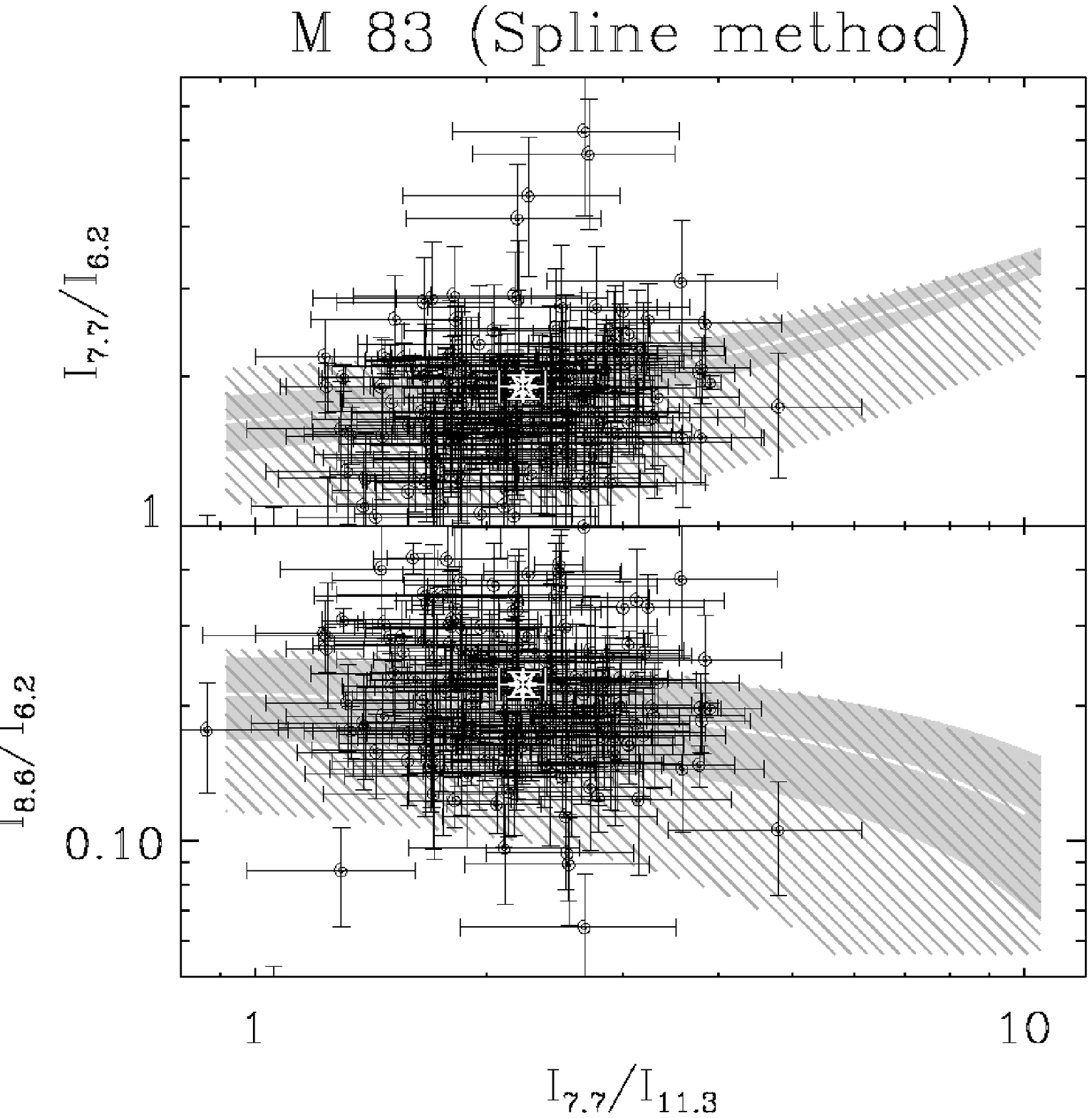} \\
  \end{tabular}
  \caption{PAH band ratios within \M{83} (continued).}
  \label{fig:cormap2m83}
\end{figure*}

\begin{figure*}[htbp]
  \centering
  \includegraphics[width=\textwidth]{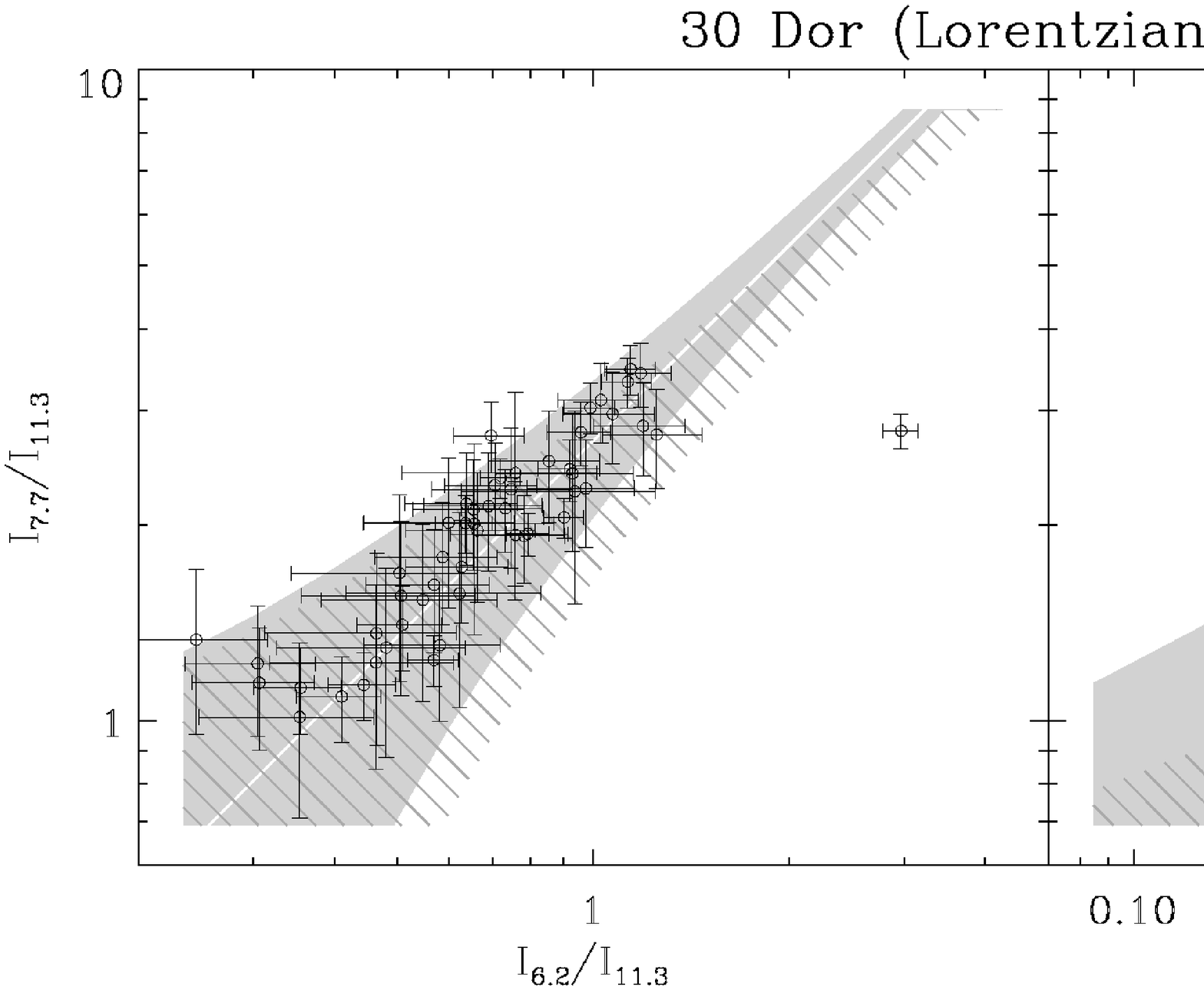} \\
  \includegraphics[width=\textwidth]{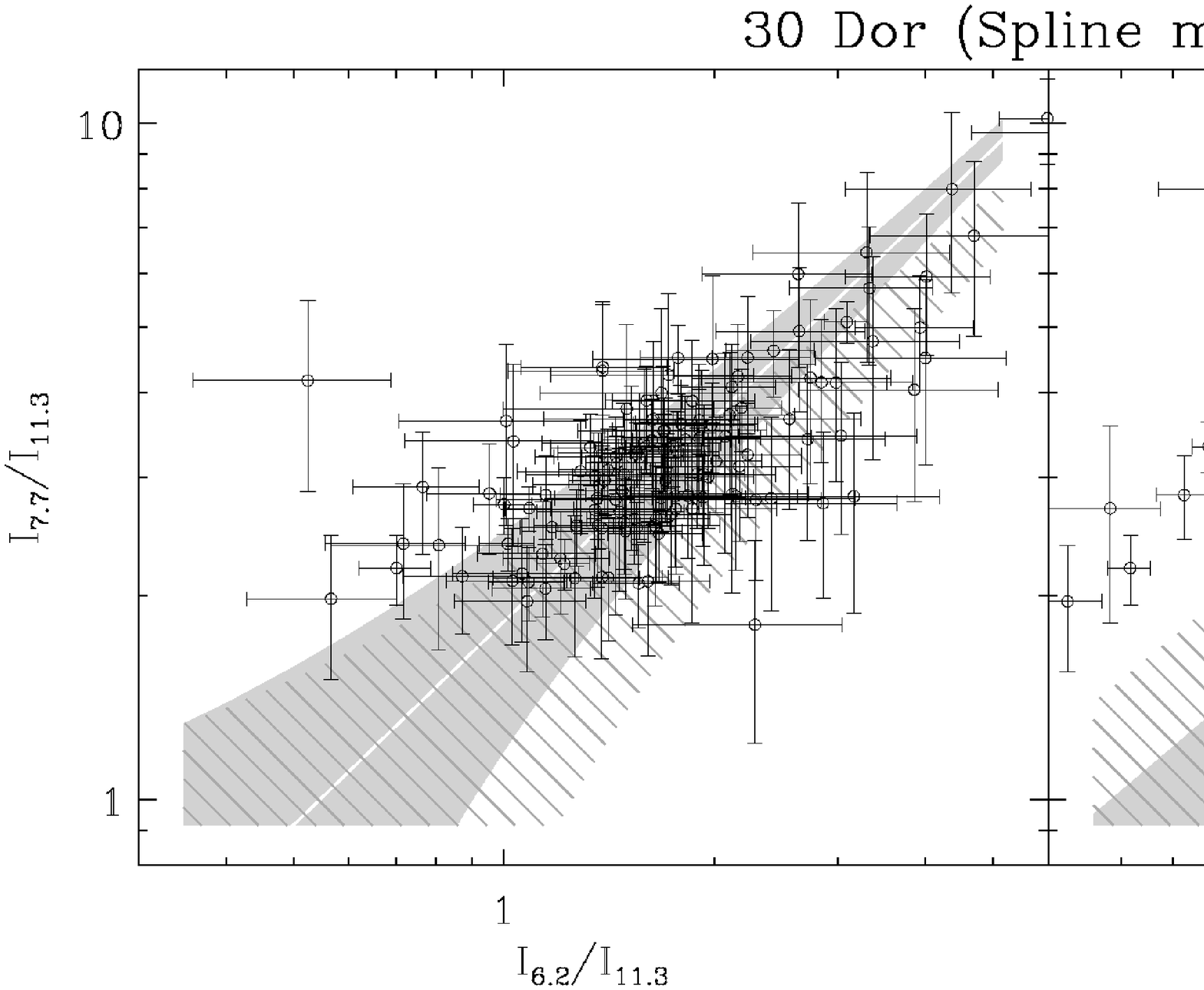}
  \caption{PAH band ratios within \xxxdor.
           The same symbol conventions are adopted as in 
           \reffig{fig:cormap1m82}.}
  \label{fig:cormap130dor}
\end{figure*}

\begin{figure*}[htbp]
  \centering
  \begin{tabular}{cc}
    \includegraphics[width=0.48\textwidth]{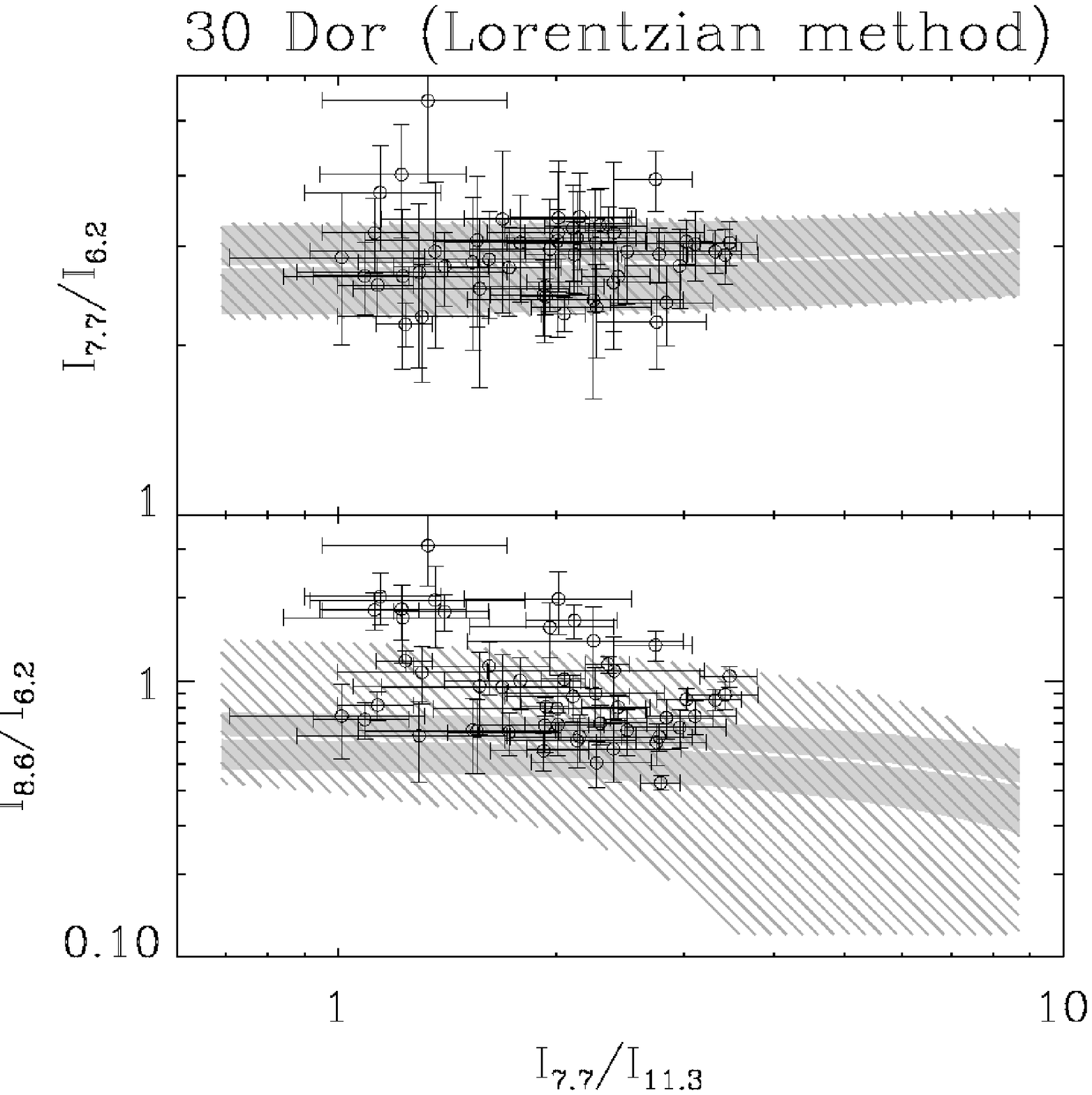} &
    \includegraphics[width=0.48\textwidth]{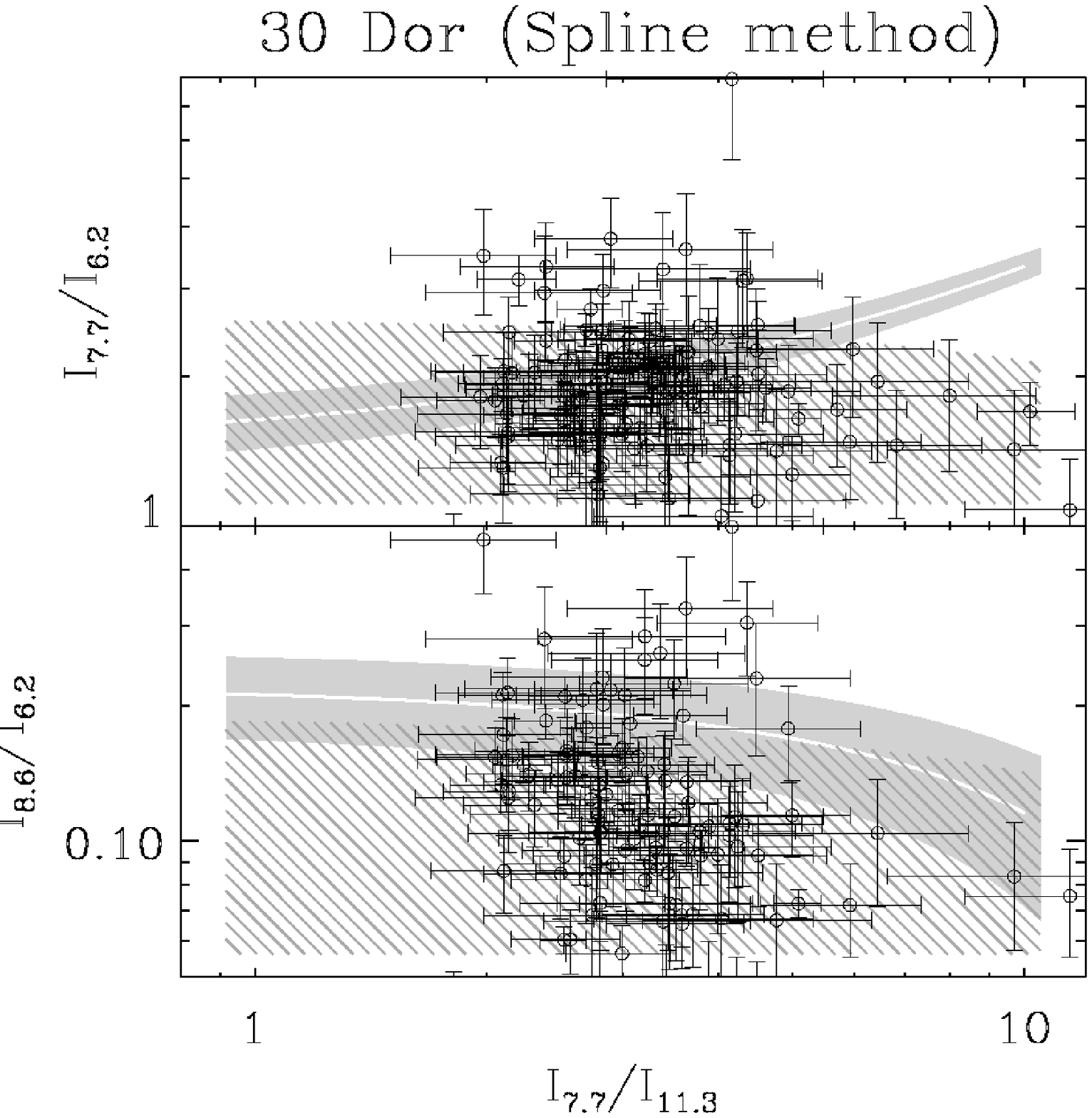} \\
  \end{tabular}
  \caption{PAH band ratios within \xxxdor\ (continued).}
  \label{fig:cormap230dor}
\end{figure*}

\begin{figure*}[htbp]
  \centering
  \includegraphics[width=\textwidth]{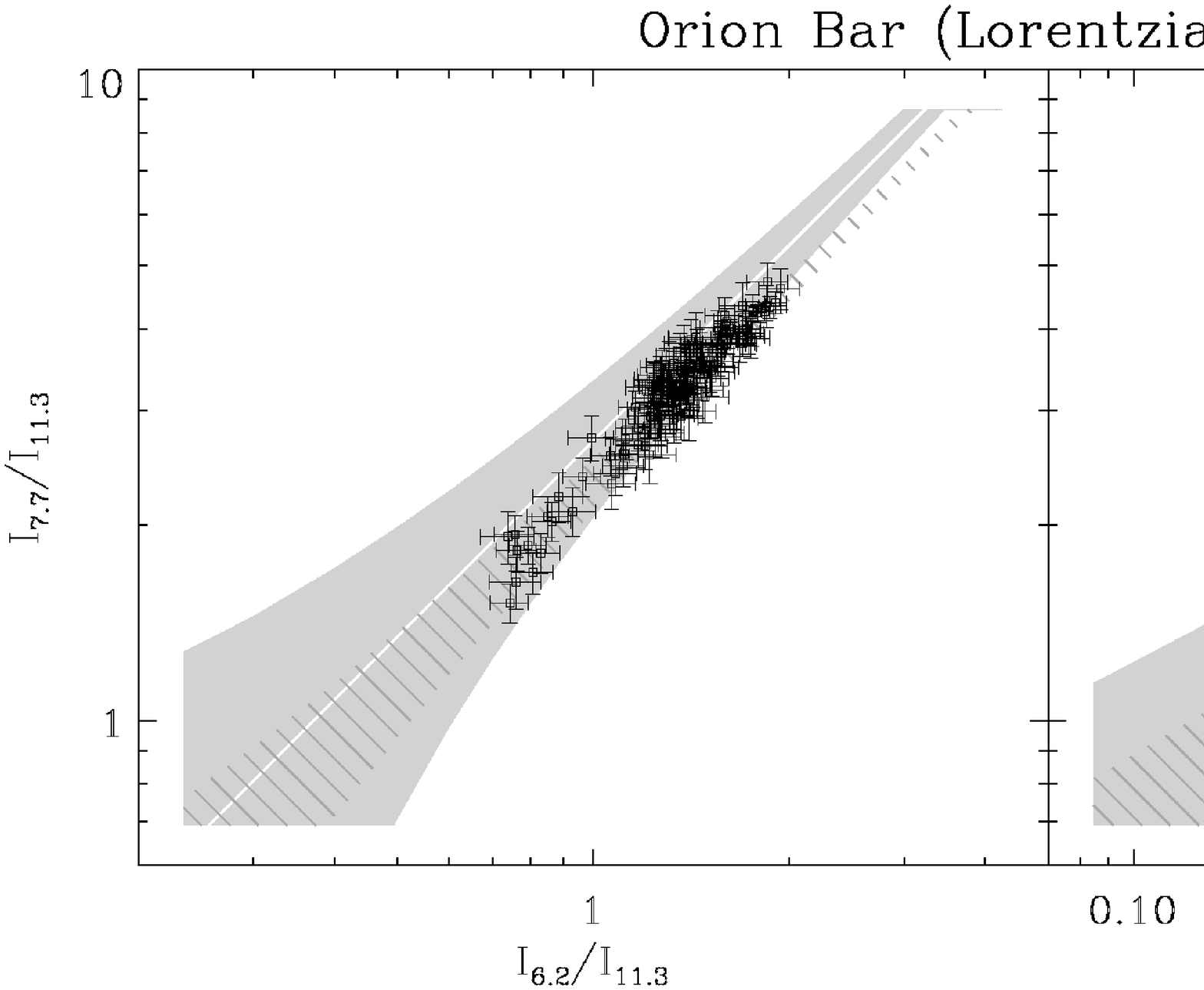} \\
  \includegraphics[width=\textwidth]{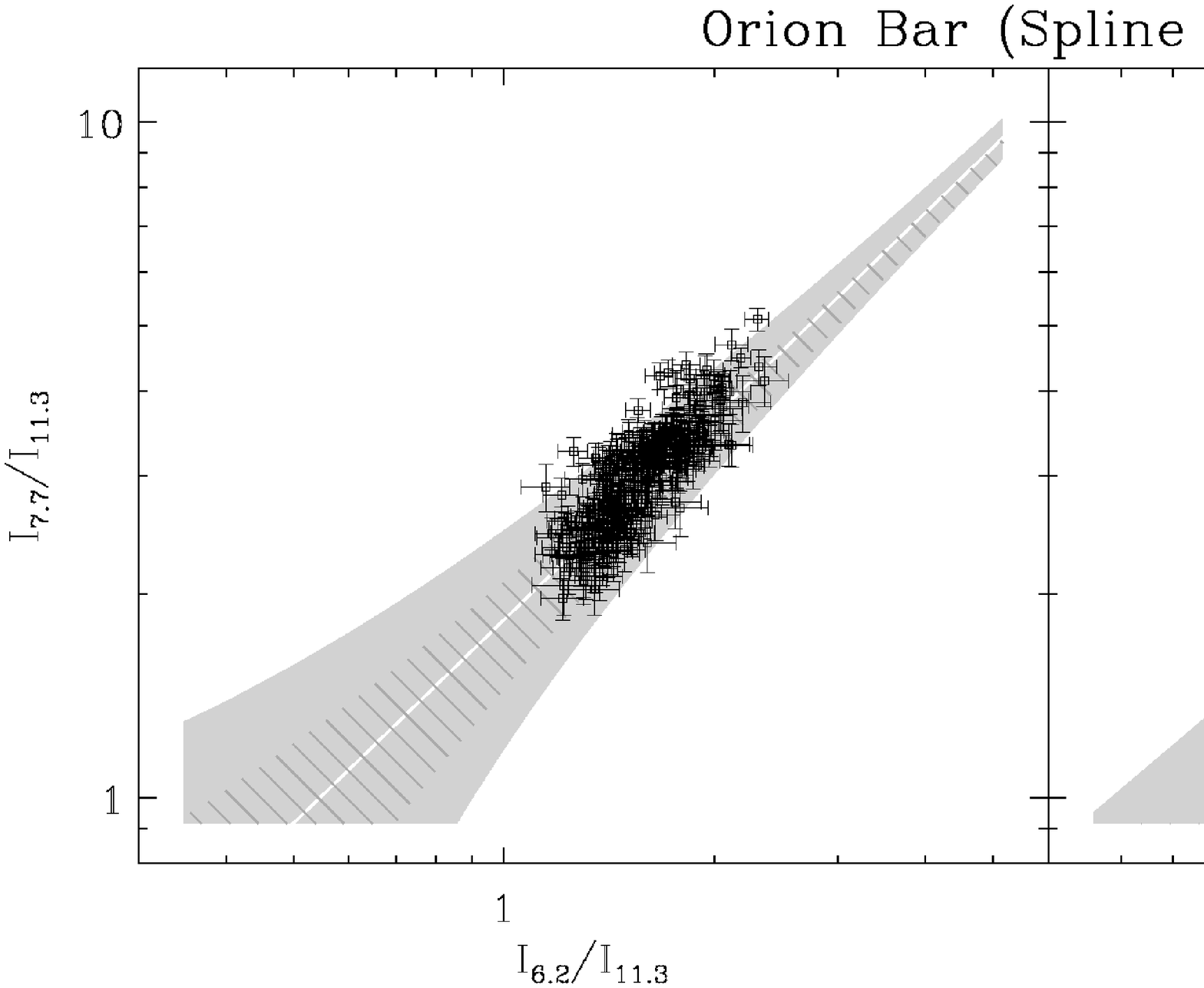}
  \caption{PAH band ratios within the \orb.
           The same symbol conventions are adopted as in 
           \reffig{fig:cormap1m82}.}
  \label{fig:cormap1orionBar}
\end{figure*}

\begin{figure*}[htbp]
  \centering
  \begin{tabular}{cc}
    \includegraphics[width=0.48\textwidth]{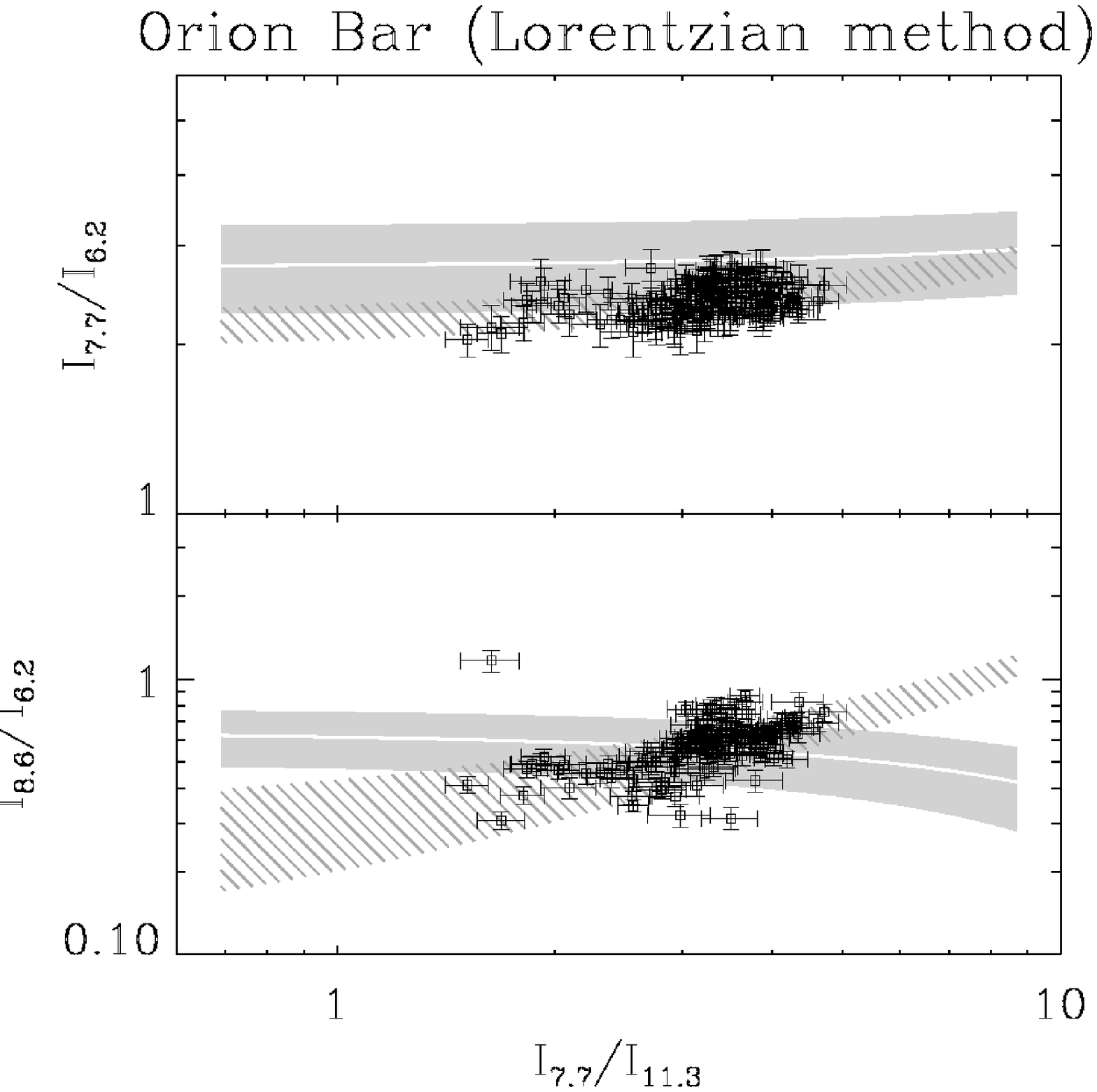} &
    \includegraphics[width=0.48\textwidth]{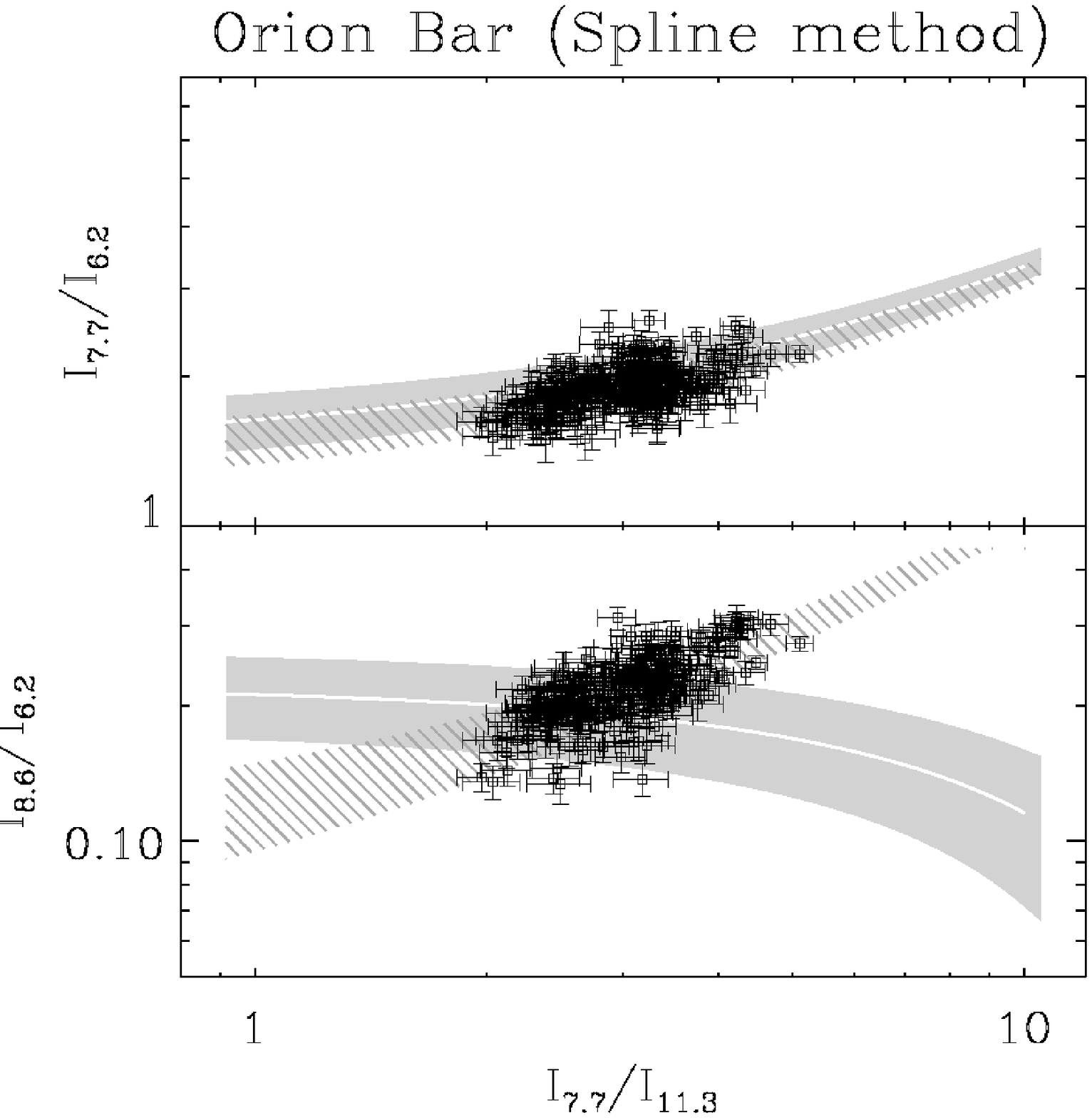} \\
  \end{tabular}
  \caption{PAH band ratios within \orb\ (continued).}
  \label{fig:cormap2orionBar}
\end{figure*}

\begin{figure}[htbp]
  \centering
  \includegraphics[width=0.48\textwidth]{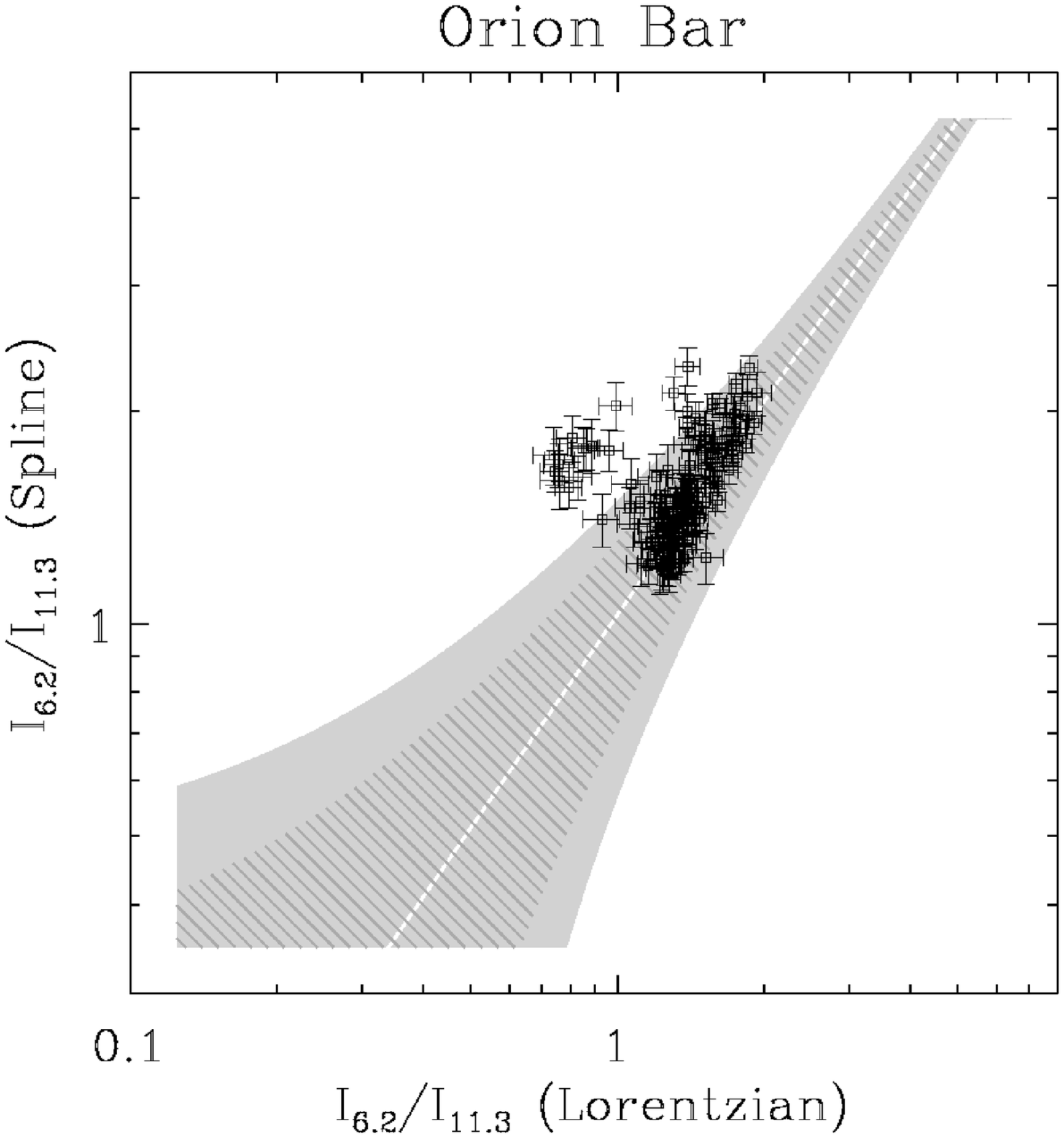}
  \caption{Comparison between the two methods in \orb.
           The same symbol conventions are adopted as in 
           \reffig{fig:cormap1m82}.}
  \label{fig:compmaporionBar}
\end{figure}

\begin{figure*}[htbp]
  \centering
  \includegraphics[width=\textwidth]{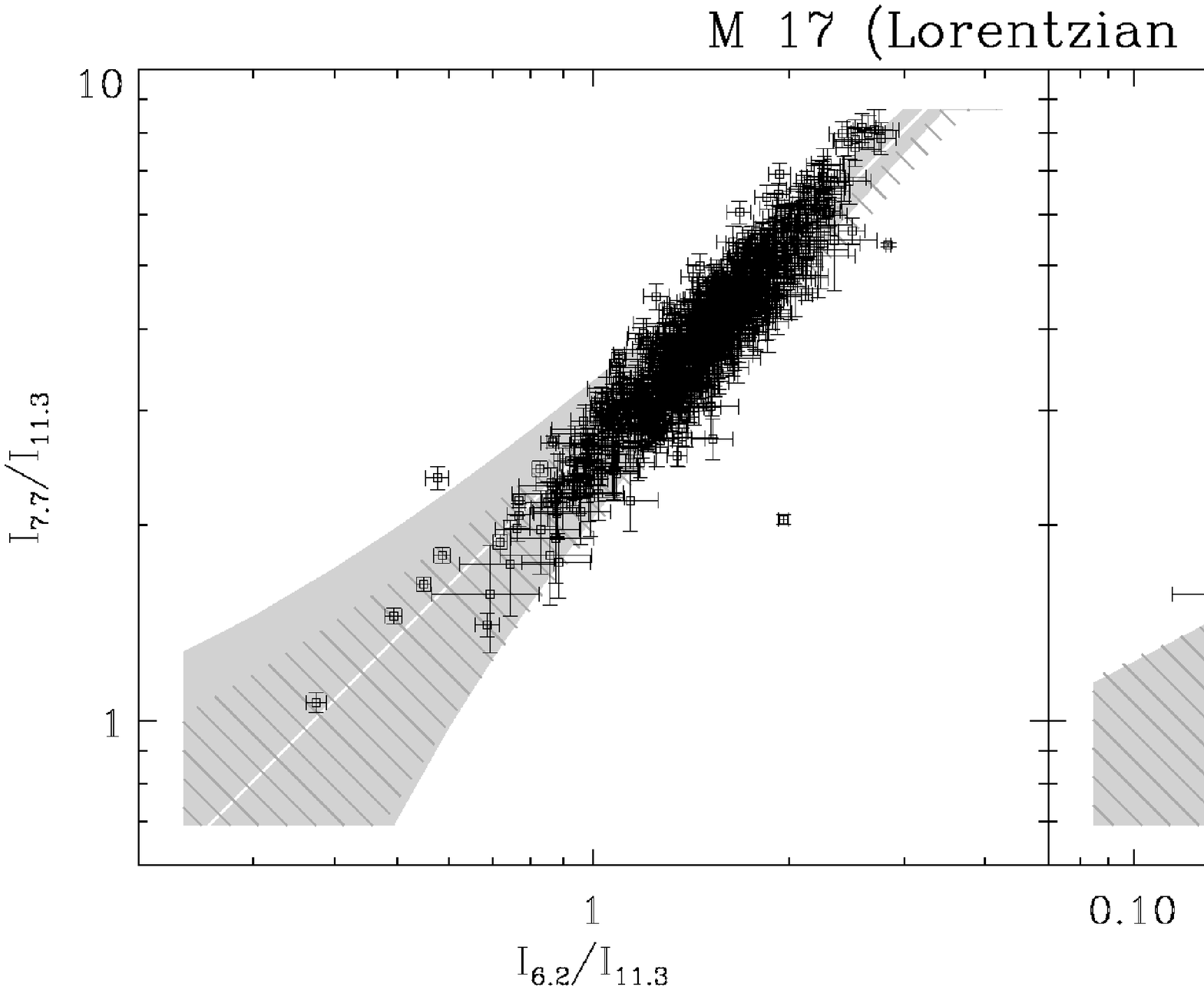} \\
  \includegraphics[width=\textwidth]{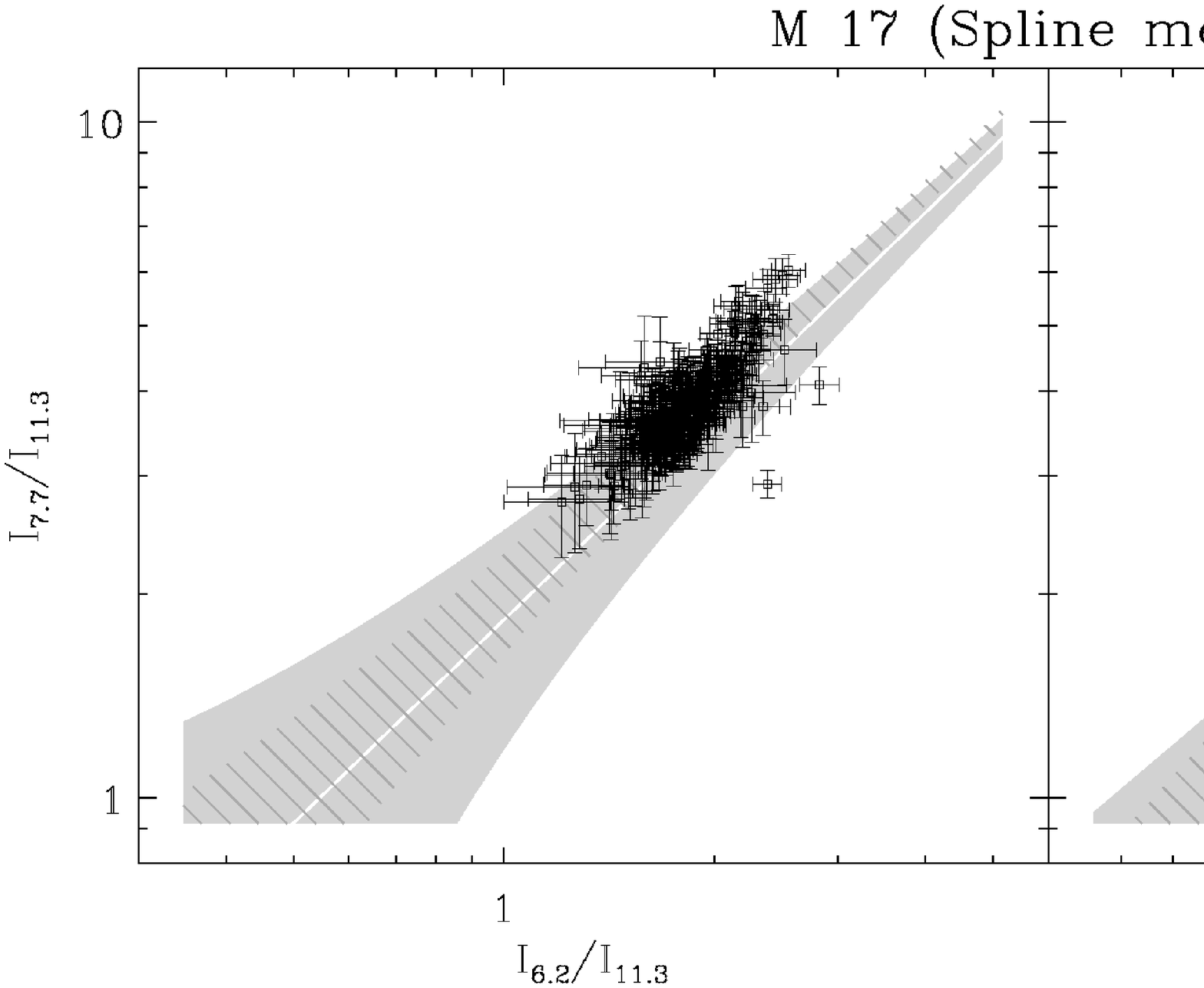}
  \caption{PAH band ratios within \M{17}.
           The same symbol conventions are adopted as in 
           \reffig{fig:cormap1m82}.}
  \label{fig:cormap1m17}
\end{figure*}

\begin{figure*}[htbp]
  \centering
  \begin{tabular}{cc}
    \includegraphics[width=0.48\textwidth]{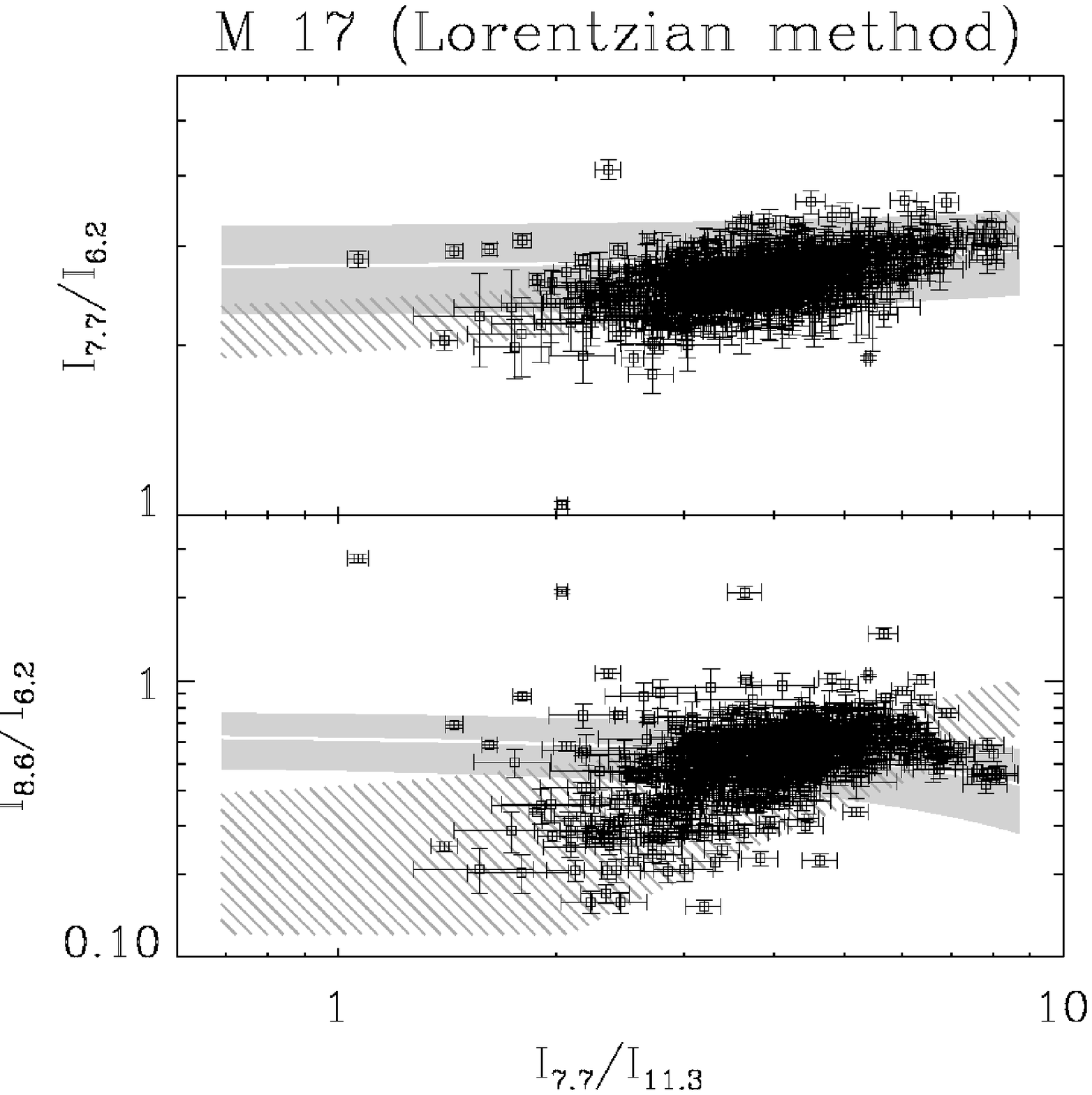} &
    \includegraphics[width=0.48\textwidth]{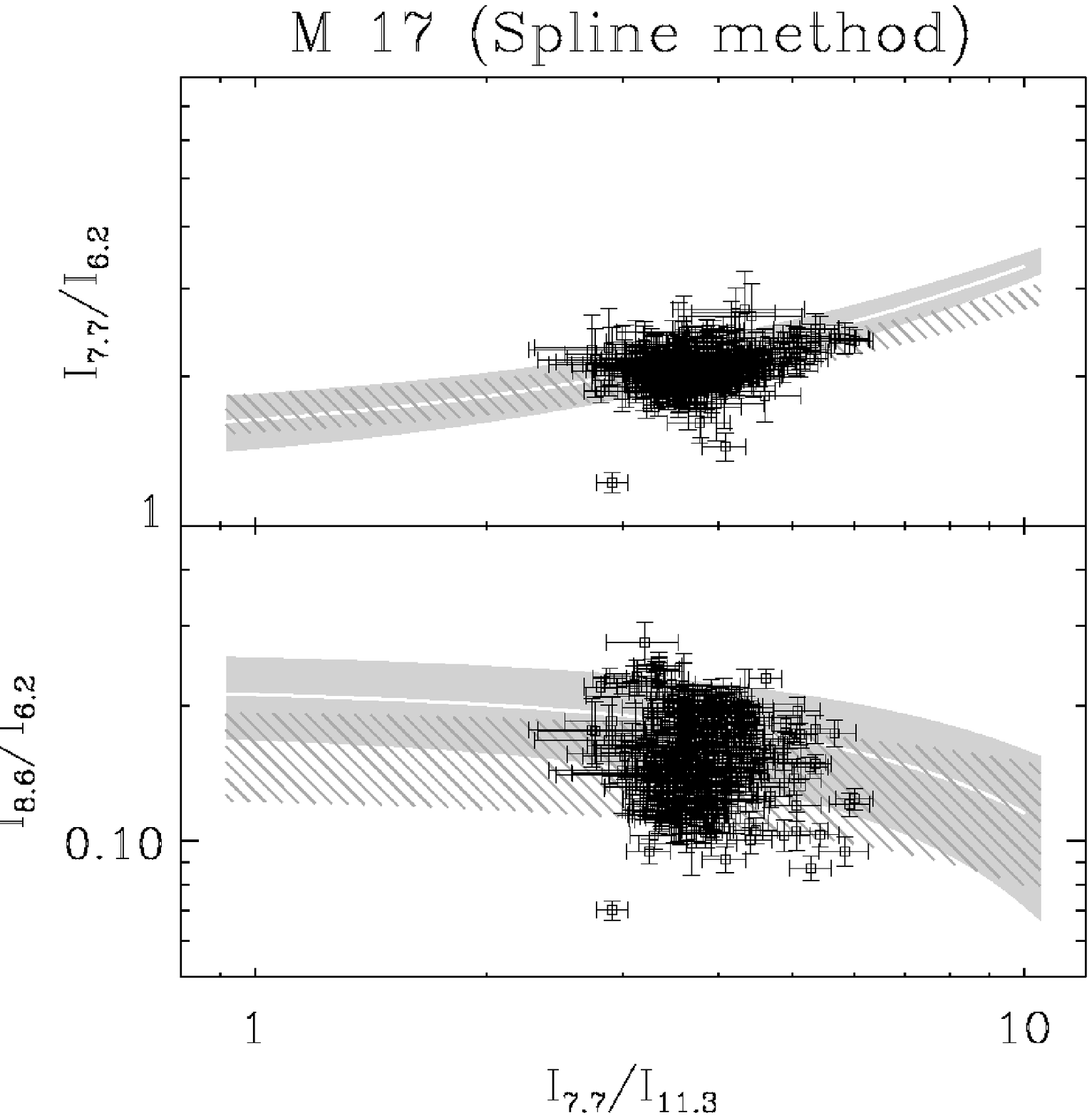} \\
  \end{tabular}
  \caption{PAH band ratios within \M{17} (continued).}
  \label{fig:cormap2m17}
\end{figure*}

\begin{figure*}[htbp]
  \centering
  \plottwo{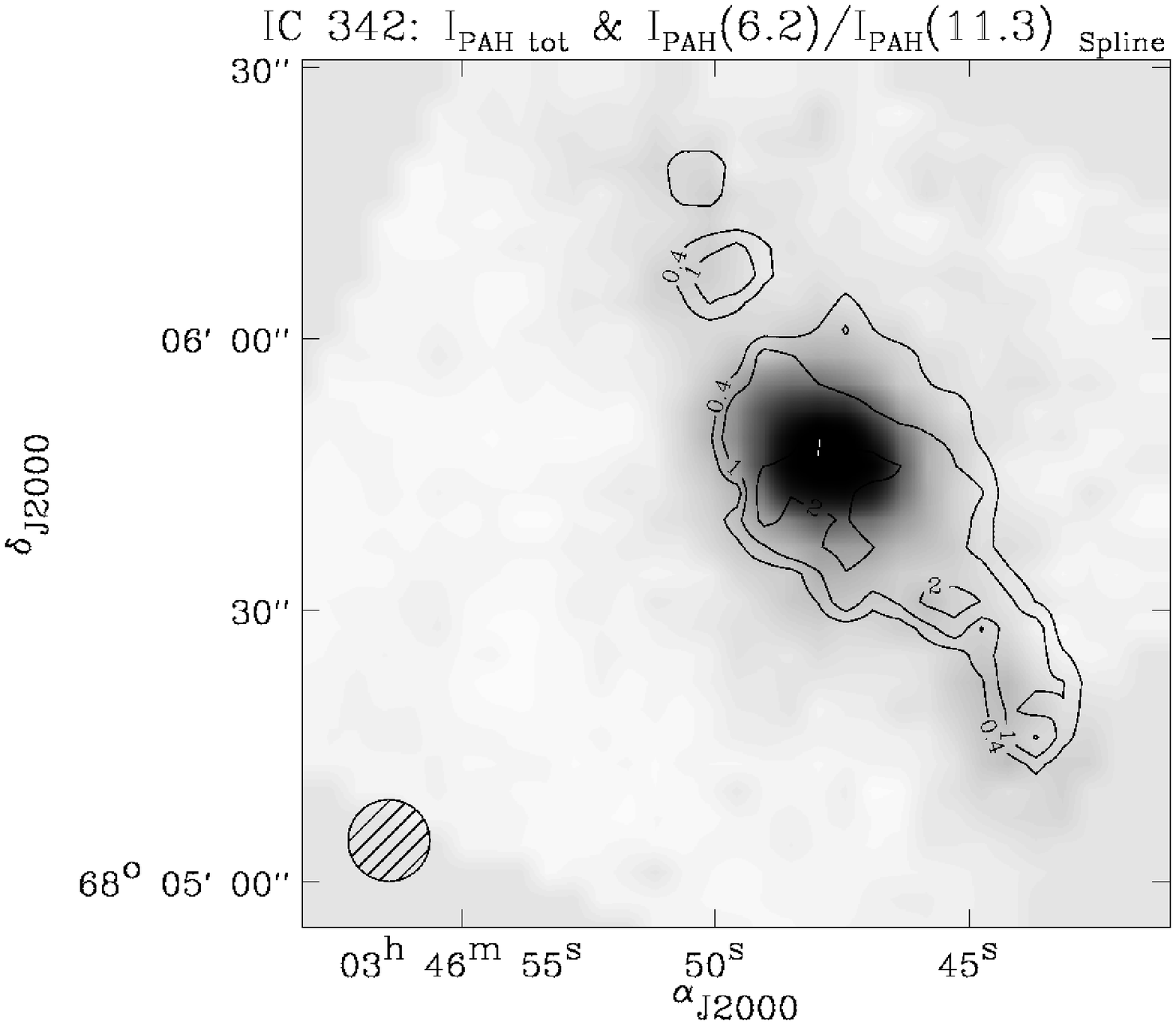}{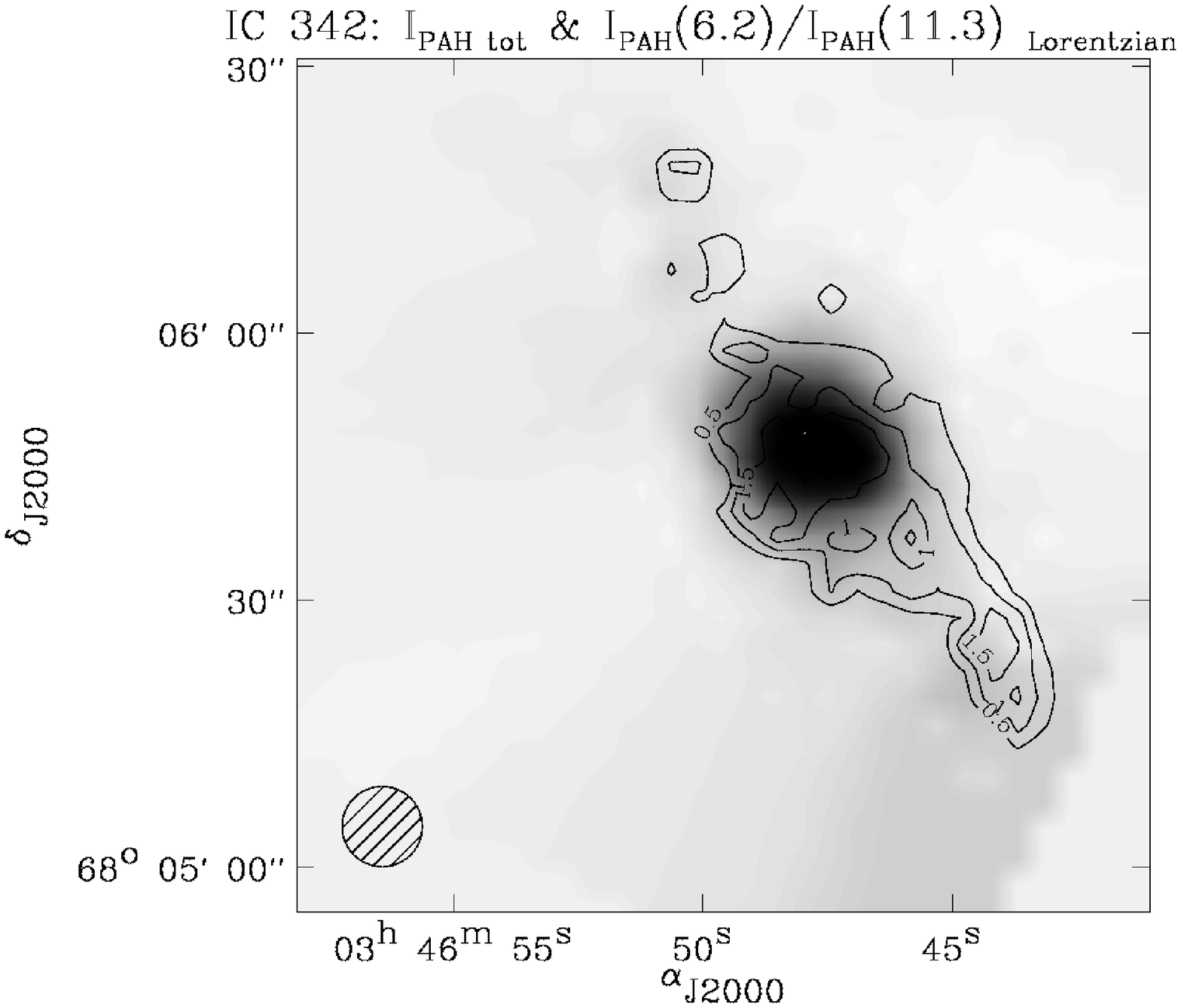}
  \caption{Spatial distribution of the PAHs in \IC{342}.
           For each method, the image is the total PAH intensity, and the 
           contours are the $\ipah{6.2}/\ipah{11.3}$ ratio.
           The shaded circle indicates the beam size.}
  \label{fig:imic342}
\end{figure*}

\begin{figure*}[htbp]
  \centering
  \plottwo{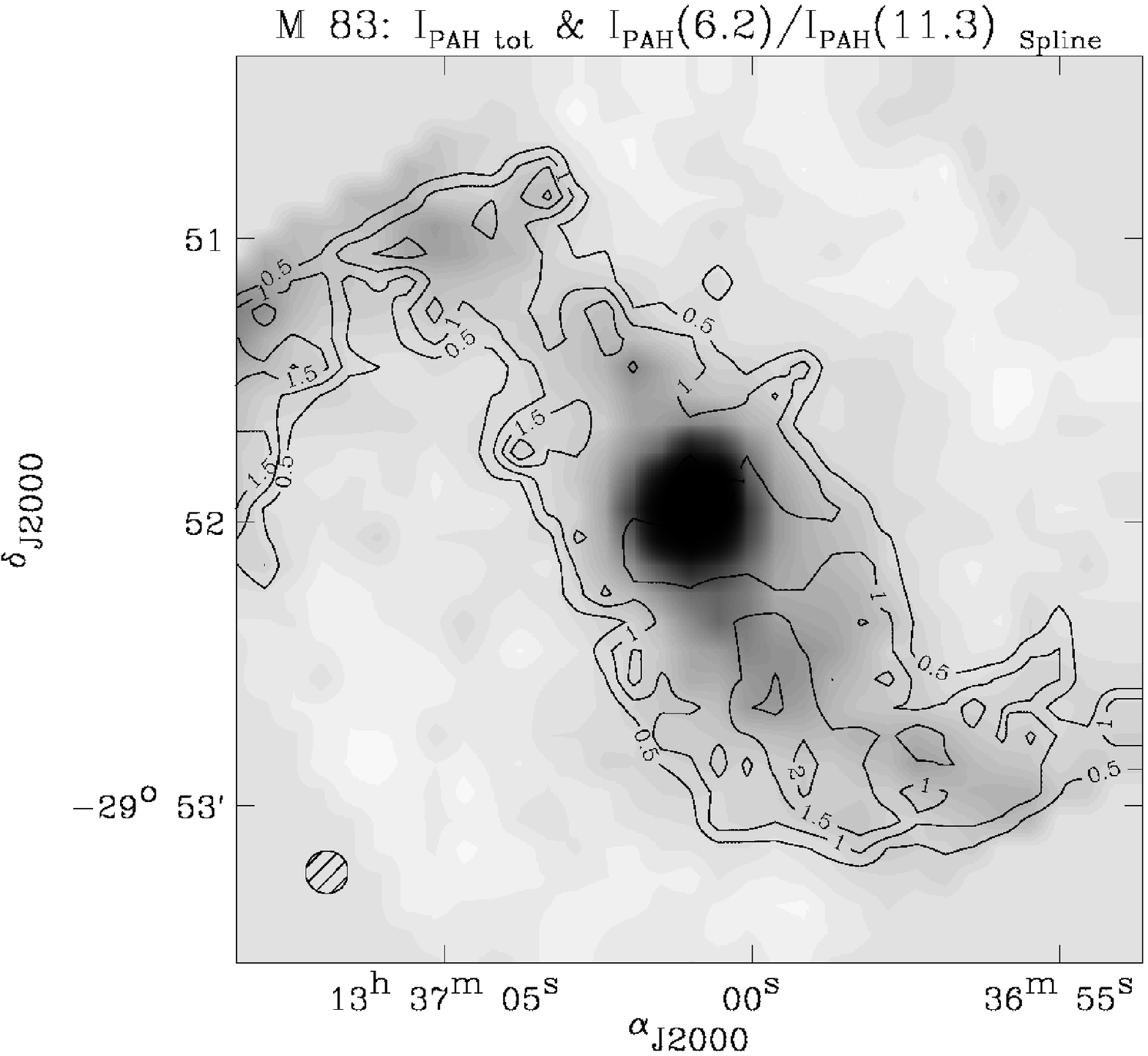}{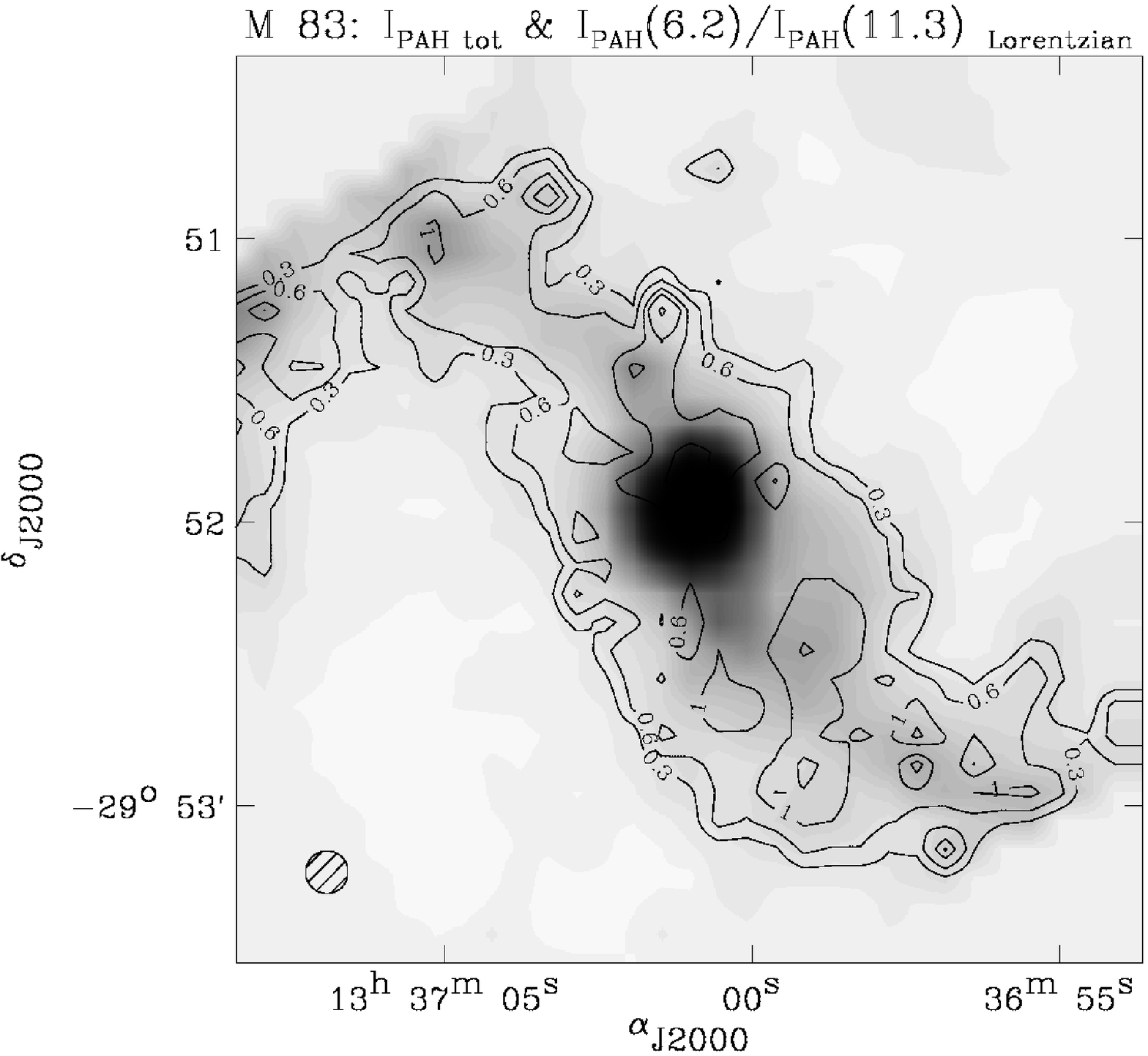}
  \caption{Spatial distribution of the PAHs in \M{83}.
           For each method, the image is the total PAH intensity, and the 
           contours are the $\ipah{6.2}/\ipah{11.3}$ ratio.
           The shaded circle indicates the beam size.}
  \label{fig:imm83}
\end{figure*}

\begin{figure*}[htbp]
  \centering
  \plottwo{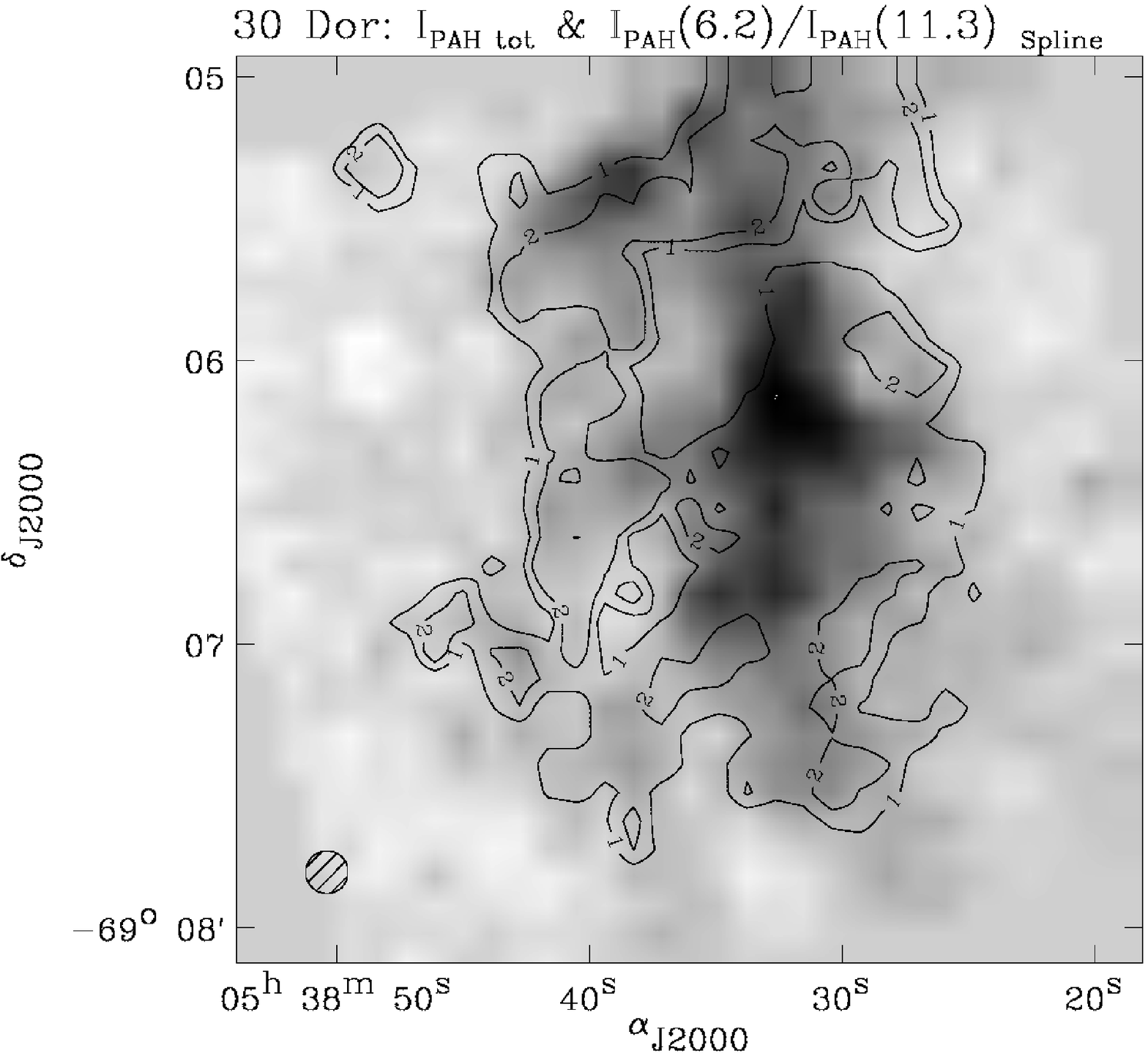}{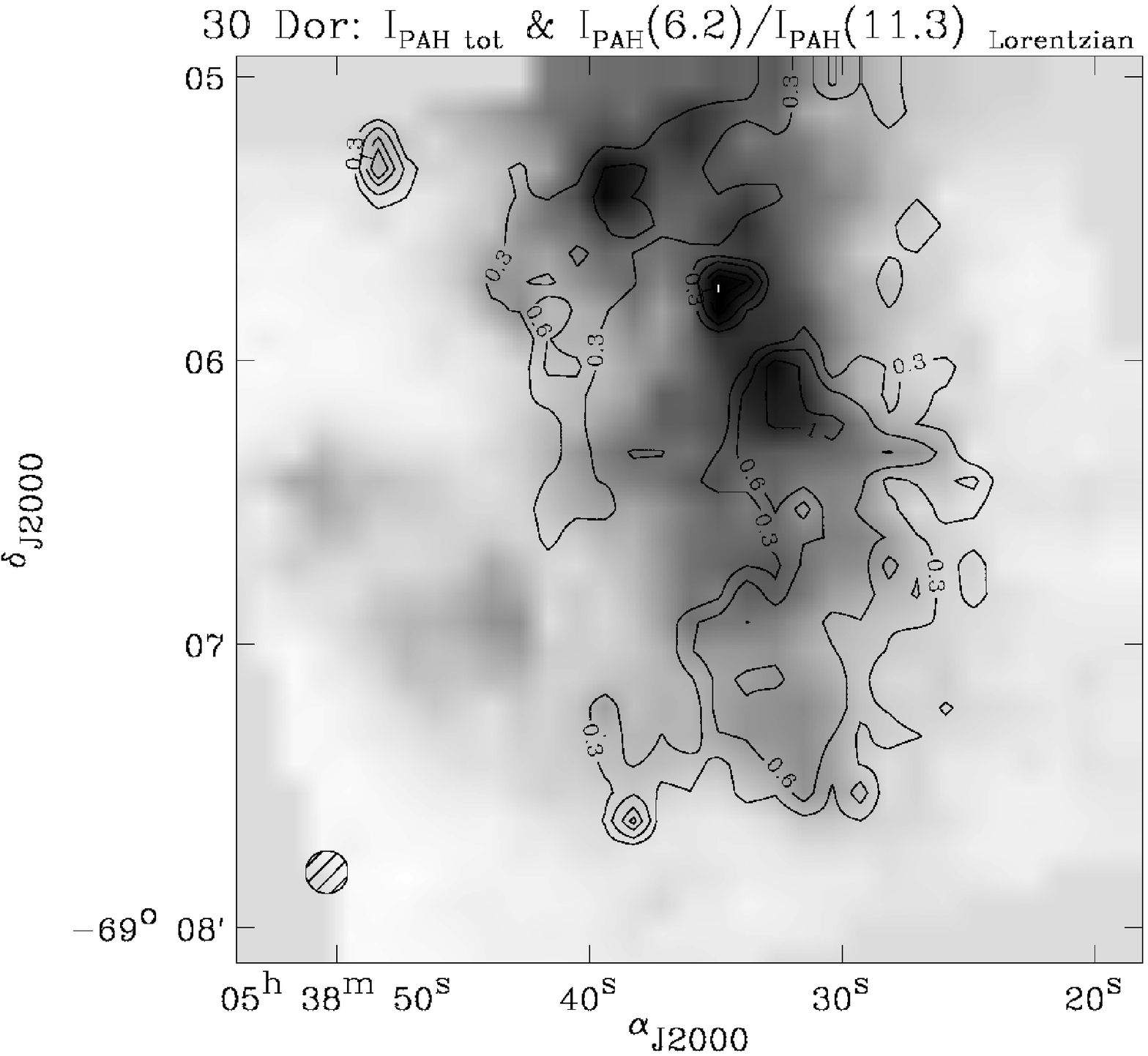}
  \caption{Spatial distribution of the PAHs in \xxxdor.
           For each method, the image is the total PAH intensity, and the 
           contours are the $\ipah{6.2}/\ipah{11.3}$ ratio.
           The shaded circle indicates the beam size.}
  \label{fig:im30dor}
\end{figure*}

\begin{figure*}[htbp]
  \centering
  \plottwo{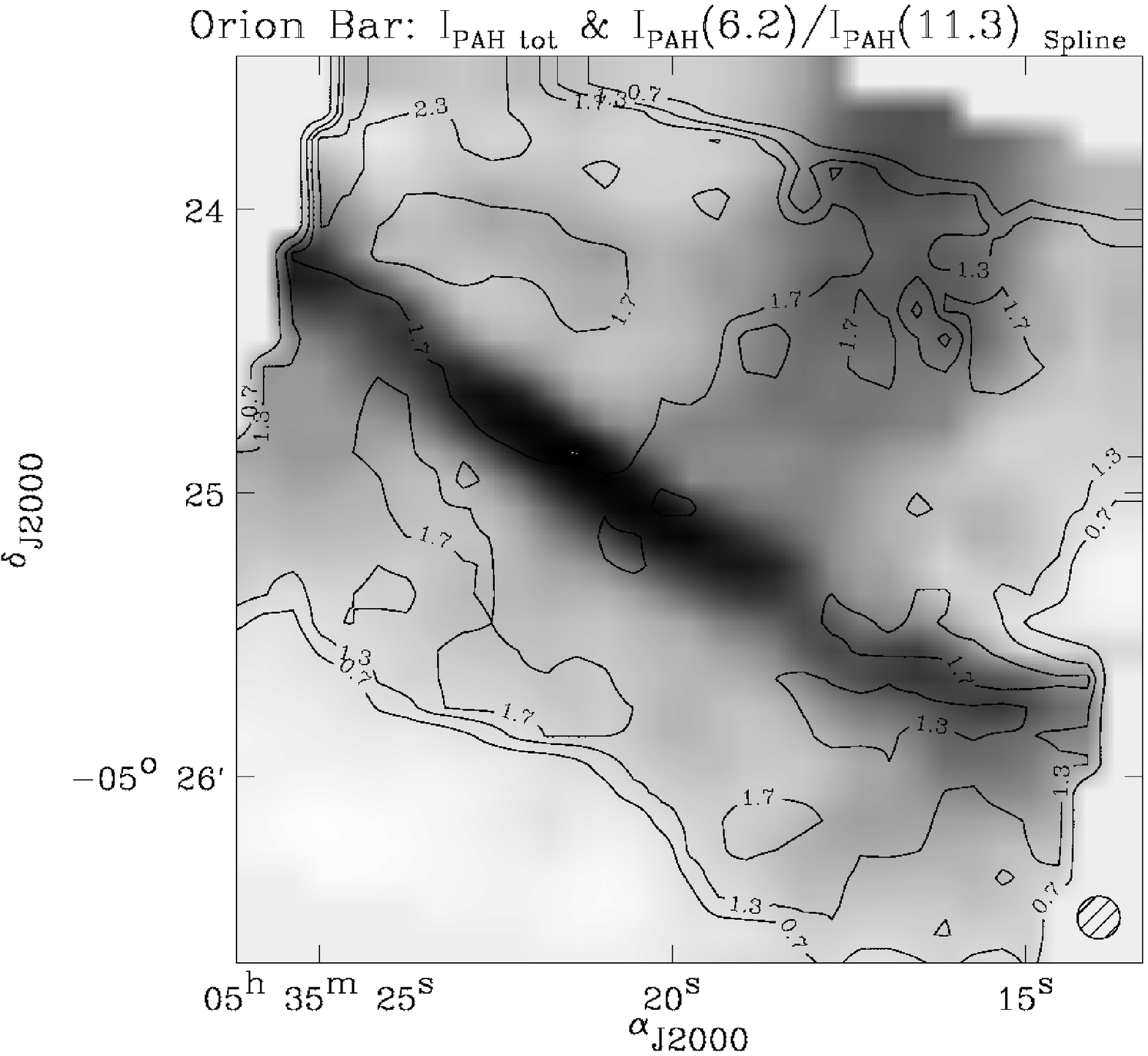}{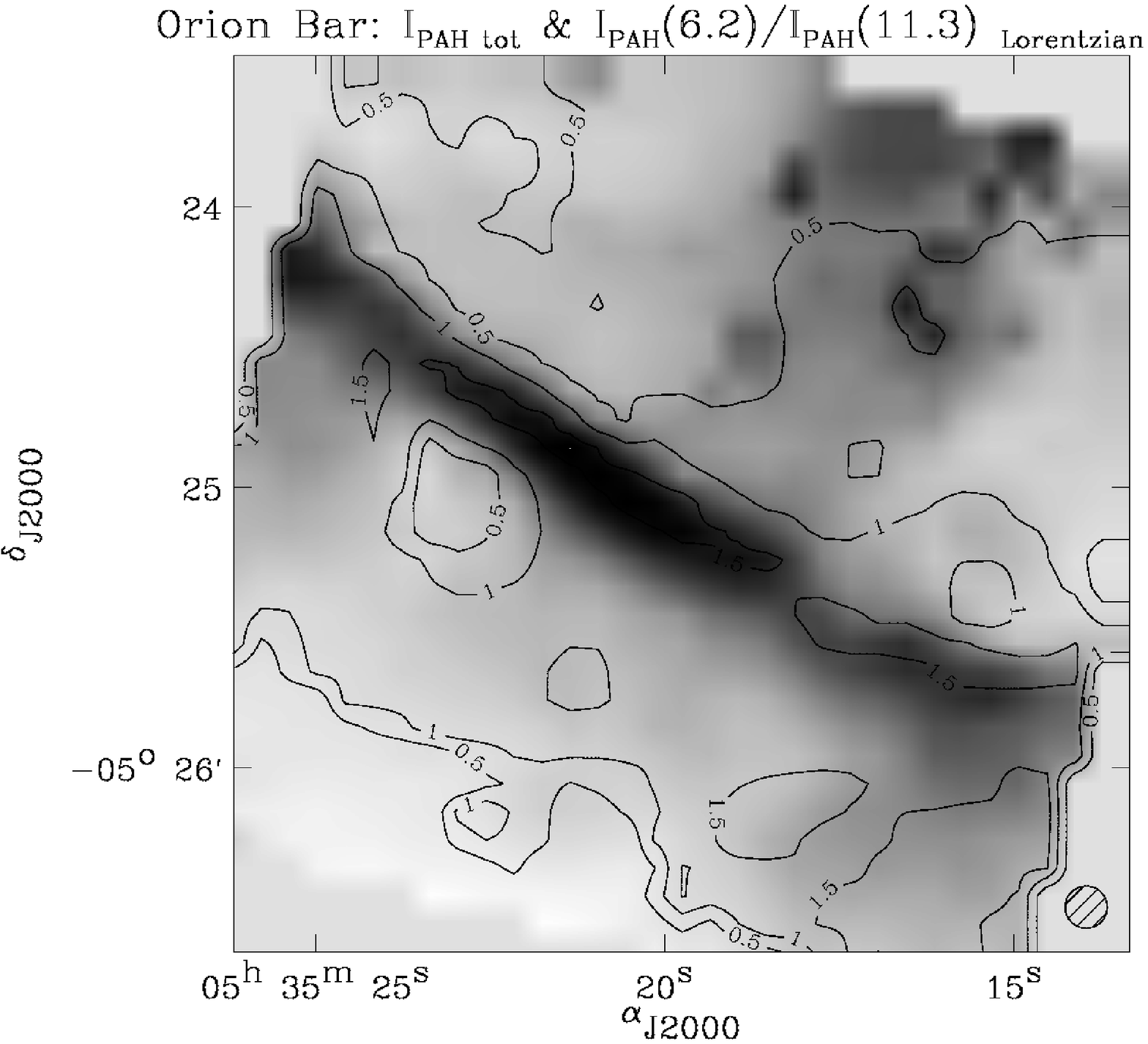}
  \caption{Spatial distribution of the PAHs in \orb.
           For each method, the image is the total PAH intensity, and the
           contours are the $\ipah{6.2}/\ipah{11.3}$ ratio.
           The shaded circle indicates the beam size.}
  \label{fig:imorionBar}
\end{figure*}

\begin{figure*}[htbp]
  \centering
  \plottwo{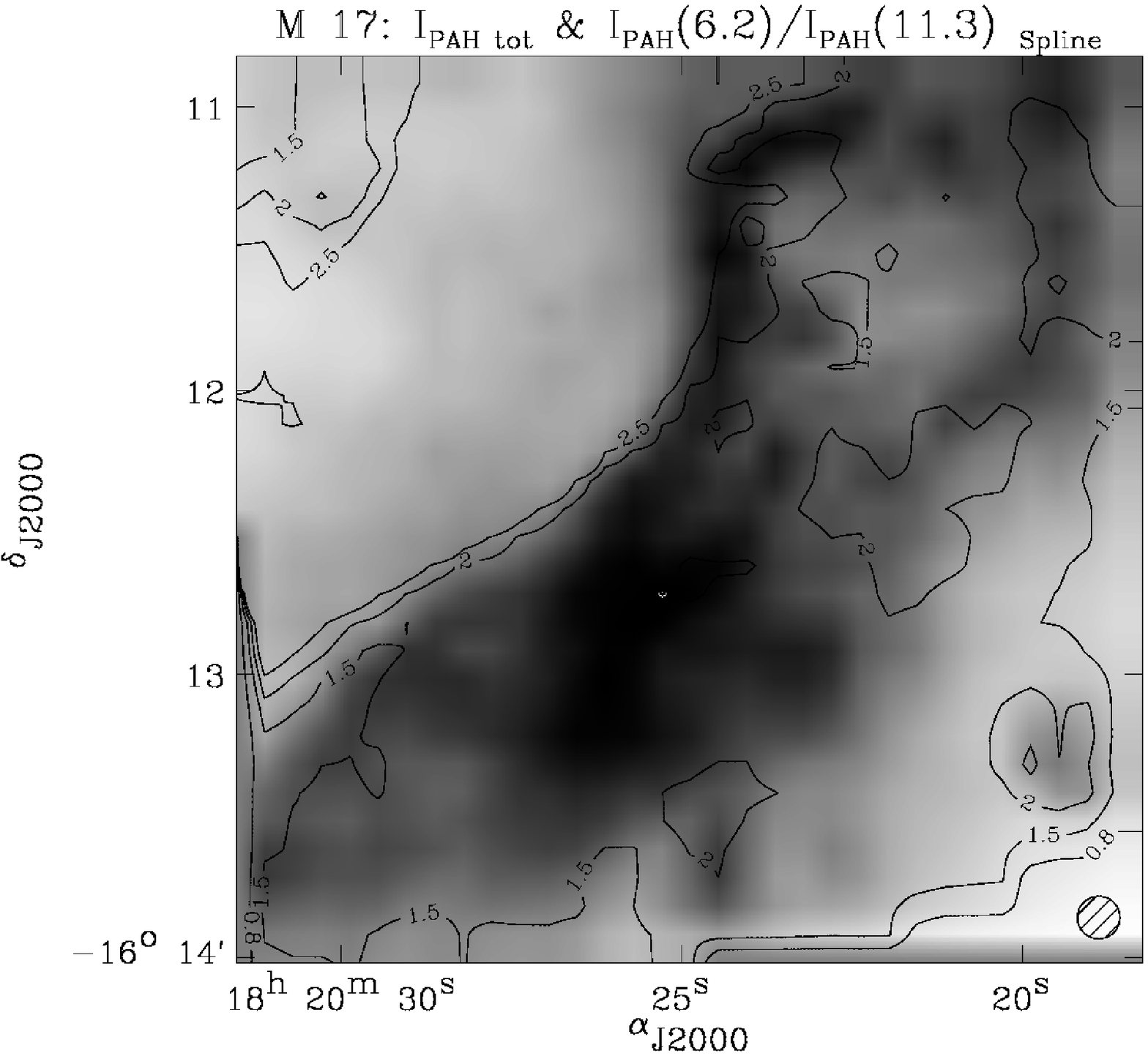}{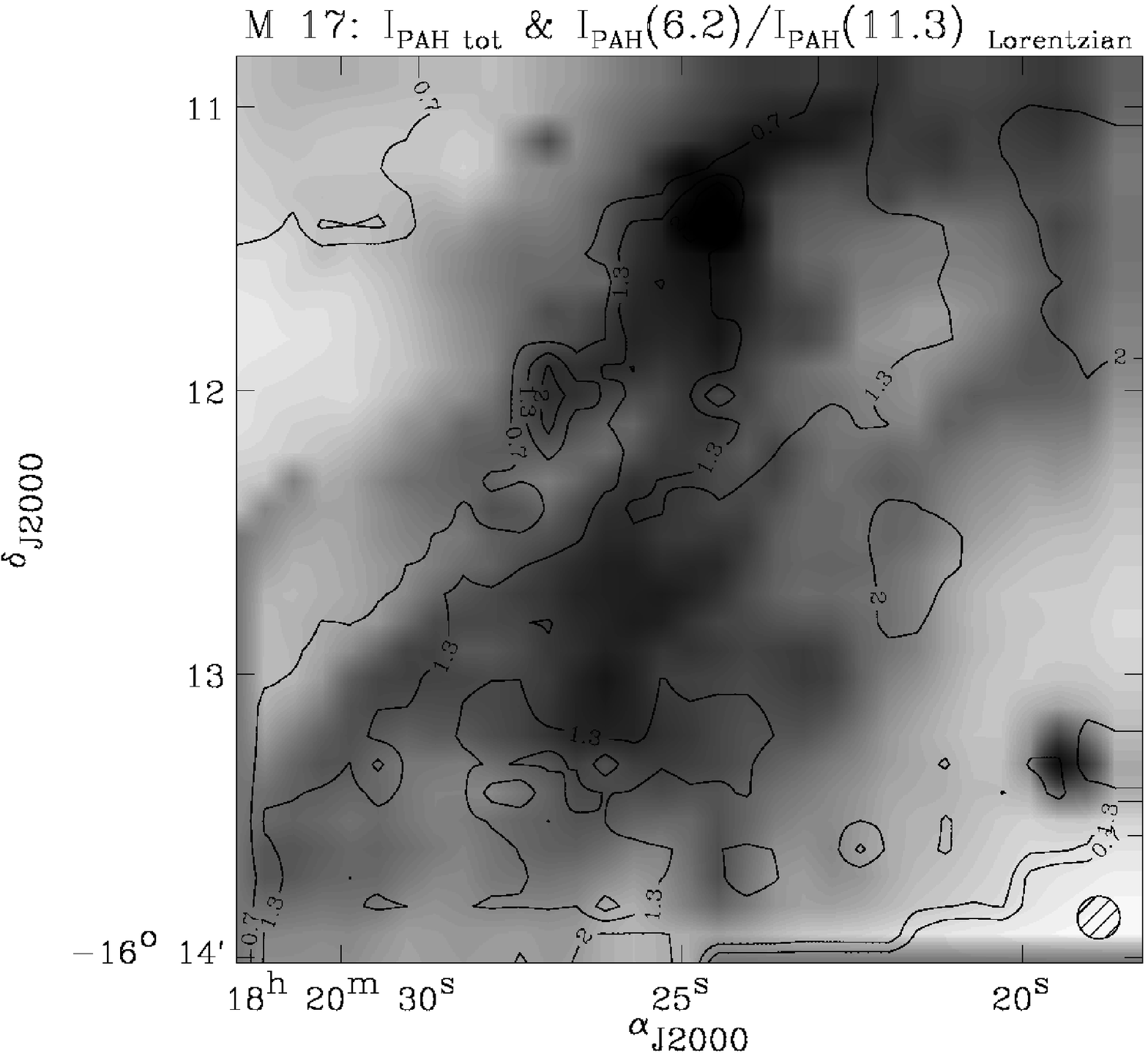}
  \caption{Spatial distribution of the PAHs in \M{17}.
           For each method, the image is the total PAH intensity, and the
           contours are the $\ipah{6.2}/\ipah{11.3}$ ratio.
           The shaded circle indicates the beam size.}
  \label{fig:imm17}
\end{figure*}

\begin{description}
  \item[\M{82}.]
    (\reffigs{fig:cormap1m82}-\ref{fig:cormap2m82})
    \reffig{fig:cormap1m82} shows that the correlations inside \M{82} are 
    remarkably well articulated. 
    The correlation coefficients (\reftab{tab:correl}) are $0.97$, for the
    correlations between $\ipah{6.2}/\ipah{11.3}$ and $\ipah{7.7}/\ipah{11.3}$.
    The band ratios inside \M{82} follow the general trends 
    observed for the integrated spectra very well.
    The centroid of the $7.7\mic$ feature is essentially fixed throughout \M{82}
    (\reftab{tab:correl}).
    The ratio $\ipah{6.2}/\ipah{11.3}$ follows the spatial distribution of the 
    PAH emission very well, which is maximum in the central bar,
    where the star formation occurs.
    With good confidence ($\rm S/N>6$), we detect low ratios in the 
    outer regions.
  \item[\IC{342}.]
    (\reffigs{fig:cormap1ic342}-\ref{fig:cormap2ic342})
    The morphology of the PAH emission (\reffig{fig:imic342}) follows the 
    \COio\ emission which traces 
    out the nuclear ring and ridge associated with the stellar bar 
    \citet{sakamoto99}.
    The spectra of this galaxy are much noisier than those of \M{82}.
    The correlations are very well defined, with a linear correlation
    coefficient of $\simeq0.8$ (\reftab{tab:correl}).
    The deviation of the relation involving the $8.6\mic$ feature with 
    the \frmet\ (middle right panel of \reffig{fig:cormap1ic342}), compared
    to the general correlation, is due to a systematic over-estimation of the 
    attenuation, in the central region.
    The average centroid of the $7.7\mic$ feature is identical for \IC{342} and
    \M{82} ($\langle\lambda_{7.7}\rangle\simeq 7.68\mic$; 
    \reftab{tab:correl}).
  \item[\M{51}.]
    (\reffigs{fig:cormap1m51}-\ref{fig:cormap2m51})
    This galaxy is more quiescent than \M{82} and \IC{342}.
    Its band ratios show less variation (\reffig{fig:cormap1m51}) 
    throughout the galaxy, although they are in good agreement with the
    general correlations of \reffig{fig:corglo1}.
    The correlation coefficients are $\simeq0.6$ (\reftab{tab:correl})
    slightly lower than those for \M{82} and \IC{342}, essentially because of
    the lower signal-to-noise ratio of the spectra.
    The spatial distribution of the $\ipah{6.2}/\ipah{11.3}$ ratio 
    (\reffig{fig:imm51}) follows the PAH emission remarkably well enhanced 
    along the spiral arms and the bulge.
    This band ratio reaches its maximum on the central ring and decreases 
    somewhat in the center.
  \item[\M{83}.]
    (\reffigs{fig:cormap1m83}-\ref{fig:cormap2m83})
    \M{83} is a spiral galaxy which is similar in several aspects to \M{51}.
    Its band ratios (\reffig{fig:cormap1m83}) and the spatial distribution
    of its PAH emission (\reffig{fig:imm83}) have the same 
    properties qualitatively.
    The average centroid of the $7.7\mic$ is identical to that for \M{51} and
    \M{83} ($\langle\lambda_{7.7}\rangle\simeq 7.72\mic$; 
    \reftab{tab:correl}).
  \item[\xxxdor.]
    (\reffigs{fig:cormap130dor}-\ref{fig:cormap230dor})
    The band ratio correlations measured inside \xxxdor\ are significantly
    dispersed (\reffig{fig:cormap130dor}).
    The results of the 2 different feature extraction methods are not 
    particularly consistent. 
    This is certainly due to the very low PAH-to-continuum ratio.
    Indeed, \reffig{fig:im30dor} shows that the spatial distribution of the 
    PAH emission, obtained with the \frmet, is different from that obtained
    with the \nlmet.
    In particular, the right panel of \reffig{fig:im30dor} shows 3 bright
    spots along the bar of \xxxdor, that are not seen on the left panel.
    After verification, these are fits of such low PAH-to-continuum ratio
    pixels that the widths of the PAH bands are not constrained at all.
    The \frmet\ determines very wide bands whose integrated intensities
    are significantly overestimated.
    For the same reason, the highest values of the $\ipah{6.2}/\ipah{11.3}$ 
    ratio, with the \nlmet\ ($\simeq2$; \reffig{fig:im30dor}), 
    correspond to low values, with the \frmet\ ($\lesssim0.3$).
  \item[\M{17}.]
    (\reffigs{fig:cormap1m17}-\ref{fig:cormap2m17})
    We encounter the same trouble in the \hii\ region part of \M{17}
    (upper left corner of the images in \reffig{fig:imm17}), as in 
    \xxxdor.
    For the same reasons, the $\ipah{6.2}/\ipah{11.3}$ ratio is maximum in 
    this region
    with the \nlmet\ ($\simeq2.5$; \reffig{fig:imm17}), while it is minimum 
    with the \frmet\ ($\lesssim0.7$).
    However, the PDR part has relatively high PAH-to-continuum ratios.
    The band ratio correlations are well defined inside \M{17}, except for
    $8.6\mic$, with the \nlmet\ (\reffig{fig:cormap1m17}).
  \item[The \orb.]
    (\reffigs{fig:cormap1orionBar}-\ref{fig:cormap2orionBar})
    The PAH-to-continuum ratio is on average higher in this region than in 
    \M{17}.
    Thus the fits are better, even along the ionization front.
    The correlation between band ratios (\reffig{fig:cormap1orionBar}) are
    relatively good with linear correlation coefficients around 0.9 
    (\reftab{tab:correl}).
    The plot of the consistency between the two methods
    (\reffig{fig:compmaporionBar}) shows a few outlying points.
    These fits are those of the pixels located inside the \hii\ region 
    (upper part of the images in \reffig{fig:imorionBar}).
    The spatial distribution of the $\ipah{6.2}/\ipah{11.3}$ band ratio
    decreases on average from the north, where the ionizing stars are located,
    to the south with the two methods.
    The average centroid of the $7.7\mic$ feature is similar in \M{17}
    and the \orb\ ($\langle\lambda_{7.7}\rangle\simeq 7.68\mic$; 
    \reftab{tab:correl}).
\end{description}

\begin{deluxetable}{llrrrrrrrr}
  \tabletypesize{\tiny}
  \rotate
  \tablecolumns{10}
  \tablewidth{0pc}
  \tablecaption{Results of the spectral decomposition.}
  \tablehead{
    \colhead{Name} & \colhead{} & \colhead{$\ipah{6.2}$} 
    & \colhead{$\ipah{7.7}$} 
    & \colhead{$\ipah{8.6}$} & \colhead{$\ipah{11.3}$}
    & \colhead{$\rm I_{PAH}$} & \colhead{$\rm I_{cont}$} 
    & \colhead{$\lambda_{7.7}$} & \colhead{Relia-} \\
    \colhead{}     & \colhead{}       & \colhead{[$10^{-15}\;\rm W\,m^{-2}$]} 
    & \colhead{[$10^{-15}\;\rm W\,m^{-2}$]}             
    & \colhead{[$10^{-15}\;\rm W\,m^{-2}$]}           
    & \colhead{[$10^{-15}\;\rm W\,m^{-2}$]}
    & \colhead{[$10^{-15}\;\rm W\,m^{-2}$]}
    & \colhead{[$10^{-15}\;\rm W\,m^{-2}$]}     
    & \colhead{[$\mu m$]} & \colhead{bility} \\}
  \startdata
    \haro{11}      & $\mathcal{S}$ & $0.90\pm0.09$ 
    & $1.81\pm0.18$     & $0.23\pm0.02$ & $3.78\pm2.15$
    & $3.84\pm2.15$     & $51.1\pm5.1$ 
    & \nodata & Low \\
                   & $\mathcal{L}$ & $1.59\pm0.16$ 
    & $5.23\pm0.52$     & $0.32\pm0.03$ & $1.28\pm0.13$
    & $11.2\pm0.6$      & $59.8\pm6.0$ 
    & $7.634\pm0.025$ & Low \\
    \smcb      & $\mathcal{S}$ & $16.2\pm2.4$ 
    & $24.1\pm3.6$     & $3.09\pm0.46$ & $16.0\pm2.4$
    & $73.8\pm35.7$     & $76.8\pm11.5$ 
    & \nodata & High \\
               & $\mathcal{L}$ & $33.8\pm5.1$ 
    & $70.2\pm1.1$     & $16.1\pm2.4$ & $51.0\pm7.6$
    & $205\pm14$     & $15.1\pm2.3$ 
    & $7.647\pm0.025$ & High \\
    \ngc{253}      & $\mathcal{S}$ & $148\pm4$ 
    & $293\pm11$     & $28.5\pm3.7$ & $105\pm4$
    & $604\pm346$     & $(1.74\pm0.09)\E{3}$ 
    & \nodata & High \\
                   & $\mathcal{L}$ & $532\pm28$ 
    & $(1.54\pm0.06)\E{3}$     & $36.4\pm1.7$ & $313\pm13$
    & $(3.42\pm0.08)\E{3}$     & $(1.49\pm0.07)\E{3}$ 
    & $7.707\pm0.025$ & High \\
    \ngc{253}~p      & $\mathcal{S}$ & $48.5\pm2.2$ 
    & $98.0\pm6.3$     & $10.2\pm0.2$ & $29.1\pm1.9$
    & $195\pm114$     & $823\pm44$ 
    & \nodata & Low \\
                   & $\mathcal{L}$ & $15.1\pm8.3$ 
    & $481\pm22$     & $113\pm6$ & $114\pm6$
    & $(1.09\pm0.03)\E{3}$     & $(1.00\pm0.05)\E{3}$ 
    & $7.702\pm0.025$ & High \\
    \ngc{253}~e      & $\mathcal{S}$ & $102\pm3$ 
    & $199\pm8$     & $19.1\pm0.6$ & $78.6\pm2.6$
    & $420\pm238$     & $974\pm66$ 
    & \nodata & High \\
                   & $\mathcal{L}$ & $324\pm22$ 
    & $(1.00\pm0.05)\E{3}$     & $228\pm13$ & $225\pm11$
    & $(2.21\pm0.06)\E{3}$     & $689\pm40$ 
    & $7.711\pm0.025$ & High \\
    \smcn      & $\mathcal{S}$ & $10.3\pm0.5$ 
    & $14.0\pm0.6$     & $0.34\pm0.05$ & $6.09\pm0.11$
    & $37.0\pm19.1$     & $290\pm41$ 
    & \nodata & Low \\
                   & $\mathcal{L}$ & $82.7\pm20.6$ 
    & $99.7\pm19.1$     & $57.5\pm12.9$ & $64.5\pm8.0$
    & $340\pm34$     & $203\pm28$ 
    & $7.619\pm0.025$ & Low \\
    \ngc{520}      & $\mathcal{S}$ & $8.38\pm0.37$ 
    & $19.6\pm0.7$     & $1.00\pm0.05$ & $4.50\pm0.10$
    & $36.6\pm22.0$     & $45.4\pm4.3$ 
    & \nodata & High \\
                   & $\mathcal{L}$ & $38.9\pm2.9$ 
    & $107\pm5$     & $29.2\pm1.9$ & $26.3\pm1.7$
    & $240\pm7$     & $72.3\pm5.2$ 
    & $7.748\pm0.025$ & High \\
    \ngc{613}      & $\mathcal{S}$ & $7.92\pm0.26$ 
    & $13.5\pm1.0$     & $1.49\pm0.01$ & $4.89\pm0.05$
    & $30.2\pm16.5$     & $74.0\pm31.9$ 
    & \nodata & Low \\
                   & $\mathcal{L}$ & $19.3\pm9.2$ 
    & $40.4\pm10.4$     & $7.41\pm3.16$ & $15.9\pm3.6$
    & $112\pm18$     & $55.4\pm26.4$ 
    & $7.693\pm0.025$ & Low \\
    \ngc{613}~p      & $\mathcal{S}$ & $4.03\pm0.12$ 
    & $8.35\pm0.48$     & $0.72\pm0.05$ & $3.88\pm0.20$
    & $18.3\pm10.1$     & $35.9\pm10.0$ 
    & \nodata & High \\
                   & $\mathcal{L}$ & $12.4\pm3.5$ 
    & $32.4\pm5.2$     & $6.94\pm1.67$ & $12.7\pm1.6$
    & $79.8\pm7.5$     & $33.9\pm10.8$ 
    & $7.703\pm0.025$ & High \\
    \ngc{891}      & $\mathcal{S}$ & $93.1\pm9.3$ 
    & $181\pm18$     & $17.0\pm1.7$ & $55.0\pm5.5$
    & $361\pm212$     & $304\pm30$ 
    & \nodata & High \\
                   & $\mathcal{L}$ & $207\pm21$ 
    & $605\pm61$     & $124\pm12$ & $148\pm15$
    & $(1.34\pm0.07)\E{3}$     & $87.1\pm8.7$ 
    & $7.705\pm0.025$ & High \\
    \ngc{1068}      & $\mathcal{S}$ & $55.5\pm5.9$ 
    & $126\pm35$     & $20.1\pm6.1$ & $59.3\pm33.6$
    & $287\pm152$     & $(4.80\pm0.37)\E{3}$ 
    & \nodata & Low \\
                   & $\mathcal{L}$ & $181\pm10$ 
    & $586\pm85$     & $131\pm16$ & $216\pm11$
    & $(1.33\pm0.09)\E{3}$     & $(4.75\pm0.23)\E{3}$ 
    & $7.706\pm0.025$ & Low \\
    \ngc{1068}~p      & $\mathcal{S}$ & $\lesssim10.2$ 
    & $\lesssim45$     & $5.1\pm4.8$ & $\lesssim31$
    & $573\pm274$     & $(3.59\pm0.34)\E{3}$ 
    & \nodata & Low \\
                   & $\mathcal{L}$ & $209\pm15$ 
    & $352\pm80$     & $86.0\pm15.0$ & $97.3\pm6.0$
    & $902\pm83$     & $(3.52\pm0.19)\E{3}$ 
    & $7.699\pm0.025$ & Low \\
    \ngc{1068}~e      & $\mathcal{S}$ & $48.9\pm7.1$ 
    & $101\pm35$     & $14.3\pm6.0$ & $51.1\pm36.4$
    & $237\pm125$     & $(1.01\pm0.30)\E{3}$ 
    & \nodata & Low \\
                   & $\mathcal{L}$ & $313\pm43$ 
    & $537\pm202$     & $108\pm40$ & $115\pm13$
    & $(1.28\pm0.21)\E{3}$     & $859\pm136$ 
    & $7.699\pm0.025$ & High \\
    \ngc{1097}      & $\mathcal{S}$ & $27.7\pm0.1$ 
    & $50.6\pm0.01$     & $5.25\pm0.15$ & $27.0\pm0.8$
    & $116\pm64$     & $229\pm32$ 
    & \nodata & High \\
                   & $\mathcal{L}$ & $56.4\pm6.9$ 
    & $166\pm12$     & $31.0\pm3.1$ & $61.7\pm4.7$
    & $410\pm18$     & $127\pm18$ 
    & $7.707\pm0.025$ & High \\
    \ngc{1097}~p      & $\mathcal{S}$ & $18.7\pm0.15$ 
    & $37.7\pm1.2$     & $4.25\pm0.32$ & $19.1\pm1.0$
    & $82.8\pm46.4$     & $156\pm13.8$ 
    & \nodata & High \\
                   & $\mathcal{L}$ & $42.8\pm3.2$ 
    & $134\pm7$     & $29.5\pm2.0$ & $48.4\pm3.1$
    & $319\pm9$     & $88.9\pm8.0$ 
    & $7.703\pm0.025$ & High \\
    \ngc{1140}      & $\mathcal{S}$ & $0.92\pm0.05$ 
    & $1.31\pm0.01$     & $0.07\pm0.02$ & $0.62\pm0.01$
    & $3.34\pm1.73$     & $6.96\pm2.16$ 
    & \nodata & Low \\
                   & $\mathcal{L}$ & $1.63\pm0.66$ 
    & $3.65\pm0.81$     & $0.45\pm0.13$ & $1.45\pm0.17$
    & $8.29\pm1.17$     & $5.57\pm1.71$ 
    & $7.684\pm0.025$ & Low \\
    \ngc{1365}      & $\mathcal{S}$ & $38.0\pm0.5$ 
    & $76.5\pm2.4$     & $9.21\pm0.27$ & $34.9\pm1.3$
    & $177\pm94$     & $367\pm31$ 
    & \nodata & High \\
                   & $\mathcal{L}$ & $89.3\pm7.1$ 
    & $253\pm13$     & $52.7\pm3.8$ & $79.5\pm4.4$
    & $566\pm16$     & $242\pm20$ 
    & $7.710\pm0.025$ & High \\
    \ngc{1365}~p      & $\mathcal{S}$ & $26.7\pm1.1$ 
    & $54.1\pm1.9$     & $6.22\pm0.21$ & $22.3\pm1.2$
    & $117\pm65$     & $253\pm15$ 
    & \nodata & High \\
                   & $\mathcal{L}$ & $57.7\pm3.1$ 
    & $182\pm8$     & $42.3\pm2.2$ & $53.2\pm2.5$
    & $148\pm10$     & $71.1\pm16.2$ 
    & $7.713\pm0.025$ & High \\
    \ngc{1365}~e      & $\mathcal{S}$ & $11.4\pm0.02$ 
    & $22.4\pm1.0$     & $2.98\pm0.10$ & $12.6\pm0.7$
    & $60.5\pm29.6$     & $113\pm24$ 
    & \nodata & High \\
                   & $\mathcal{L}$ & $18.3\pm4.9$ 
    & $61.2\pm7.6$     & $11.7\pm2.4$ & $30.8\pm3.1$
    & $148\pm10$     & $71.1\pm16.2$ 
    & $7.708\pm0.025$ & High \\
    \IC{342}      & $\mathcal{S}$ & $36.8\pm0.7$ 
    & $62.1\pm1.6$     & $9.69\pm0.27$ & $26.7\pm0.8$
    & $141\pm78$     & $270\pm15$ 
    & \nodata & High \\
                   & $\mathcal{L}$ & $62.2\pm3.9$ 
    & $182\pm8$     & $48.9\pm2.6$ & $60.7\pm2.6$
    & $441\pm10$     & $158\pm8$ 
    & $7.686\pm0.025$ & High \\
    \IC{342}~p      & $\mathcal{S}$ & $16.5\pm0.6$ 
    & $28.3\pm1.1$     & $3.90\pm0.16$ & $11.3\pm0.5$
    & $62.9\pm34.9$     & $135\pm6$ 
    & \nodata & High \\
                   & $\mathcal{L}$ & $30.4\pm1.6$ 
    & $78.0\pm3.2$     & $15.2\pm0.8$ & $26.9\pm1.2$
    & $188\pm4$     & $87.5\pm3.9$ 
    & $7.678\pm0.025$ & High \\
    \IC{342}~e      & $\mathcal{S}$ & $20.2\pm0.3$ 
    & $33.8\pm0.7$     & $5.79\pm0.14$ & $15.4\pm0.4$
    & $78.4\pm42.7$     & $135\pm12$ 
    & \nodata & High \\
                   & $\mathcal{L}$ & $32.8\pm3.3$ 
    & $105\pm6$     & $34.4\pm2.5$ & $33.5\pm1.8$
    & $257\pm8$     & $70.1\pm5.4$ 
    & $7.692\pm0.025$ & High \\
    \ngc{1569}      & $\mathcal{S}$ & $6.82\pm0.03$ 
    & $13.2\pm0.1$     & $0.50\pm0.08$ & $5.38\pm0.12$
    & $28.1\pm15.8$     & $114\pm24$ 
    & \nodata & Low \\
                   & $\mathcal{L}$ & $20.2\pm9.8$ 
    & $35.7\pm10.3$     & $3.21\pm1.35$ & $13.9\pm2.2$
    & $102\pm29$     & $91.4\pm14.9$ 
    & $7.669\pm0.025$ & Low \\
    \ngc{1569}~e      & $\mathcal{S}$ & $5.43\pm0.03$ 
    & $10.7\pm0.1$     & $0.33\pm0.08$ & $4.02\pm0.10$
    & $22.5\pm12.7$     & $55.8\pm23.9$ 
    & \nodata & Low \\
                   & $\mathcal{L}$ & $13.3\pm9.6$ 
    & $29.5\pm12.3$     & $3.06\pm1.98$ & $10.3\pm2.5$
    & $69.7\pm38.2$     & $46.2\pm15.2$ 
    & $7.679\pm0.025$ & Low \\
    \ngc{1808}      & $\mathcal{S}$ & $86.8\pm8.7$ 
    & $147\pm15$     & $15.0\pm1.5$ & $59.1\pm5.9$
    & $321\pm181$     & $497\pm50$ 
    & \nodata & High \\
                   & $\mathcal{L}$ & $180\pm18$ 
    & $487\pm49$     & $102\pm10$ & $185\pm19$
    & $(1.24\pm0.06)\E{3}$     & $274\pm27$ 
    & $7.710\pm0.025$ & High \\
    \orb~D8      & $\mathcal{S}$ & $338\pm3$ 
    & $632\pm7$     & $77.8\pm4.4$ & $301\pm12$
    & $(1.45\pm0.78)\E{3}$     & $(9.80\pm0.11)\E{3}$ 
    & \nodata & Low \\
                   & $\mathcal{L}$ & $590\pm13$ 
    & $912\pm5$     & $73.8\pm0.8$ & $501\pm12$
    & $(2.79\pm0.03)\E{3}$     & $(9.25\pm0.08)\E{3}$ 
    & $7.645\pm0.025$ & Low \\
    \orb~D5      & $\mathcal{S}$ & $532\pm3$ 
    & $(1.05\pm0.01)\E{3}$     & $131\pm2$ & $470\pm3$
    & $(2.33\pm1.28)\E{3}$     & $(5.84\pm0.13)\E{3}$ 
    & \nodata & High \\
                   & $\mathcal{L}$ & $981\pm18$ 
    & $(2.49\pm0.02)\E{3}$     & $580\pm7$ & $737\pm7$
    & $(5.97\pm0.04)\E{3}$     & $(4.94\pm0.07)\E{3}$ 
    & $7.670\pm0.025$ & High \\
    \orb~D2      & $\mathcal{S}$ & $107\pm10$ 
    & $186\pm1$     & $21.0\pm0.1$ & $102\pm8$
    & $468\pm240$     & $(2.67\pm0.12)\E{3}$ 
    & \nodata & Low \\
                   & $\mathcal{L}$ & $233\pm10$ 
    & $491\pm8$     & $70.1\pm1.6$ & $182\pm5$
    & $(1.58\pm0.03)\E{3}$     & $(2.18\pm0.10)\E{3}$ 
    & $7.680\pm0.025$ & High \\
    \xxxdor      & $\mathcal{S}$ & $205\pm1$ 
    & $405\pm8$     & $12.4\pm0.3$ & $98.2\pm3.5$
    & $78.4\pm46.6$     & $(1.39\pm0.06)\E{4}$ 
    & \nodata & Low \\
                   & $\mathcal{L}$ & $(1.12\pm0.07)\E{3}$ 
    & $(1.68\pm0.07)\E{3}$     & $781\pm36$ & $(1.04\pm0.04)\E{3}$
    & $(5.73\pm0.13)\E{3}$     & $(1.23\pm0.05)\E{4}$ 
    & $7.616\pm0.025$ & Low \\
    \xxxdor~p      & $\mathcal{S}$ & $25.3\pm1.2$ 
    & $56.0\pm0.1$     & $0.73\pm0.16$ & $11.0\pm0.4$
    & $99.7\pm62.6$     & $(2.84\pm0.11)\E{3}$ 
    & \nodata & Low \\
                   & $\mathcal{L}$ & $184\pm23$ 
    & $264\pm18$     & $122\pm8$ & $134\pm6$
    & $873\pm36$     & $(2.61\pm0.10)\E{3}$ 
    & $7.711\pm0.025$ & Low \\
    \xxxdor~e      & $\mathcal{S}$ & $164\pm1$ 
    & $315\pm4$     & $11.7\pm0.2$ & $81.2\pm2.9$
    & $624\pm366$     & $(9.62\pm0.40)\E{3}$ 
    & \nodata & Low \\
                   & $\mathcal{L}$ & $839\pm61$ 
    & $(1.34\pm0.07)\E{3}$     & $479\pm26$ & $798\pm33$
    & $(4.30\pm0.12)\E{3}$     & $(8.49\pm0.35)\E{3}$ 
    & $7.624\pm0.025$ & High \\
    \ngc{2023}      & $\mathcal{S}$ & $54.1\pm2.3$ 
    & $125\pm5$     & $8.25\pm0.11$ & $58.2\pm0.5$
    & $263\pm149$     & $263\pm55$ 
    & \nodata & High \\
                   & $\mathcal{L}$ & $120\pm9$ 
    & $284\pm8$     & $60.9\pm3.7$ & $106\pm7$
    & $705\pm19$     & $99.7\pm14$ 
    & $7.743\pm0.025$ & High \\
    \hen      & $\mathcal{S}$ & $6.74\pm0.67$ 
    & $14.7\pm1.5$     & $2.01\pm0.20$ & $5.90\pm0.59$
    & $31.1\pm17.4$     & $109\pm11$ 
    & \nodata & Low \\
                   & $\mathcal{L}$ & $14.4\pm1.4$ 
    & $36.1\pm3.6$     & $5.50\pm0.55$ & $14.2\pm1.4$
    & $85.9\pm4.3$     & $125\pm13$ 
    & $7.662\pm0.025$ & High \\
    \M{82}      & $\mathcal{S}$ & $992\pm36$ 
    & $(2.01\pm0.07)\E{3}$     & $231\pm8$ & $602\pm22$
    & $(4.09\pm2.34\E{3}$     & $(6.74\pm0.29)\E{3}$ 
    & \nodata & High \\
                   & $\mathcal{L}$ & $(2.68\pm0.10)\E{3}$ 
    & $(7.62\pm0.27)\E{3}$     & $(1.80\pm0.07)\E{3}$ & $(2.23\pm0.09)\E{3}$
    & $(1.72\pm0.03)\E{4}$     & $(5.82\pm0.24)\E{3}$ 
    & $7.697\pm0.025$ & High \\
    \M{82}~p      & $\mathcal{S}$ & $369\pm18$ 
    & $790\pm33$     & $88.4\pm3.5$ & $195\pm11$
    & $(1.53\pm0.90\E{3}$     & $(3.16\pm0.12)\E{3}$ 
    & \nodata & High \\
                   & $\mathcal{L}$ & $983\pm42$ 
    & $(2.92\pm0.11)\E{3}$     & $697\pm26$ & $701\pm27$
    & $(6.35\pm0.12)\E{3}$     & $(3.02\pm0.11)\E{3}$ 
    & $7.695\pm0.025$ & High \\
    \M{82}~e      & $\mathcal{S}$ & $819\pm30$ 
    & $(1.64\pm0.06)\E{3}$     & $192\pm7$ & $527\pm20$
    & $(3.39\pm1.93\E{3}$     & $(5.22\pm0.24)\E{3}$ 
    & \nodata & High \\
                   & $\mathcal{L}$ & $(1.99\pm0.08)\E{3}$ 
    & $(5.71\pm0.21)\E{3}$     & $(1.21\pm0.04)\E{3}$ & $(1.26\pm0.05)\E{3}$
    & $(1.21\pm0.02)\E{4}$     & $(2.82\pm0.13)\E{3}$ 
    & $7.701\pm0.025$ & High \\
    \ngc{3256}      & $\mathcal{S}$ & $24.7\pm0.3$ 
    & $45.5\pm4.2$     & $5.29\pm0.02$ & $19.4\pm0.2$
    & $97.4\pm55.6$     & $193\pm27$ 
    & \nodata & High \\
                   & $\mathcal{L}$ & $65.0\pm8.9$ 
    & $183\pm14$     & $48.6\pm5.9$ & $57.9\pm4.5$
    & $411\pm19$     & $181\pm20$ 
    & $7.701\pm0.025$ & High \\
    \ngc{3256}~p      & $\mathcal{S}$ & $6.16\pm0.40$ 
    & $11.0\pm0.3$     & $1.05\pm0.02$ & $4.37\pm0.25$
    & $23.1\pm13.4$     & $52.1\pm10.2$ 
    & \nodata & High \\
                   & $\mathcal{L}$ & $18.3\pm3.9$ 
    & $45.4\pm5.8$     & $9.88\pm2.31$ & $13.6\pm1.7$
    & $105\pm10$     & $51.7\pm7.3$ 
    & $7.703\pm0.025$ & High \\
    \ngc{3256}~e      & $\mathcal{S}$ & $14.5\pm1.5$ 
    & $24.5\pm0.1$     & $3.19\pm0.01$ & $12.3\pm0.1$
    & $55.7\pm32.1$     & $101\pm26$ 
    & \nodata & High \\
                   & $\mathcal{L}$ & $38.0\pm9.3$ 
    & $104\pm13$     & $29.2\pm6.0$ & $37.1\pm4.6$
    & $245\pm19$     & $89.8\pm18.8$ 
    & $7.699\pm0.025$ & High \\
    \mrk{33}      & $\mathcal{S}$ & $2.51\pm0.25$ 
    & $5.12\pm0.51$     & $0.31\pm0.03$ & $3.14\pm0.31$
    & $11.6\pm6.5$     & $27.8\pm2.8$ 
    & \nodata & High \\
                   & $\mathcal{L}$ & $3.88\pm0.39$ 
    & $14.2\pm0.14$     & $2.35\pm0.24$ & $4.94\pm0.49$
    & $30.4\pm1.6$     & $25.4\pm2.5$ 
    & $7.721\pm0.025$ & High \\
    \arp{299}      & $\mathcal{S}$ & $9.42\pm0.72$ 
    & $16.8\pm1.0$     & $1.93\pm0.01$ & $6.10\pm0.20$
    & $37.2\pm20.4$     & $223\pm27$ 
    & \nodata & Low \\
                   & $\mathcal{L}$ & $28.0\pm6.6$ 
    & $43.7\pm5.7$     & $4.99\pm0.82$ & $13.2\pm1.6$
    & $147\pm18$     & $413\pm36$ 
    & $7.669\pm0.025$ & Low \\
    \um{448}      & $\mathcal{S}$ & $0.83\pm0.08$ 
    & $1.39\pm0.14$     & $0.27\pm0.03$ & $0.83\pm0.08$
    & $3.59\pm1.85$     & $11.4\pm1.1$ 
    & \nodata & High \\
                   & $\mathcal{L}$ & $1.34\pm0.13$ 
    & $4.92\pm0.49$     & $0.90\pm0.09$ & $1.52\pm0.15$
    & $10.0\pm0.6$     & $9.95\pm1.0$ 
    & $7.670\pm0.025$ & High \\
    \IR{12331}      & $\mathcal{S}$ & $38.0\pm4.2$ 
    & $69.4\pm1.9$     & $7.11\pm1.20$ & $23.2\pm0.7$
    & $160\pm84$     & $869\pm62$ 
    & \nodata & Low \\
                   & $\mathcal{L}$ & $115\pm7$ 
    & $247\pm6$     & $34.5\pm3.2$ & $38.5\pm4.1$
    & $624\pm20$     & $765\pm28$ 
    & $7.696\pm0.025$ & High \\
    \ngc{4945}      & $\mathcal{S}$ & $65.7\pm1.3$ 
    & $182\pm7$     & $11.6\pm0.3$ & $40.7\pm0.3$
    & $310\pm198$     & $393\pm35$ 
    & \nodata & High \\
                   & $\mathcal{L}$ & $260\pm16$ 
    & $(1.15\pm0.05)\E{3}$     & $186\pm11$ & $222\pm12$
    & $(2.15\pm0.05)\E{3}$     & $389\pm29$ 
    & $7.713\pm0.025$ & High \\
    \ngc{4945}~p      & $\mathcal{S}$ & $29.6\pm0.1$ 
    & $94.2\pm3.9$     & $3.72\pm0.03$ & $10.0\pm0.1$
    & $149\pm100$     & $156\pm12$ 
    & \nodata & High \\
                   & $\mathcal{L}$ & $322\pm15$ 
    & $(1.15\pm0.05)\E{3}$     & $195\pm11$ & $208\pm13$
    & $(2.22\pm0.05)\E{3}$     & $370\pm23$ 
    & $7.709\pm0.025$ & High \\
    \ngc{4945}~e      & $\mathcal{S}$ & $36.5\pm0.7$ 
    & $87.7\pm3.6$     & $7.89\pm0.19$ & $30.7\pm0.1$
    & $164\pm100$     & $237\pm31$ 
    & \nodata & High \\
                   & $\mathcal{L}$ & $90.1\pm9.6$ 
    & $413\pm23$     & $87.5\pm7.7$ & $99.1\pm6.4$
    & $826\pm28$     & $136\pm16$ 
    & $7.713\pm0.025$ & High \\
    \cenA      & $\mathcal{S}$ & $17.0\pm0.2$ 
    & $26.0\pm0.1$     & $2.66\pm0.14$ & $13.6\pm0.2$
    & $61.1\pm34.0$     & $219\pm33$ 
    & \nodata & Low \\
                   & $\mathcal{L}$ & $30.2\pm5.8$ 
    & $112\pm13$     & $27.2\pm1.3$ & $50.3\pm5.0$
    & $282\pm16$     & $219\pm30$ 
    & $7.696\pm0.025$ & High \\
    \cenA~e      & $\mathcal{S}$ & $14.3\pm0.1$ 
    & $22.7\pm0.1$     & $2.07\pm0.12$ & $10.7\pm0.1$
    & $51.3\pm29.0$     & $97.2\pm32.0$ 
    & \nodata & Low \\
                   & $\mathcal{L}$ & $23.8\pm7.4$ 
    & $87.3\pm14.3$     & $21.8\pm5.8$ & $36.7\pm5.7$
    & $219\pm19$     & $64.3\pm20.4$ 
    & $7.695\pm0.025$ & Low \\
    \M{51}      & $\mathcal{S}$ & $64.1\pm0.9$ 
    & $123\pm3$     & $12.9\pm0.4$ & $57.2\pm1.7$
    & $264\pm151$     & $435\pm49$ 
    & \nodata & High \\
                   & $\mathcal{L}$ & $125\pm10$ 
    & $398\pm20$     & $76.0\pm5.1$ & $131\pm8$
    & $843\pm25$     & $251\pm35$ 
    & $7.716\pm0.025$ & High \\
    \M{51}~p      & $\mathcal{S}$ & $22.6\pm0.5$ 
    & $42.7\pm0.8$     & $4.93\pm0.1$ & $21.0\pm0.5$
    & $93.6\pm52.9$     & $161\pm17$ 
    & \nodata & High \\
                   & $\mathcal{L}$ & $46.0\pm3.9$ 
    & $145\pm8$     & $28.8\pm2.1$ & $49.5\pm2.7$
    & $326\pm10$     & $82.4\pm10.6$ 
    & $7.721\pm0.025$ & High \\
    \M{51}~e      & $\mathcal{S}$ & $34.9\pm0.4$ 
    & $68.4\pm1.0$     & $6.76\pm0.19$ & $30.7\pm0.8$
    & $145\pm83$     & $231\pm41$ 
    & \nodata & High \\
                   & $\mathcal{L}$ & $61.6\pm8.4$ 
    & $212\pm15$     & $41.7\pm4.4$ & $71.5\pm5.4$
    & $436\pm19$     & $141\pm32$ 
    & $7.713\pm0.025$ & High \\
    \M{83}      & $\mathcal{S}$ & $152\pm3$ 
    & $291\pm9$     & $33.9\pm1.4$ & $130\pm5$
    & $656\pm357$     & $901\pm174$ 
    & \nodata & High \\
                   & $\mathcal{L}$ & $304\pm22$ 
    & $810\pm48$     & $147\pm14$ & $343\pm30$
    & $(1.79\pm0.06)\E{3}$     & $622\pm154$ 
    & $7.708\pm0.025$ & High \\
    \M{83}~p      & $\mathcal{S}$ & $60.4\pm2.8$ 
    & $117\pm5$     & $14.2\pm0.5$ & $52.9\pm2.7$
    & $260\pm143$     & $458\pm28$ 
    & \nodata & High \\
                   & $\mathcal{L}$ & $163\pm8$ 
    & $430\pm18$     & $103\pm5$ & $119\pm5$
    & $931\pm22$     & $311\pm19$ 
    & $7.707\pm0.025$ & High \\
    \M{83}~e      & $\mathcal{S}$ & $99.4\pm1.9$ 
    & $187\pm6$     & $21.1\pm0.9$ & $84.3\pm3.2$
    & $426\pm230$     & $501\pm156$ 
    & \nodata & High \\
                   & $\mathcal{L}$ & $163\pm16$ 
    & $540\pm40$     & $105\pm14$ & $247\pm28$
    & $(1.21\pm0.06)\E{3}$     & $180\pm94$ 
    & $7.712\pm0.025$ & High \\
    Circinus      & $\mathcal{S}$ & $79.1\pm0.5$ 
    & $211\pm6$     & $24.8\pm0.6$ & $89.4\pm5.5$
    & $421\pm244$     & $(22.8\pm1.0)\E{3}$ 
    & \nodata & Low \\
                   & $\mathcal{L}$ & $347\pm18$ 
    & $(1.05\pm0.04)\E{3}$     & $173\pm8$ & $494\pm20$
    & $(2.55\pm0.05)\E{3}$     & $(3.58\pm0.14)\E{3}$ 
    & $7.684\pm0.025$ & High \\
    Circinus~e      & $\mathcal{S}$ & $64.2\pm0.4$ 
    & $161\pm4$     & $20.4\pm0.4$ & $70.5\pm5.1$
    & $330\pm188$     & $945\pm67$ 
    & \nodata & Low \\
                   & $\mathcal{L}$ & $261\pm19$ 
    & $774\pm38$     & $187\pm11$ & $385\pm19$
    & $(1.95\pm0.05)\E{3}$     & $(1.13\pm0.06)\E{3}$ 
    & $7.688\pm0.025$ & High \\
    \arp{220}      & $\mathcal{S}$ & $0.48\pm0.03$ 
    & $6.33\pm0.37$     & $(3.63\pm1.72)\E{-2}$ & $1.50\pm0.02$
    & $11.5\pm7.1$     & $531\pm214$ 
    & \nodata & Low \\
                   & $\mathcal{L}$ & $\lesssim59$ 
    & $68.5\pm23.7$     & $43.7\pm32.4$ & $36.2\pm21.1$
    & $259\pm76$     & $258\pm62$ 
    & $7.704\pm0.025$ & Low \\
    \arp{220}~p      & $\mathcal{S}$ & $0.63\pm0.14$ 
    & $4.00\pm0.05$     & $\lesssim4.4\E{-2}$ & $0.90\pm0.01$
    & $6.64\pm4.23$     & $34.9\pm9.9$ 
    & \nodata & Low \\
                   & $\mathcal{L}$ & $31.6\pm21.7$ 
    & $68.7\pm17.6$     & $78.3\pm38.7$ & $57.0\pm28.8$
    & $334\pm73$     & $247\pm41$ 
    & $7.715\pm0.025$ & Low \\
    \IR{15384}      & $\mathcal{S}$ & $315\pm2$ 
    & $608\pm6$     & $51.4\pm2.8$ & $142\pm3$
    & $(1.22\pm0.70)\E{3}$     & $(3.60\pm0.07)\E{3}$ 
    & \nodata & High \\
                   & $\mathcal{L}$ & $711\pm13$ 
    & $(2.13\pm0.02)\E{3}$     & $304\pm0.07$ & $305\pm4$
    & $(4.32\pm0.03)\E{3}$     & $(2.89\pm0.03)\E{3}$ 
    & $7.674\pm0.025$ & High \\
    \ngc{6240}      & $\mathcal{S}$ & $4.22\pm0.06$ 
    & $9.05\pm0.01$     & $0.91\pm0.05$ & $3.33\pm0.04$
    & $18.2\pm10.6$     & $51.2\pm6.1$ 
    & \nodata & High \\
                   & $\mathcal{L}$ & $14.1\pm2.6$ 
    & $43.2\pm4.0$     & $6.42\pm1.05$ & $17.6\pm1.5$
    & $104\pm6$     & $96.0\pm8.5$ 
    & $7.725\pm0.025$ & High \\
    \ngc{6240}~p      & $\mathcal{S}$ & $2.88\pm0.13$ 
    & $5.78\pm0.17$     & $0.49\pm0.05$ & $1.94\pm0.02$
    & $11.4\pm6.8$     & $32.6\pm3.0$ 
    & \nodata & High \\
                   & $\mathcal{L}$ & $10.3\pm1.3$ 
    & $30.1\pm2.0$     & $3.78\pm0.43$ & $9.21\pm0.63$
    & $66.3\pm2.8$     & $59.4\pm4.4$ 
    & $7.722\pm0.025$ & High \\
    \IR{18317}      & $\mathcal{S}$ & $292\pm17$ 
    & $527\pm4$     & $44.7\pm0.1$ & $122\pm5$
    & $(1.07\pm0.62)\E{3}$     & $(3.70\pm0.87)\E{3}$ 
    & \nodata & Low \\
                   & $\mathcal{L}$ & $671\pm14$ 
    & $(1.99\pm0.02)\E{3}$     & $256\pm3$ & $275\pm4$
    & $(4.18\pm0.03)\E{3}$     & $(2.93\pm0.05)\E{3}$ 
    & $7.681\pm0.025$ & High \\
    \ngc{6946}      & $\mathcal{S}$ & $81.4\pm2.8$ 
    & $156\pm6$     & $16.0\pm0.5$ & $62.1\pm2.4$
    & $322\pm188$     & $433\pm47$ 
    & \nodata & High \\
                   & $\mathcal{L}$ & $173\pm9$ 
    & $527\pm24$     & $88.7\pm5.6$ & $165\pm11$
    & $(1.10\pm0.03)\E{3}$     & $192\pm24$ 
    & $7.713\pm0.025$ & High \\
    \ngc{6946}~p      & $\mathcal{S}$ & $24.8\pm1.8$ 
    & $51.7\pm3.3$     & $4.62\pm0.04$ & $17.6\pm1.3$
    & $100\pm60$     & $162\pm10$ 
    & \nodata & High \\
                   & $\mathcal{L}$ & $77.9\pm4.1$ 
    & $211\pm8$     & $42.3\pm2.1$ & $58.4\pm4.3$
    & $462\pm11$     & $85.5\pm4.6$ 
    & $7.714\pm0.025$ & High \\
    \ngc{6946}~e      & $\mathcal{S}$ & $56.8\pm1.8$ 
    & $105\pm4$     & $11.4\pm0.4$ & $44.6\pm1.8$
    & $222\pm128$     & $273\pm39$ 
    & \nodata & High \\
                   & $\mathcal{L}$ & $105\pm6$ 
    & $327\pm17$     & $52.6\pm4.0$ & $119\pm9$
    & $687\pm21$     & $122\pm22$ 
    & $7.712\pm0.025$ & High \\
    \ngc{7027}      & $\mathcal{S}$ & $1.00\pm0.10$ 
    & $1.82\pm0.18$     & $0.14\pm0.02$ & $1.52\pm0.15$
    & $6.84\pm0.68$     & $33.4\pm3.3$ 
    & \nodata & Low \\
                   & $\mathcal{L}$ & $1.38\pm0.14$ 
    & $4.05\pm0.41$     & $0.48\pm0.05$ & $4.34\pm0.43$
    & $17.0\pm0.7$     & $27.0\pm2.7$ 
    & $7.774\pm0.025$ & High \\
    \IR{22308}      & $\mathcal{S}$ & $102\pm2$ 
    & $236\pm3$     & $25.0\pm1.0$ & $85.1\pm1.3$
    & $485\pm273$     & $906\pm75$ 
    & \nodata & High \\
                   & $\mathcal{L}$ & $191\pm5$ 
    & $572\pm10.9$     & $100\pm2$ & $153\pm10$
    & $(1.31\pm0.02)\E{3}$     & $629\pm26$ 
    & $7.684\pm0.025$ & High \\
    \IR{23030}      & $\mathcal{S}$ & $75.4\pm8.7$ 
    & $136\pm3$     & $17.4\pm0.4$ & $54.0\pm2.3$
    & $301\pm166$     & $(1.36\pm0.06)\E{3}$ 
    & \nodata & Low \\
                   & $\mathcal{L}$ & $135\pm5$ 
    & $188\pm5$     & $21.5\pm0.6$ & $129\pm5$
    & $736\pm15$     & $(1.13\pm0.03)\E{3}$ 
    & $7.649\pm0.025$ & High \\
    \IR{23133}      & $\mathcal{S}$ & $233\pm6$ 
    & $496\pm2$     & $66.8\pm0.4$ & $158\pm1$
    & $997\pm575$     & $(2.88\pm0.10)\E{3}$ 
    & \nodata & High \\
                   & $\mathcal{L}$ & $493\pm16$ 
    & $(1.18\pm0.02)\E{3}$     & $216\pm2$ & $(2.71\pm0.03)\E{3}$
    & $(2.71\pm0.03)\E{3}$     & $(2.46\pm0.04)\E{3}$ 
    & $7.674\pm0.025$ & High \\
    \IR{23128}      & $\mathcal{S}$ & $0.82\pm0.18$ 
    & $2.43\pm0.12$     & $0.27\pm0.09$ & $1.14\pm0.02$
    & $6.35\pm3.09$     & $26.4\pm5.5$ 
    & \nodata & Low \\
                   & $\mathcal{L}$ & $5.24\pm1.95$ 
    & $13.1\pm2.6$     & $2.24\pm0.63$ & $3.38\pm0.54$
    & $35.7\pm5.2$     & $35.4\pm5.5$ 
    & $7.681\pm0.025$ & Low \\
    \IR{23128}~p      & $\mathcal{S}$ & $0.79\pm0.02$ 
    & $1.69\pm0.07$     & $0.25\pm0.01$ & $0.74\pm0.01$
    & $4.23\pm2.10$     & $11.9\pm2.6$ 
    & \nodata & Low \\
                   & $\mathcal{L}$ & $1.91\pm0.56$ 
    & $7.23\pm1.06$     & $2.57\pm0.56$ & $1.47\pm0.20$
    & $20.8\pm1.6$     & $6.66\pm0.90$ 
    & $7.688\pm0.025$ & Low \\
    \ngc{7714}      & $\mathcal{S}$ & $3.17\pm0.32$ 
    & $6.00\pm0.60$     & $0.90\pm0.09$ & $2.78\pm0.28$
    & $13.8\pm7.4$     & $43.5\pm4.4$ 
    & \nodata & High \\
                   & $\mathcal{L}$ & $5.15\pm0.51$ 
    & $14.9\pm1.5$     & $2.75\pm0.28$ & $5.34\pm0.53$
    & $34.8\pm1.7$     & $36.4\pm3.6$ 
    & $7.681\pm0.025$ & High \\
  \enddata
  \label{tab:intens}
  \tablecomments{We consider that a given fit has a ``high'' reliability, if 
                 $S/N\geq6$ at $\lambda=7.7\mic$, 
                 and $\rm I_{PAH}/I_{cont}\geq0.3$ with the \nlmet,
                 and $\rm I_{PAH}/I_{cont}\geq0.5$ with the \frmet.
                 Otherwise, the reliability is ``low''.}
\end{deluxetable}
\clearpage

\begin{figure}[htbp]
  \centering
  \includegraphics[width=0.48\textwidth]{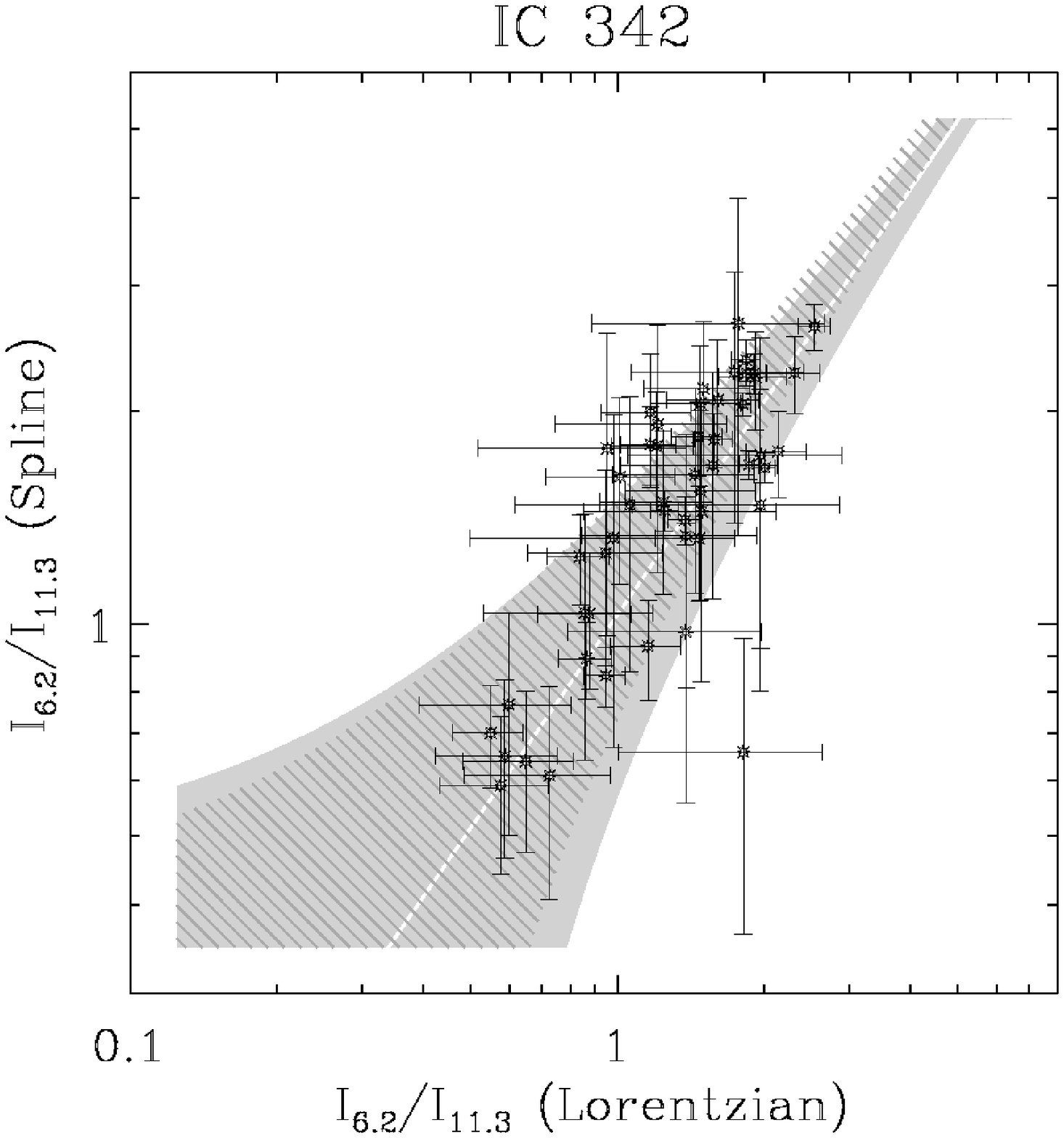}
  \caption{Comparison between the two methods in \IC{342}.
           The same symbol conventions are adopted as in 
           \reffig{fig:cormap1m82}.}
  \label{fig:compmapic342}
\end{figure}
\clearpage
\begin{figure}[htbp]
  \centering
  \includegraphics[width=0.48\textwidth]{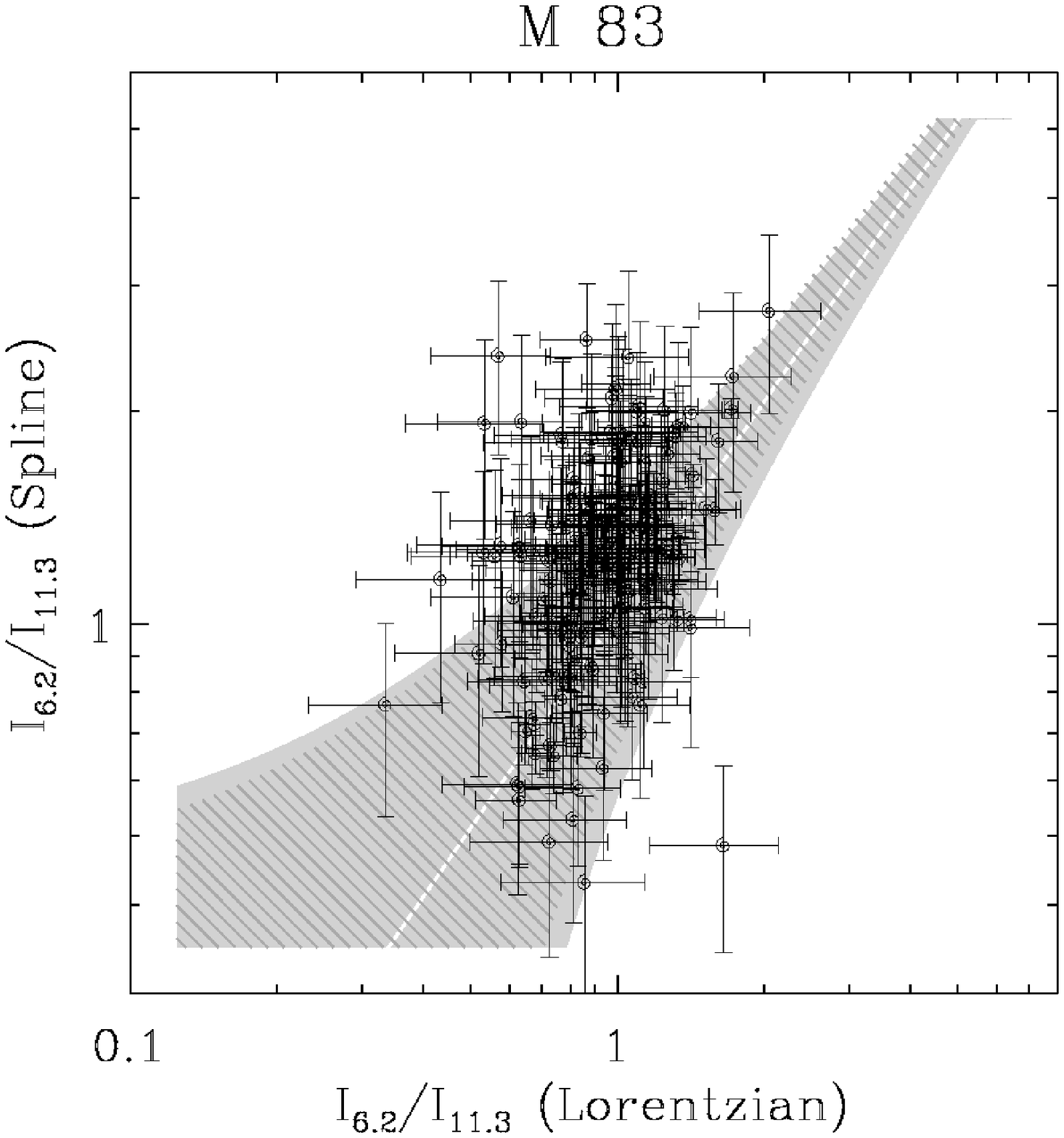}
  \caption{Comparison between the two methods in \M{83}.
           The same symbol conventions are adopted as in 
           \reffig{fig:cormap1m82}.}
  \label{fig:compmapm83}
\end{figure}
\clearpage
\begin{figure}[htbp]
  \centering
  \includegraphics[width=0.48\textwidth]{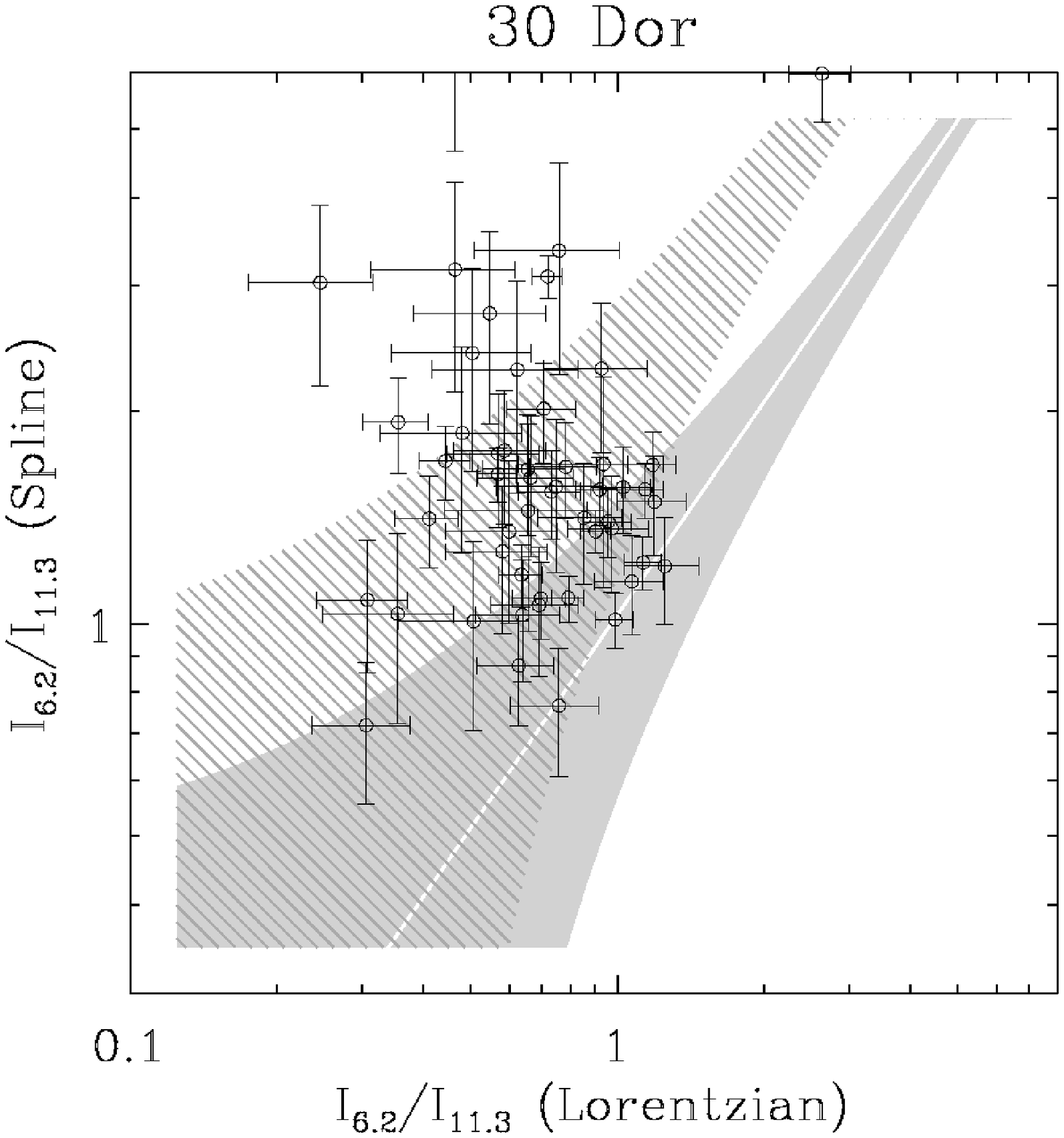}
  \caption{Comparison between the two methods in \xxxdor.
           The same symbol conventions are adopted as in 
           \reffig{fig:cormap1m82}.}
  \label{fig:compmap30dor}
\end{figure}
\clearpage
\clearpage
\begin{figure}[htbp]
  \centering
  \includegraphics[width=0.48\textwidth]{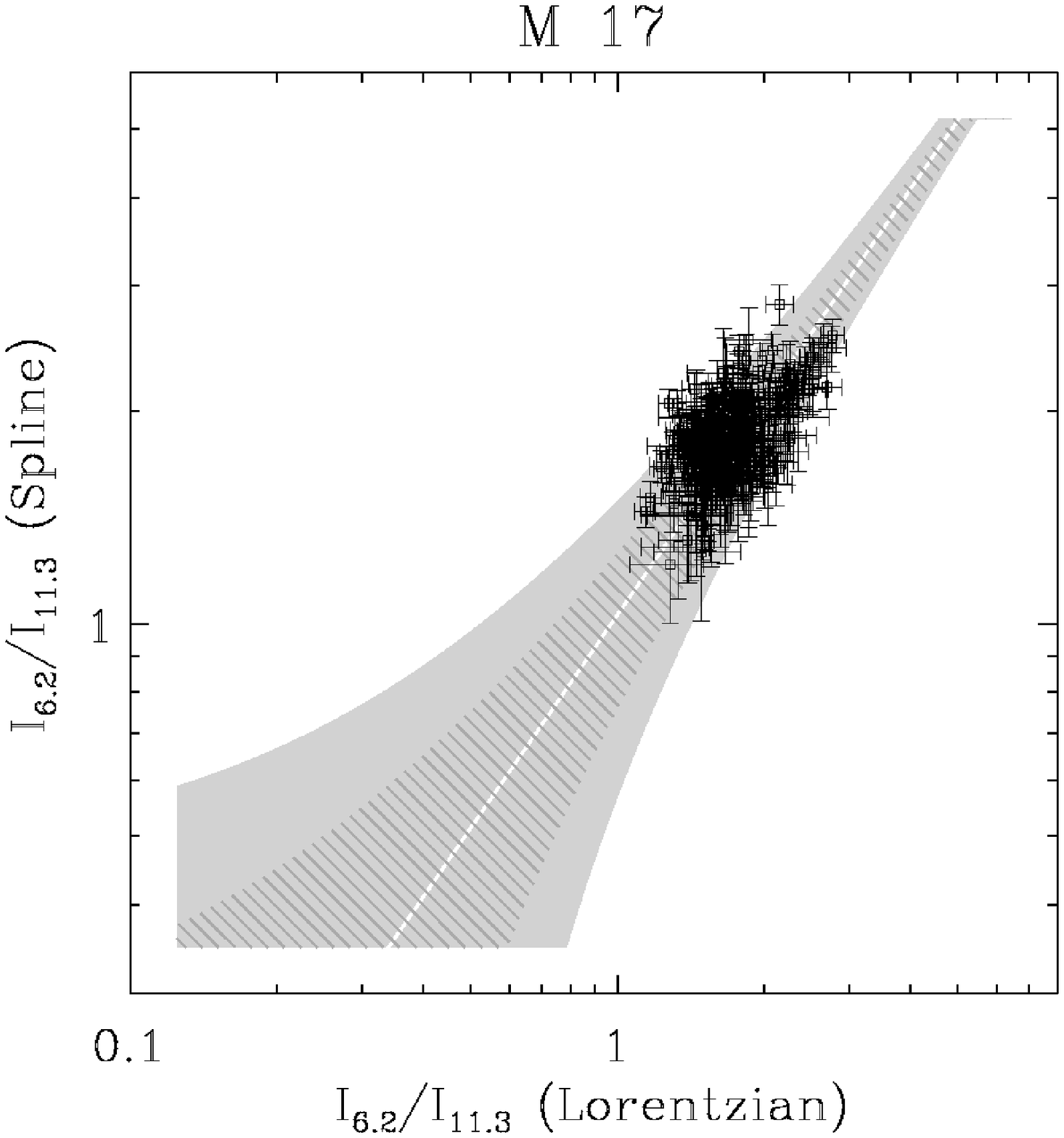}
  \caption{Comparison between the two methods in \M{17}.
           The same symbol conventions are adopted as in 
           \reffig{fig:cormap1m82}.}
  \label{fig:compmapm17}
\end{figure}
\clearpage


\acknowledgments

We are grateful to Lou Allamandola and Henrik Spoon for in-depth useful 
discussions and detailed comments on our work.
We thank the anonymous referee of this paper for comments that improved its
quality.
This work was performed, while F.~G. held a National Research 
Council research associateship award at NASA GSFC, 
and later a NASA Postdoctoral Program fellowship at NASA GSFC.
This study is based essentially on observations made with ISO, an ESA project 
with instruments funded by ESA Member States (especially the PI countries: 
France, Germany, the Netherlands and the United Kingdom) and with the participation of ISAS and NASA.
It is also based in part on observations made with the \spitzST, 
which is operated by the Jet Propulsion Laboratory, California 
Institute of Technology under a contract with NASA.

{\it Facilities:} \facility{ISO (CAM)}, 
                  \facility{ISO (SWS)}, 
                  \facility{Spitzer (IRS)}.

\bibliographystyle{/Users/Fred/Documents/Astro/TeXstyle/Packages_AAS/aas}
\bibliography{/Users/Fred/Documents/Astro/TeXstyle/references}

\end{document}